\documentclass[longauth]{aa} % for the long lists of affiliations 
\usepackage{graphicx}
\usepackage{lscape}
\usepackage{longtable}
\usepackage{txfonts}
\usepackage{enumitem}
\usepackage{pifont}
\usepackage{hyperref}
\hypersetup{colorlinks,linkcolor=red,citecolor=blue,urlcolor=blue}
\def\teff{$T_\mathrm{eff}$} % effective temperature
\def\logg{$\log g$}
\def\feh{[Fe/H]}

\begin{document}

   \title{The Gaia-ESO Survey: The DR5 analysis of the medium-resolution GIRAFFE and high-resolution UVES spectra of FGK-type stars\thanks{Based on observations collected at the ESO telescopes under programme 188.B3002, the Gaia-ESO Public Spectroscopic Survey.}}

   \subtitle{}

    % intial authors list...
    \author{C. C. Worley\inst{\ref{ioa},\ref{ucan}}
    \fnmsep\thanks{ccworley@ast.cam.ac.uk, clare.worley@canterbury.ac.nz}
    \and R. Smiljanic\inst{\ref{ncac}}
    % Nodes list...
    \and L. Magrini\inst{\ref{oarcetri}}
    \and A. Frasca\inst{\ref{ocatania}}
    \and E. Franciosini\inst{\ref{oarcetri}}
    \and D. Montes\inst{\ref{ucmadrid}}
    \and D. K. Feuillet\inst{\ref{olund2}}
    \and H. M. Tabernero\inst{\ref{cda-vc}}
    \and J. I. Gonz{\'a}lez Hern{\'a}ndez\inst{\ref{iac}}
    \and S. Villanova\inst{\ref{uconcepcion}}
    \and \v{S}. Mikolaitis\inst{\ref{uvilnius-ao}}
    \and K. Lind\inst{\ref{ustockholm}}
    \and G. Tautvai{\v s}ien{\. e}\inst{\ref{uvilnius}}
    \and A. R. Casey\inst{\ref{umonash}}
    \and A. J. Korn\inst{\ref{uuppsala-oa}}
    \and P. Bonifacio\inst{\ref{oparis-m}}
    \and C. Soubiran\inst{\ref{ubordeaux}}
    \and E. Caffau\inst{\ref{oparis-m}}
    \and G. Guiglion\inst{\ref{mpia}}
    \and T. Merle\inst{\ref{ulbruxelles}}
    % PO list...
    \and A. Hourihane\inst{\ref{ioa}}
    \and A. Gonneau\inst{\ref{ioa}}
    \and P. Fran{\c c}ois\inst{\ref{oparis},\ref{upjv}}
    \and S. Randich\inst{\ref{oarcetri}}
    \and G. Gilmore\inst{\ref{ioa},\ref{crete}}
    % Reduction list...
    \and J. R. Lewis$^\dagger$\inst{\ref{ioa}}
    \and D. N. A. Murphy\inst{\ref{ioa}}
    % Radial Velocity list...
    \and R. D. Jeffries\inst{\ref{ukeele}}
    \and S. E. Koposov\inst{\ref{ifa-roe},\ref{ioa}}
    % WG Leads list...
    \and R. Blomme\inst{\ref{obelgium}}
    \and A. C. Lanzafame\inst{\ref{ucatania}}
    \and T. Bensby\inst{\ref{olund1}}
    \and A. Bragaglia\inst{\ref{obologna}}
    \and E. J. Alfaro\inst{\ref{iaa}}
    \and N. A. Walton\inst{\ref{ioa}}
    % Steering Committee list...
    \and A. Vallenari\inst{\ref{opadova}}
    \and T. Prusti\inst{\ref{esa}}
    % Builders list...
    \and K. Biazzo\inst{\ref{oroma}}
    \and P. Jofr{\'e}\inst{\ref{udp}}
    \and S. Zaggia\inst{\ref{opadova}}
    \and U. Heiter\inst{\ref{uuppsala-oa}}
    % Rest of Co-A list...
    \and E. Marfil\inst{\ref{hamstrn},\ref{ucmadrid}}
    \and F. Jim{\'e}nez-Esteban\inst{\ref{cda-vc}}
    \and M. L. Guti{\'e}rrez Albarr{\'a}n\inst{\ref{ucmadrid}}
    \and L. Morbidelli\inst{\ref{oarcetri}}
    }    
    \institute{Institute of Astronomy, University of Cambridge, Madingley Road, Cambridge CB3 0HA, United Kingdom \\ \email{ccworley@ast.cam.ac.uk} \label{ioa}
    \and School of Physical and Chemical Sciences --- Te Kura Mat\={u}, University of Canterbury, Private Bag 4800, Christchurch 8140, New Zealand \label{ucan}
    \and Nicolaus Copernicus Astronomical Center, Polish Academy of Sciences, ul. Bartycka 18, 00-716, Warsaw, Poland\label{ncac}
    \and INAF - Osservatorio Astrofisico di Arcetri, Largo E. Fermi, 5, 50125, Firenze, Italy\label{oarcetri}
    \and Observational Astrophysics, Division of Astronomy and Space Physics, Department of Physics and Astronomy, Uppsala University, Box 516, 75120 Uppsala, Sweden\label{uuppsala-oa}
    \and ROB - Royal Observatory of Belgium, Ringlaan 3, B-1180 Brussels, Belgium\label{obelgium}
    \and Dipartimento di Fisica e Astronomia, Sezione Astrofisica, Universit{\'a} di Catania, via S. Sofia 78, 95123, Catania, Italy\label{ucatania}
    \and School of Physics \& Astronomy, Monash University, Wellington Road, Clayton 3800, Victoria, Australia\label{umonash}
    \and INAF - Osservatorio Astrofisico di Catania, Via S. Sofia 78, 95123 Catania, Italy\label{ocatania}
    \and GEPI, Observatoire de Paris, Universit{\'e} PSL, CNRS,  5 Place Jules Janssen, 92190 Meudon, France\label{oparis-m}
    \and Laboratoire d'astrophysique de Bordeaux, Univ. Bordeaux, CNRS, B18N, all{\'e}e Geoffroy Saint-Hilaire, 33615 Pessac, France\label{ubordeaux}
    \and Departamento de F{\'i}sica de la Tierra y Astrof{\'i}sica \& IPARCOS-UCM (Instituto de F{\'i}sica de Partículas y del Cosmos de la UCM), Facultad de Ciencias F{\'i}sicas, Universidad Complutense de Madrid, E-28040 Madrid, Spain\label{ucmadrid}
    \and Lund Observatory, Division of Astrophysics, Department of Physics, Lund University, Box 43, SE-22100 Lund, Sweden\label{olund1}
    \and Lund Observatory, Department of Geology, Department of Physics, Lund University, Box 43, SE-22100 Lund, Sweden\label{olund2}
    \and Institute of Theoretical Physics and Astronomy, Vilnius University, Sauletekio av. 3, LT-10257 Vilnius, Lithuania\label{uvilnius}
    \and Department of Astronomy, Stockholm University, AlbaNova University Center, SE-106 91 Stockholm, Sweden\label{ustockholm}
    \and Max-Planck-Institut f{\"u}r Astronomie, K{\"o}nigstuhl 17, D-69117 Heidelberg, Germany\label{mpia}
    \and Centro de Astrobiolog{\'{\i}}a, CSIC-INTA, Camino bajo del castillo s/n, E-28692, Villanueva de la Can{\~a}da, Madrid, Spain.\label{cda-vc}
    \and Instituto de Astrof{\'i}sica de Canarias, V{\'i}a L{\'a}ctea s/n, E-38205 La Laguna, Tenerife, Spain\label{iac}
    \and Departamento de Astronom{\'i}a, Casilla 160-C, Universidad de Concepci{\'o}n, Concepci{\'o}n, Chile\label{uconcepcion}
    \and Astronomical  Observatory,  Institute  of  Theoretical  Physics  and  Astronomy,  Vilnius  University,  Sauletekio  av.  3,  10257  Vilnius, Lithuania\label{uvilnius-ao}
    \and Institut d'Astronomie et d'Astrophysique, Universit{\'e} Libre de Bruxelles, CP 226, Boulevard du Triomphe, B-1050 Bruxelles, Belgium\label{ulbruxelles}
    \and Astrophysics Group, Keele University, Keele, Staffordshire ST5 5BG, United Kingdom\label{ukeele}
    \and Institute for Astronomy, Royal Observatory, University of Edinburgh, Blackford Hill, Edinburgh EH9 3HJ, UK\label{ifa-roe}
    \and GEPI, Observatoire de Paris, PSL Research University, CNRS, Univ. Paris Diderot,  Sorbonne Paris Cit{\'e}, 61 avenue de l'Observatoire, 75014, Paris, France \label{oparis}
    \and Universit{\'e} de Picardie Jules Verne, 33 rue St Leu, 80080 Amiens, France\label{upjv}
    \and INAF - Osservatorio di Astrofisica e Scienza dello Spazio, via P. Gobetti 93/3, 40129 Bologna, Italy\label{obologna}
    \and Instituto de Astrof{\'i}sica de Andaluc{\'i}a, CSIC, Glorieta de la Astronom{\'i}a s/n, Granada 18008, Spain\label{iaa}
    \and Institute of Astrophysics, FORTH, Crete\label{crete}
    \and INAF - Osservatorio Astronomico di Padova, Vicolo dell’Osservatorio 5, I-35122, Padova, Italy\label{opadova}
    \and European Space Agency (ESA), European Space Research and Technology Centre (ESTEC), Keplerlaan 1, 2201 AZ Noordwijk, The Netherlands\label{esa}
    \and INAF - Osservatorio Astronomico di Roma, Via Frascati 33, I-00040 Monte Porzio Catone (Roma), Italy\label{oroma}
    \and N{\'u}cleo Milenio ERIS \& Instituto de Estudios Astrof{\'i}sicos, Universidad Diego Portales, Ej{\'e}rcito 441, Santiago, Chile\label{udp}
    \and Hamburger Sternwarte, Gojenbergsweg 112, 21029 Hamburg, Germany\label{hamstrn}
    }
    \date{Received July 25, 2023; accepted November 17, 2023}

% \abstract{}{}{}{}{} 
% 5 {} token are mandatory
 
  \abstract{
  The Gaia-ESO Survey is an European Southern Observatory (ESO) public spectroscopic survey that targeted $10^5$ stars in the Milky Way covering the major populations of the disk, bulge and halo. The observations were made using FLAMES on the VLT obtaining both UVES high ($R\sim47,000$) and GIRAFFE medium ($R\sim20,000$) resolution spectra.
  
  The analysis of the Gaia-ESO spectra was the work of multiple analysis teams (nodes) within five working groups (WG). The homogenisation of the stellar parameters within WG11 (high resolution observations of FGK stars) and the homogenisation of the stellar parameters within WG10 (medium resolution observations of FGK stars) is described here.  In both cases, the homogenisation was carried out using a bayesian Inference method developed specifically for the Gaia-ESO Survey by WG11. The method was also used for the chemical abundance homogenisation within WG11, however, the WG10 chemical abundance data set was too sparsely populated so basic corrections for each node analysis were employed for the homogenisation instead.

  The WG10 homogenisation primarily used the cross-match of stars with WG11 as the reference set in both the stellar parameter and chemical abundance homogenisation. In this way the WG10 homogenised results have been placed directly onto the WG11 stellar parameter and chemical abundance scales. The reference set for the metal-poor end was sparse which limited the effectiveness of the homogenisation in that regime.

  For WG11, the total number of stars for which stellar parameters were derived was 6,231 with typical uncertainties for \teff{}, \logg{} and \feh{} of 32~K, 0.05 and 0.05 respectively. One or more chemical abundances out of a possible 39 elements were derived for 6,188 of the stars.

  For WG10, the total number of stars for which stellar parameters were derived was 76,675 with typical uncertainties for \teff{}, \logg{} and \feh{} of 64~K, 0.15 and 0.07 respectively. One or more chemical abundances out of a possible 30 elements were derived for 64,177 of the stars.       
  
  }
  

   \keywords{Surveys -- Stars: abundances -- Stars: fundamental parameters -- Methods: statistical}
\authorrunning{Worley et al.}
\titlerunning{The Gaia-ESO Survey: medium and high resolution FGK analyses}

   \maketitle
%
%-------------------------------------------------------------------

\section{Introduction}
The Gaia-ESO Survey is a European Southern Observatory (ESO) public spectroscopic survey designed to observe $10^5$ stars. It sampled the main populations of the Milky Way: the disk, bulge, and halo. The observing programme and the science goals of the Gaia-ESO Survey are described in \cite{gilmore2022} and \cite{randich2022}. The Gaia-ESO observing campaign was undertaken using the Fibre Large Array Multi Element Spectrograph (FLAMES) multi-object, intermediate and high resolution spectrograph of the Very Large Telescope (VLT) \citep{FLAMES}, using both the GIRAFFE spectrograph ($R \simeq 20,000$) and the Ultraviolet and Visual Echelle Spectrograph (UVES) ($R \simeq 50,000$).  As such, approximately 100 stars at medium resolution and six stars at high resolution were observed in each fibre configuration.

The observed data were divided between five working groups (WGs) as follows: WG10, FGK medium-resolution stars; WG11, FGK high-resolution stars \citep{Smiljanic14}; WG12, pre-main sequence \citep{lanzafame2015}; WG13, OBA stars \citep{blomme2022}; and WG14, non-standard objects and quality flags (Van Eck et al. in prep). Each WG was composed of multiple analysis teams from which the results were combined into a per star catalogue per WG. These results were then combined by the top-level working group (WG15) to produce the final Gaia-ESO per star catalogue. This is described in \cite{hourihane2023}.

The WG10 homogenisation of the analyses of the medium-resolution FGK stars for the final Gaia-ESO data release is described in this work. The observations at medium resolution were made in four of the available GIRAFFE setups: HR10, HR21, HR15N, and HR9B, for which the characteristics are given in Table~\ref{tab:setup_info}. Throughout this paper, `SETUP' refers to the four setups used in the Gaia-ESO Survey. In total there were 92,348 stars (158,809 spectra) observed at medium resolution that were analysed within WG10.

%% Table nodes_setups
\begin{table}[htbp]
   \setlength{\tabcolsep}{1pt}
   \centering
   \scriptsize{
  \caption{Overview of WG10 spectral dataset}
    \begin{tabular}{llrrrr}
    \hline
          &       & HR10 & HR21 & HR15N & HR9B \\
    \hline
    \hline
    Wavelength (\AA)$^{a}$ &       & 5,339-5,619 & 8,484-9,001 & 6,470-6,790 & 5,143-5,356 \\
    Resolution$^{a}$ &       & 21,500 & 18,000 & 19,200 & 31,750 \\
    No. Stars &       & 59,722 & 66,542 & 25,785 & 3,473 \\
    No. Spectra &       & 60,579 & 67,519 & 41,759 & 4,561 \\
        \multicolumn{6}{l}{a. \url{https://www.eso.org/sci/facilities/paranal/instruments/flames/inst/specs1.html}} \\
          &       &       &       &       &  \\
        Programme & GES\_TYPE & \multicolumn{4}{c}{No. Spectra} \\
    \hline
    \hline
    Open clusters & GE\_CL & 492   & 491   & 35,840 & 2,630 \\
          & AR\_CL & 0     & 0     & 684   & 997 \\
    Milky Way & GE\_MW & 53,798 & 53,446 & 0     & 0 \\
    Milky Way bulge & GE\_MW\_BL & 114   & 5,707  & 0     & 0 \\
          & AR\_MW\_BL & 0     & 228   & 0     & 0 \\
    FGK benchmarks & GE\_SD\_BM & 289   & 243   & 229   & 285 \\
          & AR\_SD\_BM & 129   & 128   & 129   & 129 \\
    Warm benchmarks & GE\_SD\_BW & 87    & 70    & 87    & 105 \\
    Cool benchmarks & GE\_SD\_BC & 74    & 63    & 38    & 16 \\
    Calibrating open clusters & GE\_SD\_OC & 653   & 651   & 103   & 0 \\
          & AR\_SD\_OC & 0     & 0     & 292   & 188 \\
    Globular clusters & GE\_SD\_GC & 1,259  & 1,255  & 180   & 0 \\
          & AR\_SD\_GC & 100   & 1,738  & 852   & 0 \\
    Corot Fields & GE\_SD\_CR & 2,952  & 2,847  & 2,954  & 0 \\
    K2C3 Field & GE\_SD\_K2 & 265   & 265   & 0     & 0 \\
          & AR\_SD\_BC & 0     & 0     & 0     & 0 \\
    Peculiar star templates & GE\_SD\_PC & 14    & 13    & 14    & 0 \\
    Radial velocity standards & GE\_SD\_RV & 305   & 321   & 309   & 211 \\
    Telluric standards & GE\_SD\_TL & 10    & 17    & 10    & 0 \\
    Miscellaneous stars & GE\_SD\_MC & 38    & 36    & 38    & 0 \\
    \hline
    \end{tabular}%
    \tablefoot{Characteristics of each WG10 observed spectral range, resolution, number of spectra and number of stars. Summary of number of spectra observed for each science programme and calibration sample as labelled in GES\_TYPE.}
  \label{tab:setup_info}%
  }
\end{table}%

As is described below, the WG10 homogenisation particularly relies on the WG11 homogenisation of the analyses of the UVES observations. The WG11 homogenisation of the fourth data release is described in \cite{Smiljanic14}; however, the WG11 homogenisation was updated for this release, so we provide an updated description in Section~\ref{sec:wg11bayes}. In total there were 6,987 stars (16,350 spectra) observed at high resolution for WG11.

Section~\ref{sec:data} presents a description of the GIRAFFE spectral dataset, Sect.~\ref{sec:nodes} shows the analysis methods of the WG10 nodes, and in Sect.~\ref{sec:referencesets} we define the reference sets used in the homogenisation. In Section~\ref{sec:wg11bayes}, we outline the Bayesian inference homogenisation method developed by WG11 and describe the WG11 parameter homogenisation and chemical abundance homogenisation.  In Section~\ref{sec:param_homog} we describe the WG10 parameter homogenisation,  Sect.~\ref{sec:abun_homog} presents the WG10 chemical abundance homogenisation, and the final catalogue and conclusions are presented in Sect.~\ref{sec:conclusion}. 

\section{GIRAFFE medium-resolution spectral dataset}\label{sec:data}
Table~\ref{tab:setup_info} summarises the number of spectra for the relevant observing programmes within WG10. These are a subset of the full list of observing programmes within Gaia-ESO. (See \cite{gilmore2022,randich2022,hourihane2023} for the full list. )The GES\_TYPE is the associated code per observing programme that allows these sub-samples to be easily identified in the Gaia-ESO catalogue.

The two main science programmes are the open clusters (OCs) and the Milky Way (MW). The observing strategies for these programmes are explained in full in \cite{randich2022}, \cite{bragaglia2022}, and \cite{gilmore2022}. The OC and MW programmes contain the bulk of the spectra. The remaining programmes are part of the calibration strategy of Gaia-ESO \citep{Pancino17}.

\begin{figure*}
\centering
%/Users/charlotteworley/Documents/GES/WG10/iDR6AbundanceHomog/FiguresPaper
\includegraphics[width=0.95\linewidth]{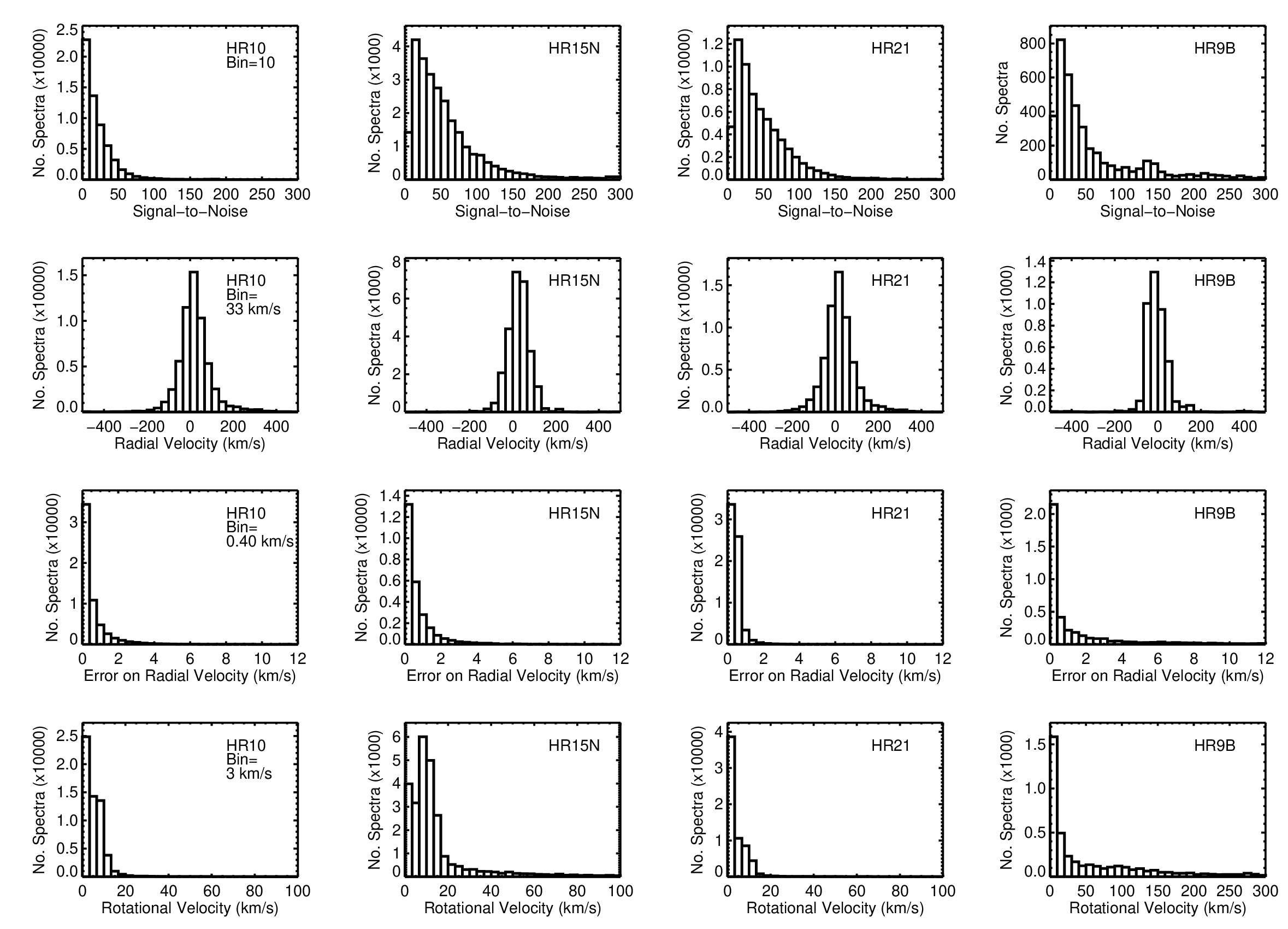}
\caption{Distribution of key values available for use in the WG10 analysis per SETUP: signal-to-noise, radial velocity, error on radial velocity, and rotational velocity. The bin size (Bin) for each parameter is given.}
\label{fig-WG10_snr_rv.ps}
\end{figure*}

The four key values provided to WG10 for use in the analysis are the signal-to-noise, the radial velocity and its uncertainty, and the rotational velocity. The calculation of these for the GIRAFFE spectra are described in \cite{gilmore2022}. The distribution of these values per SETUP are shown in Figure~\ref{fig-WG10_snr_rv.ps}. The signal-to-noise (S/N) distribution shows a large contribution of stars with a S/N less than ten. These can mainly be attributed to filler stars which were used to fill in the fibres once the observing programme targets for a field-of-view were exhausted. The majority of the stars have a rotational velocity of less than 20~kms$^{-1}$. This indicates that they are mainly slow rotating stars, as expected for the observing programmes. The radial velocity distribution is centred on zero and the bulk of the stars lie within $-200$ to $200$~km/s. The distribution of the uncertainty on the radial velocity shows the bulk of the stars have a precision better than 2~km/s. Particularly for the MW fields, the main effect is that for many of these filler stars, only the radial velocity could be reliably determined out of the set of stellar parameters.

\section{Working Group 10 node analysis methods}\label{sec:nodes}
Seven analysis teams (hereafter referred to as 'nodes') undertook either stellar parameter or chemical abundance, or both, analyses of subsets of the GIRAFFE SETUPs within WG10 for the final Gaia-ESO data release. The list of nodes and the SETUPs each node employed in which analysis phase is presented in Table~\ref{tab:nodes_setups}.

To provide a standardisation to the node analyses, the nodes were required to use the MARCS Stellar Atmosphere Models \citep{MARCS2008}, the solar abundances as from \cite{Grevesse07}, and the Gaia-ESO Line list \citep{heiterLL}. Pre-generated synthetic spectra for Gaia-ESO were also available, calculated as described in \cite{Laverny2012}. The following describes the analysis process of each node in turn.

%% Table nodes_setups
\begin{table}[htbp]
\setlength{\tabcolsep}{3pt}
  \centering
  \caption{SETUP and phase analyses carried out by each WG10 node.}
    \begin{tabular}{|l|l|c|c|c|c|c|c|c|c|}
    \hline
    SETUP & Phase & \rotatebox{90}{CAUP} &\rotatebox{90}{EPINARBO} & \rotatebox{90}{IAC} & \rotatebox{90}{Lumba} & \rotatebox{90}{MaxPlanck} & \rotatebox{90}{OACT}  &    \rotatebox{90}{Vilnius} &   \rotatebox{90}{Arcetri}\\
    \hline
    \hline
    HR10 &  Parameter  &           &          & \checkmark &          & \checkmark &         &    &        \\
    \hline
         & Abundance  &\checkmark &          &            & \checkmark & \checkmark &      &\checkmark &    \\
    \hline
    HR10|HR21 & Parameter &       &          & \checkmark & \checkmark &          &         &   &         \\
    \hline
    HR21:Bulge & Parameter &       &        & \checkmark &\checkmark   & \checkmark  &       &    &      \\
    \hline
    HR21:All & Abundance & \checkmark&      &             & \checkmark &          &       & \checkmark&    \\
    \hline
    HR15N & Parameter &        & \checkmark &       & \checkmark &      & \checkmark &   &     \\
    \hline
          & Abundance  &\checkmark& \checkmark&       & \checkmark &      &           &   &  \checkmark   \\
    \hline
    HR9B & Parameter  &           & \checkmark &           &           &            &   \checkmark & &    \\ \hline   
     & Abundance  &       & \checkmark &       &  &  &   &  &     \\
    \hline
    \end{tabular}%
  \label{tab:nodes_setups}%
\end{table}%

\subsection{Active in both the parameter and abundance phases:}
\paragraph{EPINARBO:} Equivalent widths were measured with DAOSPEC \citep{DAOSPEC}. Atmospheric parameters and abundances were determined with the Fast Automatic MOOG Analysis code \citep[FAMA,][]{FAMA}, which automatises the use of MOOG \citep{MOOG}. The HR15N and HR9B SETUPs were analysed.

\paragraph{Lumba:} The Lumba GIRAFFE analysis pipeline makes use of the Spectroscopy Made Easy code \citep[SME,][]{SME,SME.evolution} to compute on-the-fly synthetic spectra that are used to determine atmospheric parameters and chemical abundances. Departures from local thermodynamic equilibrium (LTE) line formation were included for Li, Mg, Al, Si, and Fe lines. The HR15N, HR10, and HR21 SETUPs were analysed. This pipeline has also been used for UVES analysis \cite{gavel2019} and is very similar to the pipeline used for the second and third data releases of the GALAH survey \cite{buder2018, buder2021}.

\paragraph{MaxPlanck:} The MaxPlanck node used  neural networks to determine stellar parameters and magnesium abundance \citep{Kovalev2019}. A training set of synthetic spectra was generated using the MARCS stellar atmosphere models and the Gaia-ESO line list. The MaxPlanck node investigated analysing the HR15N, HR10, and HR21 spectra (HR10+HR21 as a single analysis) but determined that the results from their analysis of HR10 were the only reliable results and thus provided results for that SETUP only. Results for the HR21 SETUP for the bulge fields and standard stars were also provided.

\subsection{Active only in the parameter phase:}
\paragraph{IAC:} The code FERRE \citep[see][and references therein]{allende2014} was used. The strategy was to search for the atmospheric parameters of the best fitting model among a grid of pre-computed synthetic spectra for each observed spectrum. The HR10 and HR21 SETUPs were analysed together as a single analysis,  and HR10 was also analysed separately. Results for the HR21 SETUP for the bulge fields and standard stars were also provided.

\paragraph{OACT:} The OACT node used the code ROTFIT \citep{frasca2003,frasca2006}. The method consists of a $\chi^2$ minimisation of the residuals between the observed spectrum and a set of reference spectra. In this case, a library of observed spectra from the ELODIE archive \citep{ELODIE} was used as reference. The HR15N and HR9B SETUPs were analysed.

\subsection{Active only in the abundance phase:}
\paragraph{Arcetri:} Equivalent widths of the Li and the nearby Fe line were measured with a Gaussian fitting. Abundances were determined with a set of curves of growth \citep{franciosini2022} determined from the Gaia-ESO stellar spectra grid. The HR15N SETUP was analysed.

\paragraph{CAUP:} Equivalent widths were measured with the Automatic Routine for line Equivalent widths in the stellar Spectra code \citep[ARES,][]{ARES1,ARES2}. Atmospheric parameters and chemical abundances were determined with MOOG \citep{MOOG}. This was carried out on the HR15N, HR10, and HR21 SETUPs.

\paragraph{Vilnius:} Equivalent widths were measured with DAOSPEC \citep{DAOSPEC}. The node developed its own wrapper to automatise the use of MOOG \citep{MOOG} for the determination of chemical abundances. The HR10 and HR21 SETUPs were analysed.

\section{Definition of reference sets}\label{sec:referencesets}
An underlying difficulty in the analysis of large stellar datasets is ensuring that the parameters and abundances that are produced are as close to the truth as is possible in our current understanding of stellar physics. The sheer number of spectra make it impossible to carry out a detailed `by hand' analysis of each spectrum,  so automated analyses must be used, such as those described previously. During their development, automated analyses are calibrated and validated against reference sets. For the homogenisation of the node results into the per star catalogue for both WG10 and WG11, the results from each node analysis were compared to known results of key reference sets to verify the node results and, where necessary, correct them onto the reference set scale prior to the node results being combined. The WG11 analysis, use of reference sets, and final homogenisation is described in Section~\ref{sec:wg11bayes} as an update to the process described in \cite{Smiljanic14}.

For WG10, there were a reasonable number of stars in common with WG11 such that the results could be combined with the FGK benchmark stars in order to construct a larger reference set for which the parameter space was more filled in. This also meant that the WG10 results would be calibrated directly onto the WG11 parameter scale.

\section{Working Group 11: Bayesian inference homogenisation method}\label{sec:wg11bayes}

The WG10 and WG11 node parameters were homogenised separately but used the same Bayesian inference method. Thus, the description below supersedes the previous WG11 homogenisation strategy described in \citet{Smiljanic14}. 

The Bayesian homogenisation for the WG11 results was first developed by Andrew R.~Casey (2014-2017, private communication)\footnote{\url{https://github.com/andycasey/ges-idr5}} and used for the Gaia-ESO internal data release 5 (iDR5). For the final data release (which corresponds to iDR6), we built upon his initial work and developed a different implementation of the method.  The homogenisation process and Bayesian modelling was written with R \citep{Rcitation}\footnote{\url{https://cran.r-project.org/}} using JAGS \citep{plummer2003jags}\footnote{`JAGS' stands for Just Another Gibbs Sampler: \url{http://mcmc-jags.sourceforge.net/}} and a number of related packages\footnote{rjags \citep{rjags}, runjags \citep{runjags}, jagstools \url{https://github.com/johnbaums/jagstools/}, and coda \citep{coda}}.

The problem we want to solve is that of finding the best estimate of a stellar parameter, given the multiple values determined by the different nodes. The parameter can be any of \teff, \logg, or \feh. For the microturbulence velocity ($\xi$), the procedure was slightly different (see Sec.\ \ref{sec:wg11.xi} below).

Let us consider that a star `$n$' is characterised by a true value of a given parameter, {\bf true.param$_{n}$}. When a node `$i$' attempts to estimate that value, the analysis returns a measurement, {\bf param$_{i.n}$}, that is affected by systematic and random errors introduced by the methodology that was used. We made the assumption that these errors are independent and can be separated if parameters are reported for enough repeat spectra. The systematic error accounts for any and all zero point offsets and biases. The random error accounts for any and all effects that are stochastic in nature. We further assumed that the bias error is itself a function of the atmospheric parameter in question and that the random errors can be described by a Gaussian distribution.

Numerically, we would write
\vspace{-0.15cm}
\begin{equation}
\mathrm{param}_{i,n} \sim dnorm(\mathrm{true.param}_{n},\, \mathrm{random.err}_{i} ) + \mathrm{bias.param}_{i,n}, \label{eq:true.param}
\end{equation}

where $dnorm$($\mu$,~$\sigma$) stands for the Gaussian distribution of mean = $\mu$ and standard deviation = $\sigma$. Equation \ref{eq:true.param} states that the measurement provided by the node $i$ is a random draw (which is the meaning of the symbol `$\sim$' in the equation) from the distribution centred on {\bf true.param$_{n}$} affected by the random error {\bf random.err$_{i}$}. This random error is postulated to be a property of the node. Further, the measurement is affected by an offset {\bf bias.param$_{i,n}$}. This bias value comes from a function that is also a property of the node and was computed at the value of the parameter that characterises the star $n$. For this bias function, we assumed that the variation of the bias in the parameter space can be described by a quadratic function:
\vspace{-0.15cm}
\begin{equation}
\mathrm{bias.param}_{i,n} = \alpha_{1} + \alpha_{2} * \mathrm{param}_{i,n} + \alpha_{3} * (\mathrm{param}_{i,n})^{2}. \label{eq:bias.param}
\end{equation}

Strictly speaking, we should write Eq.\ \ref{eq:bias.param} as a function of the true parameter and not of the parameter value measured by the node. However, that makes the problem circular (i.e. to know the true value, we need to correct for the bias, but to compute the bias, we need to use the true value). As it stands, Eq.\ \ref{eq:bias.param} should work reasonably well if the difference between the true and measured values is not too big, although exactly what that means has to be checked a posteriori. In any case, differences that cannot be accounted for by the bias will tend to inflate the random component of the error. Our tests with the final results showed that such choice for the modelling worked well (which does not mean that things could not be improved by assuming a different model).

For numerical reasons, we actually write Eq.\ \ref{eq:true.param} as
\vspace{-0.15cm}
\begin{equation}
\mathrm{param}_{i,n} \sim dnorm(\mathrm{true.param}_{n} + \mathrm{bias.param}_{i,n}, \, \mathrm{random.err}_{i}).\label{eq:bias.inside}
\end{equation}

This choice means that we assumed it is equivalent to say that the measured value was shifted by an offset or to say that when a node is affected by a certain bias, the measurement was made from a distribution centred around a `biased true parameter' ({\bf true.param$_{n}$} + {\bf bias.param}$_{i,n}$).

Because there are actually multiple nodes making the measurements, we can write the problem using a multi-dimensional normal distribution (where each node is one dimension):
\vspace{-0.15cm}
\begin{equation}
\overrightarrow{\mathrm{node.params}_n} \sim dmnorm( \overrightarrow{\mu_n},\, \Sigma_\mathrm{param}),\label{eq:multi}
\end{equation}

where $dmnorm$($\overrightarrow{\mu}$,~$\Sigma$) stands for the multi-dimensional normal distribution of mean vector = $\overrightarrow{\mu}$ and covariance matrix = $\Sigma$. The vector $\overrightarrow{\mathrm{node.params}_n}$ = (param$_{1,n}$, ..., param$_{K,n}$), that is, it combines the measurements of all $K$ different nodes for star $n$. The covariance matrix,$\Sigma_\mathrm{param}$, takes into account the random errors of each node and the correlations between their measurements. The mean vector $\overrightarrow{\mu_n}$ combines together the `mean' that we would write for each node separately in Eq.\ \ref{eq:bias.inside}; in other words, it is made of repeated entries for each node with the {\bf true.param$_{n}$} of the star and the corresponding node bias:
\vspace{-0.15cm}
\begin{equation}
\overrightarrow{\mu_n} = (\mathrm{true.param}_{n} + \mathrm{bias.param}_{1,n}, ...,\, \mathrm{true.param}_{n} + \mathrm{bias.param}_{K,n}),\label{eq:mu}
\end{equation}

\begin{table*}
	\caption{WG11 nodes that participated in the analysis of the final Gaia-ESO data release.}
	\label{tab:wg11.nodes}
	\centering
	\small
	\begin{tabular}{lll}
		\hline\hline
		Node &  Method & Data products \\
		\hline
		Arcetri  & Equivalent widths & Lithium abundances \\
		CAUP     & Equivalent widths & Stellar parameters and abundances \\
        EPINARBO & Equivalent widths & Stellar parameters and abundances\\
        IAC-AIP  & Library of synthetic spectra & Stellar parameters \\
        LUMBA    & On-the-fly spectrum synthesis & Stellar parameters and abundances\\
        Nice     & Library of synthetic spectra & Stellar parameters \\
        OACT     & Library of observed spectra & Stellar parameters and activity \\
        UCM      & Equivalent widths & Stellar parameters \\
        Vilnius  & Equivalent widths and spectrum synthesis & Stellar parameters and abundances \\
		\hline
	\end{tabular}
%	\tablefoot{	}
\end{table*}

To be able to apply the Bayesian inference to the homogenisation, we first needed to estimate the coefficients that define the bias function of each node (Eq. \ref{eq:bias.param}) and the covariance matrix (Eq. \ref{eq:multi}). Once the covariance matrix and the biases are defined, the only unknown in Eq. \ref{eq:multi} is the {\bf true.param$_{n}$} of star $n$. For these calculations, we have relied on a set of reference objects with known values of their true parameters. Of course,  how well such true values are known can be discussed. In practice, what we write for the reference objects is that the known parameter values, and their uncertainties, are priors of the true values:
\vspace{-0.15cm}
\begin{equation}
    \mathrm{true.param}_{n} \sim dnorm(\mathrm{reference.param}_n,\, \mathrm{reference.error}_n)\label{eq:ref.prior}
\end{equation}

This means that the simulation is free to adapt the true value of the parameter, within the error of the estimate given for that reference star. Each spectral SETUP analysed by the WG10 and WG11 nodes (UVES 520 and 580) was homogenised separately, in order to account for the possibility of different biases in the analysis of these different spectra. To estimate the biases and the covariance matrix, we ran a Bayesian Monte Carlo simulation using the tools mentioned above. We ran diagnostic tests to ensure that the simulations converged and that the autocorrelations were low (below 1-2\%).

For WG10 the reference sets were constructed as described in Sec. \ref{sec:referencesets}. For WG11, different sets of reference stars were used, depending on the parameter being homogenised. Details are given in the subsections below.

Nine analysis nodes participated in the WG11 analysis. They are identified in Table \ref{tab:wg11.nodes}. A summary of the methods, with references to the codes employed, is given in the companion paper by \cite{gilmore2022} and is not repeated here. A description of the methodologies can also be found in Appendix A of \cite{Smiljanic14} or, in the case of the OACT node, in \cite{lanzafame2015}.

\subsection{Working Group 11: Homogenisation of effective temperature}\label{sec:wg11.teff}

When estimating values of \teff, we used as reference the FGK benchmark stars \citep{heiter15} and the cool M-dwarf benchmark candidates from \citet[][see Table 5]{Pancino17}. However, not all of these cool benchmarks were successfully analysed, as they fall in a region of the parameter space where the WG11 methods do not work well. In total, the WG11 nodes analysed 640 individual spectra for 35 benchmark stars. In terms of \teff, those that were successfully analysed cover the interval between 3,224 and 6,635~K.

One problem we faced in our method is that nodes do not provide results for each and every spectrum. For the purposes of the MCMC simulations, the missing values were substituted with a broad non-informative uniform prior (from 3,000 to 8,000~K). As a result of the simulations, we obtained the coefficients of Eq.\ \ref{eq:bias.param} and the covariance matrix of Eq.\ \ref{eq:multi}, along with their uncertainties. 

As a second step, the bias function and covariance matrix were applied to compute the best estimate of \teff{} for each star in the sample (including to the reference stars themselves).  We note that the errors of the atmospheric parameters provided by the nodes are not used (neither for the homogenisation of \teff\ nor of the other parameters). Each node estimated their errors in a different way, some providing internal errors of the method, while others applied more complicated prescriptions. Consequently, these values cannot be directly compared. We let the comparison between reference and measured parameters define the intrinsic node random errors.

At the end of this second step, we found that the \teff\ values of the reference stars were recovered with a standard deviation of $\pm$85 K. We assumed that this value represents the external accuracy of our final \teff\ scale, even though, in truth, this value is a composition of our accuracy and the errors of the reference scale. The internal errors of our \teff\ values have a median of 65~K, with the first and third quartiles at 60 and 75~K. A comparison against the isochrones of open and globular clusters seemed to validate the final \teff\ results (See Sec.\ \ref{sec:wg11.iso}).

\subsection{Working Group 11: Homogenisation of surface gravity}\label{sec:wg11.logg}

For the WG11 $\log~g$ homogenisation, two models of the node biases had to be combined. The first model, valid for dwarfs and metal-rich giants, used as the reference set the same sample of 35 benchmark stars employed in the analysis of \teff. The benchmark stars cover the interval between 0.68 and 5.05~dex. 

However, a second model had to be built for the metal-poor giants ([Fe/H] $\leq$ $-$0.50 and $\log~g$ $\leq$ 3.50). In this case, in addition to the benchmark stars, a sample of giants with asteroseismic values of $\log~g$ was used. The sample included 62 stars with data from K2 \citep{worley2020} and 88 stars with data from CoRoT (Masseron et al., in preparation). The K2 stars have $\log~g$ with an interval between 1.74 and 3.41. The CoRoT stars have $\log~g$ with an interval between 1.75 and 2.99. The combination of the two models was found necessary to reproduce the \teff{}-\logg{} diagram of the globular clusters. Conversely, the solution with the seismic values degraded the quality of the diagrams and tests for other types of clusters and field stars. A few iterations were needed to assign the model that should be used for the stars at the edges of the parameter space division.

For the purposes of the model, the uncertainty of the seismic $\log~g$ values was fixed at a value of $\pm$0.02~dex. The missing values in $\log~g$ were substituted by a broad uniform prior (between 0.0 and 5.0~dex). At the end of the homogenisation, we found that the reference \logg\ values were recovered with a standard deviation of $\pm$0.14~dex. The internal errors have a median value of 0.15~dex, with the first and third quartiles at 0.14 and 0.18~dex.  A comparison with cluster isochrones is shown in Sec.\ \ref{sec:wg11.iso}.

We performed a number of tests attempting the use \textit{Gaia} \logg~priors as additional constraints in the Bayesian model. However, none of the attempts produced results of higher quality than the ones obtained with the approach described above.

\subsection{Working Group 11: Homogenisation of {[Fe/H]}}\label{sec:wg11.feh}

For the [Fe/H] homogenisation, the model was obtained using the results provided for 34 of the 35 FGK benchmark stars (with results for a total of 565 spectra). One benchmark (HD140283) had no reference [Fe/H] value but good values of \teff\ and \logg. The stars cover the interval between $-$2.64 and +0.35 dex (including the cool benchmarks). As the error in the reference [Fe/H], we used the standard deviation of the \ion{Fe}{i} lines given in  \citet{jofre2014}.

However, the homogenisation of metallicities using only the benchmarks was found to make the globular clusters become too metal rich. At the same time,  it was also making the known metal-rich open clusters become too metal poor. To correct this we were forced to add additional references for metallicities. These references are the open and globular clusters from Tables 7 and 8 of \citet{Pancino17}. The open clusters used were NGC 6253, NGC 6705, NGC 2477, NGC 3532, and Melotte 71. The globular clusters used were NGC 4372, NGC 5927, NGC 2808, M 15, NGC 4833, NGC 6752, NGC 104, NGC 1904, NGC 6553, NGC 1261, and M 12. For the simulation, a typical value of $\pm$0.05 was used as the reference error of the metallicities of the clusters.

Cluster members were either adopted from \citet{Pancino2017b} or a two-sigma cut around the mean of the radial velocities was used. For computing the biases, we essentially needed to select a majority of members (but not necessarily only members). This crude membership criterion was found to produce acceptable results. At the end of the homogenisation, we found that the reference [Fe/H] values were recovered with a standard deviation of $\pm$0.09 dex. The internal errors have a median value of 0.07 dex, with the first and third quartiles at 0.06 and 0.10 dex.

We find a few important remarks regarding the final WG11 metallicities should be made: i) Most of the globular cluster stars are bright giants, with \logg~$\leq$ 2.0, and are thus in an area of the parameter space where the WG11 pipelines usually do not perform very well. ii) Better agreement with literature values for the metal-rich open clusters and metal-poor globular clusters was achieved at the expense of an increased scatter in the homogenised [Fe/H] values of the benchmark stars. This seems to indicate that the two scales (literature clusters and benchmarks)
have important differences. iii) The final metallicity for three of the cool benchmark stars (GJ205, GJ436, and GJ581) was very low. An inspection after the homogenisation revealed that the input node values are always much lower than the reference values. The conclusion seems to be that the WG11 metallicities for cool stars (\teff\ $\leq$ 4000 K) are not reliable.

\subsection{Working Group 11: Homogenisation of the microturbulence}\label{sec:wg11.xi}

The homogenisation of the microturbulence is different from the other parameters since there are no benchmark values that can be used as reference. Instead, we made use of the Gaia-ESO microturbulence calibration derived using iDR5 results to write the prior for the true values of $\xi$:
\vspace{-0.15cm}
\begin{equation}
    \mathrm{true.xi}_n \sim dnorm(\mathrm{calib.xi}_n,\, 0.25),\label{eq:truexi}
\end{equation}

where {\bf calib.xi}$_{n}$ is computed using the homogenised values of \teff, \logg, and [Fe/H], and we assumed a typical uncertainty of 0.25 km s$^{-1}$. When writing Eq.\ \ref{eq:multi} for $\xi$, we did not consider biases. In the Bayesian simulation, to homogenise this parameter, both the true values and the covariance matrix were determined at the same time.
 
In essence, the Bayesian modelling of $\xi$ is an elaborated way of finding the mean of the distribution of multiple node values, with the advantage of taking into account the correlations between the nodes and of using the calibration as a prior. We remark, however, that the final results are indeed different from both the simple mean of the individual node results and from the direct application of the calibration.

\subsection{Consistency checks of the final Working Group 11 stellar parameters}\label{sec:wg11.iso}

Here, we discuss the final stellar parameters obtained within WG11. We remark that these are not necessarily the final Gaia-ESO parameters for the stars we analysed, as there is still a process of survey-wide homogenisation. This final homogenisation process is described in the companion paper by \cite{hourihane2023}.

\begin{figure*}
	\centering
	\includegraphics[width=0.85\textwidth]{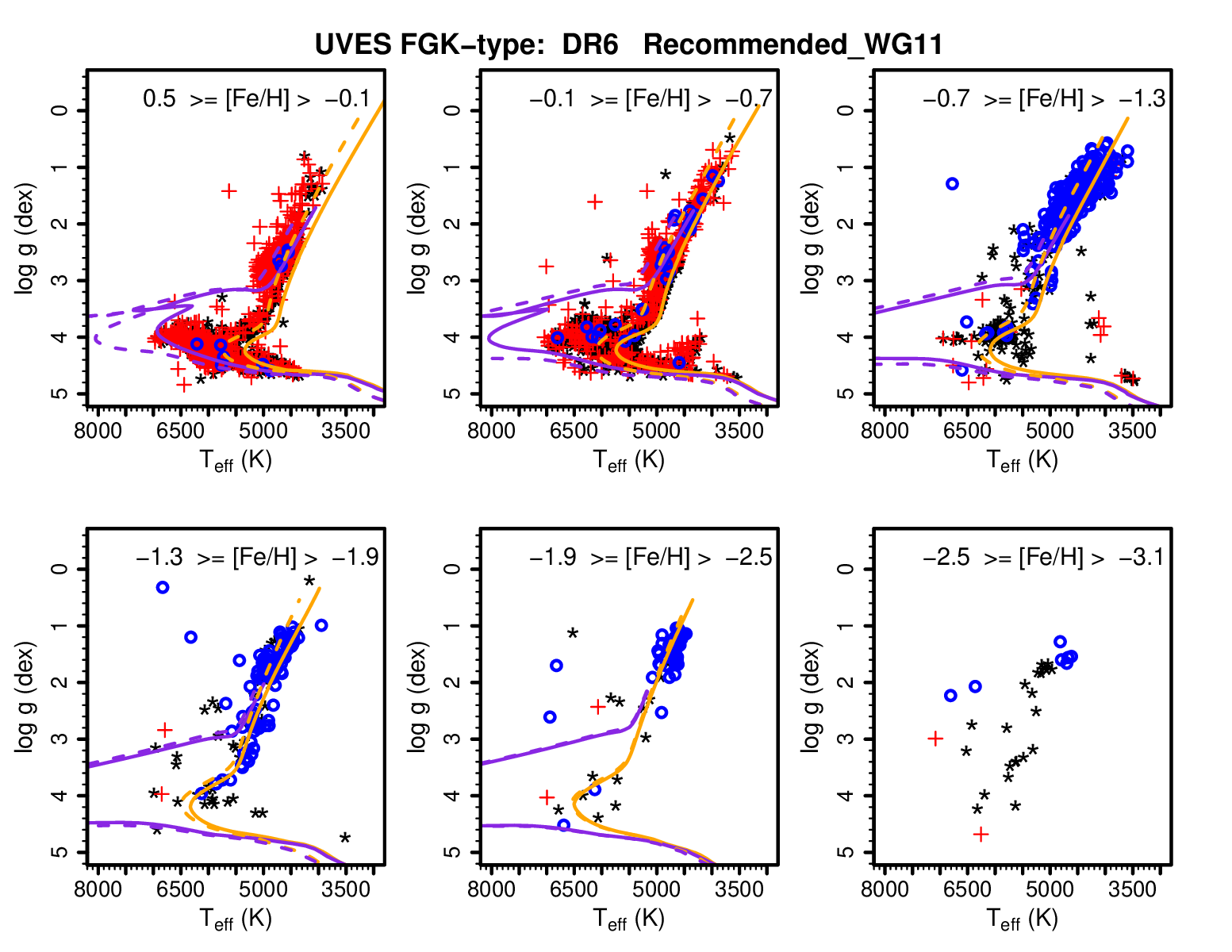}
	\caption{\small{\teff-\logg\ diagram with the WG11 recommended results. PARSEC isochrones \citep{Bressan2012} are shown for ages of 1 and 12.5 Gyr (violet and orange, respectively) and for the minimum and maximum metallicity indicated in each panel (dashed and solid lines, respectively). Red crosses are stars in open clusters, blue circles are stars in globular clusters, and the black starred symbols are the remaining stars.}}
	\label{fig:wg11.teff.logg}
\end{figure*}

Figure \ref{fig:wg11.teff.logg} shows the \teff-\logg\ diagram of the homogenised results, in bins of metallicity and in comparison to isochrones. The agreement with the location of the isochrones is in general very good. The location of the main sequence and of the red giant branch are in general well reproduced. 

At the lowest metallicity bin (where no isochrones are plotted), the scatter does seem to be excessive. There is the possibility that some of these stars are not real metal-poor stars, but artefacts of the analysis. We recall that for the cool benchmarks, the WG11 homogenised metallicities were too low. Investigation showed that some hot stars ($>$ 7000 K) included in the sample also ended up with very low metallicities. Care is therefore advised when using the results for the most metal-poor stars.

\begin{figure*}
\centering
\includegraphics[width=4.5cm]{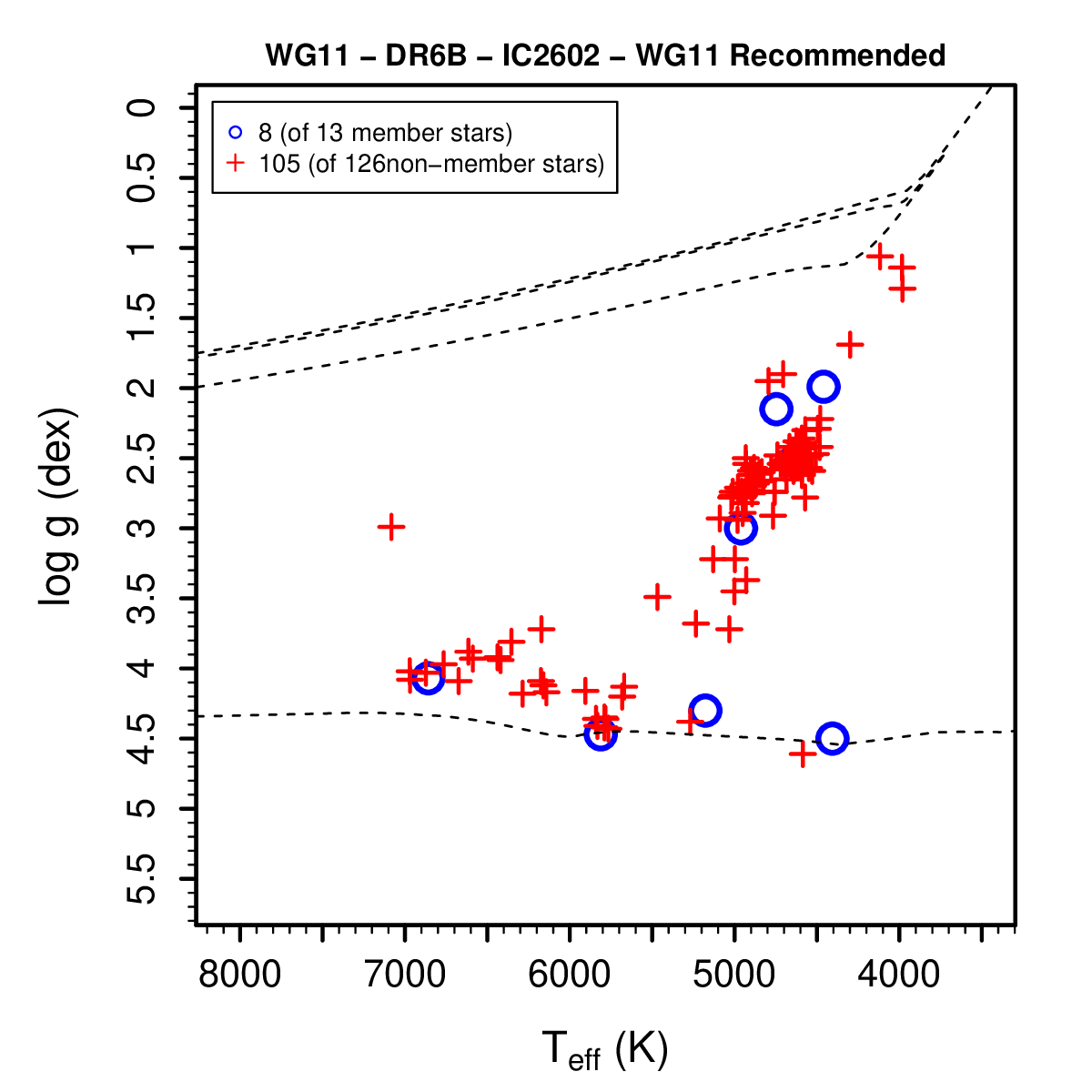}
\includegraphics[width=4.5cm]{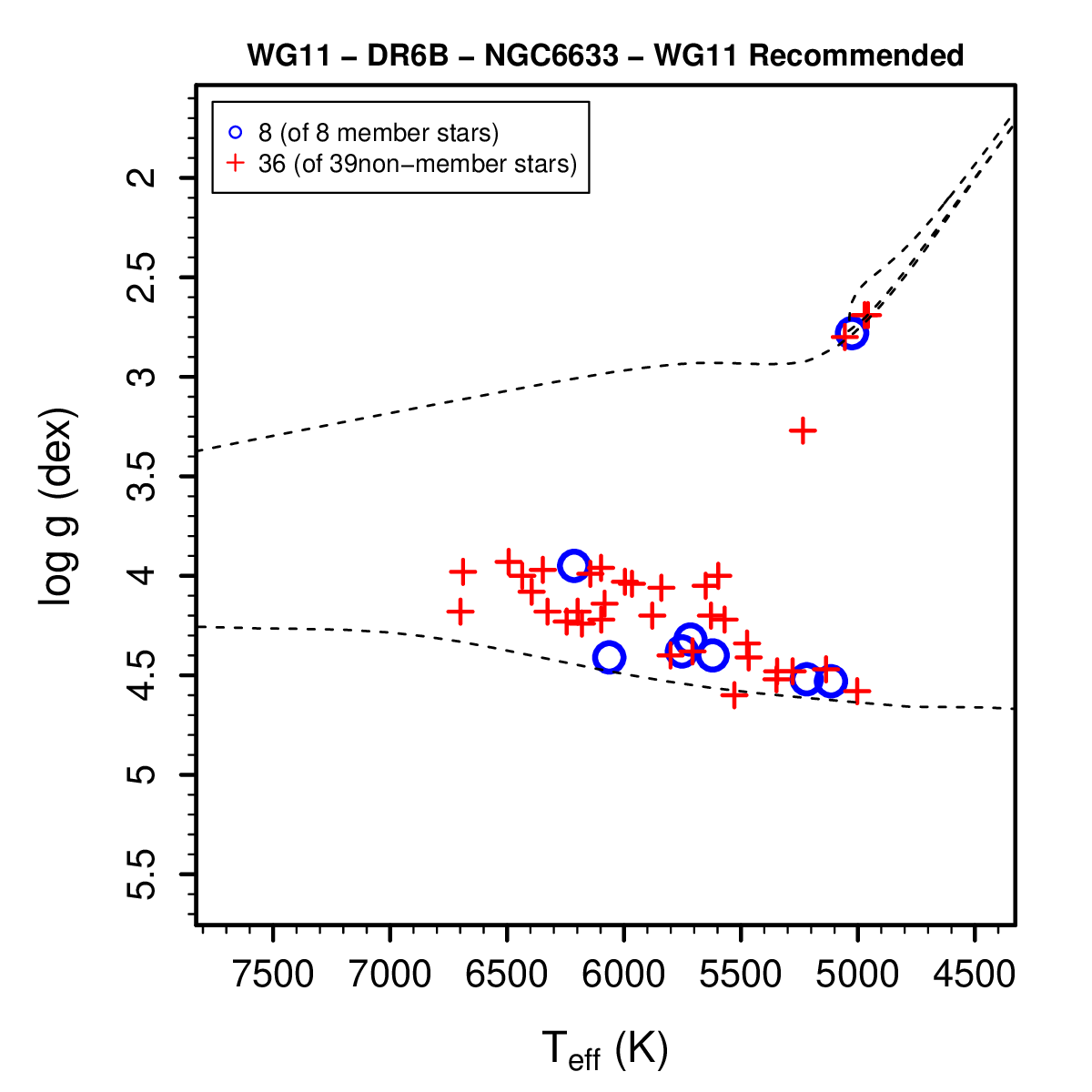}
\includegraphics[width=4.5cm]{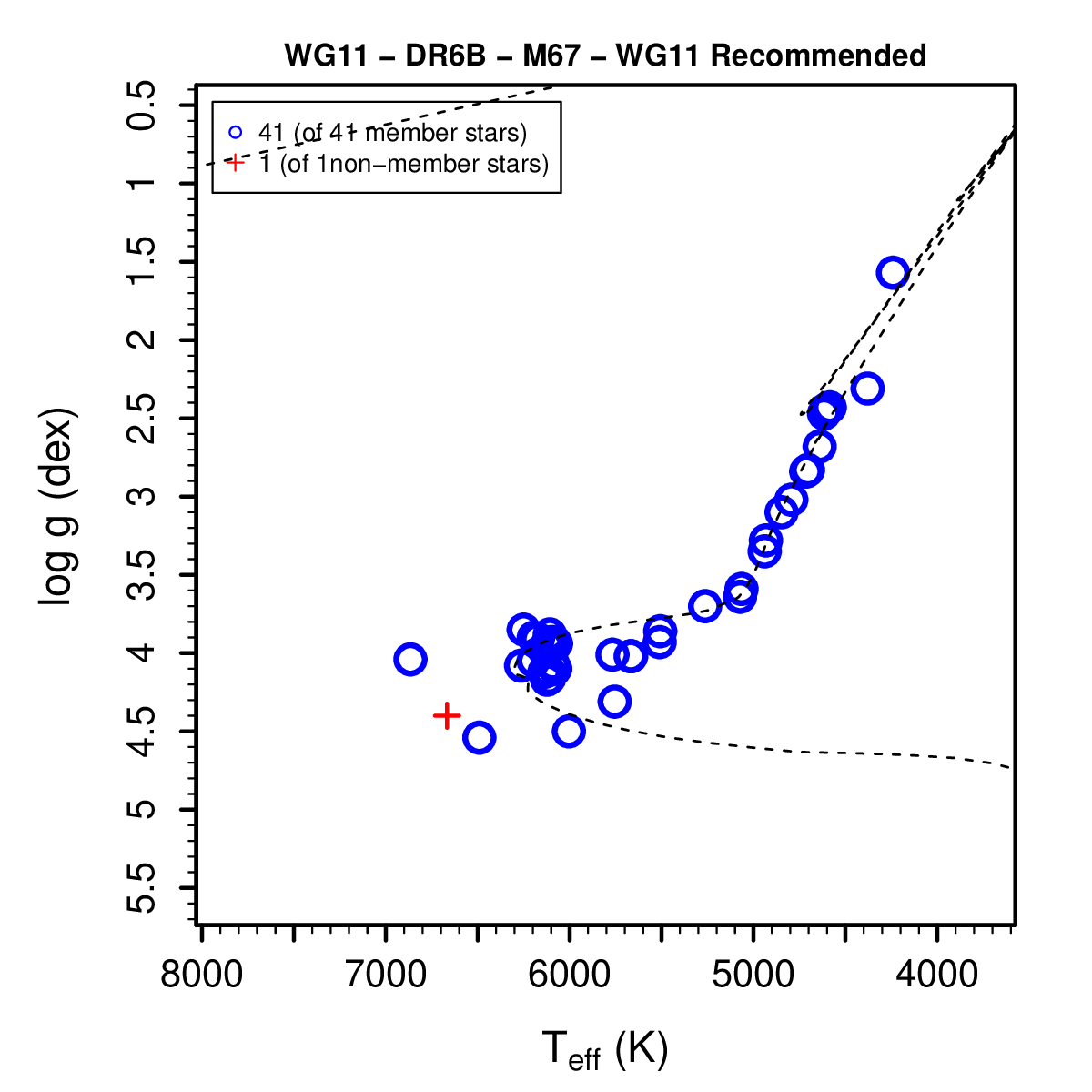}
\includegraphics[width=4.5cm]{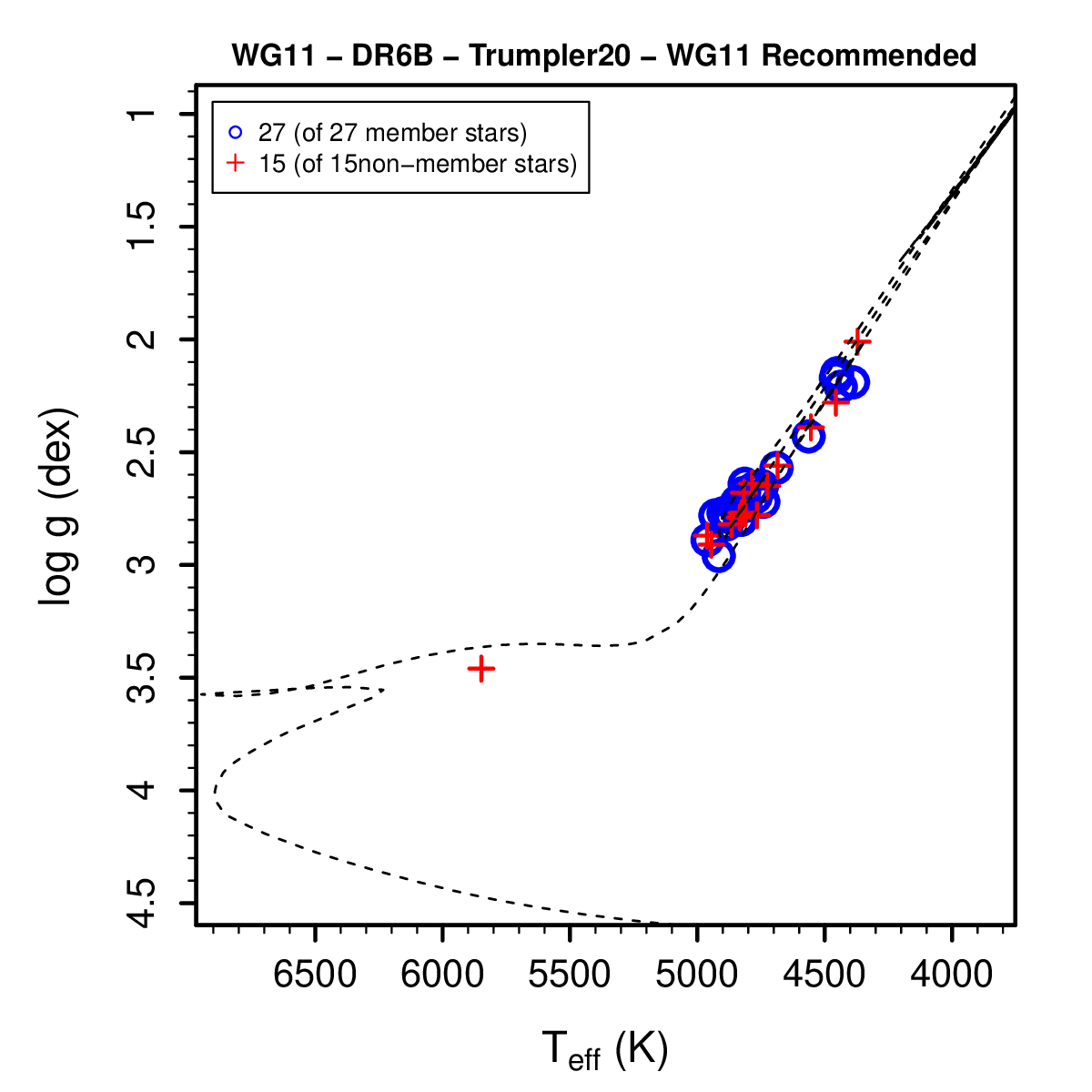}
\caption{\small{\teff-\logg{} diagrams for the open clusters IC 2602, NGC 6663, M67, and Trumpler 20.}}
\label{fig:ocs}
%\end{figure*}
%
%
%\begin{figure*}
\includegraphics[width=4.5cm]{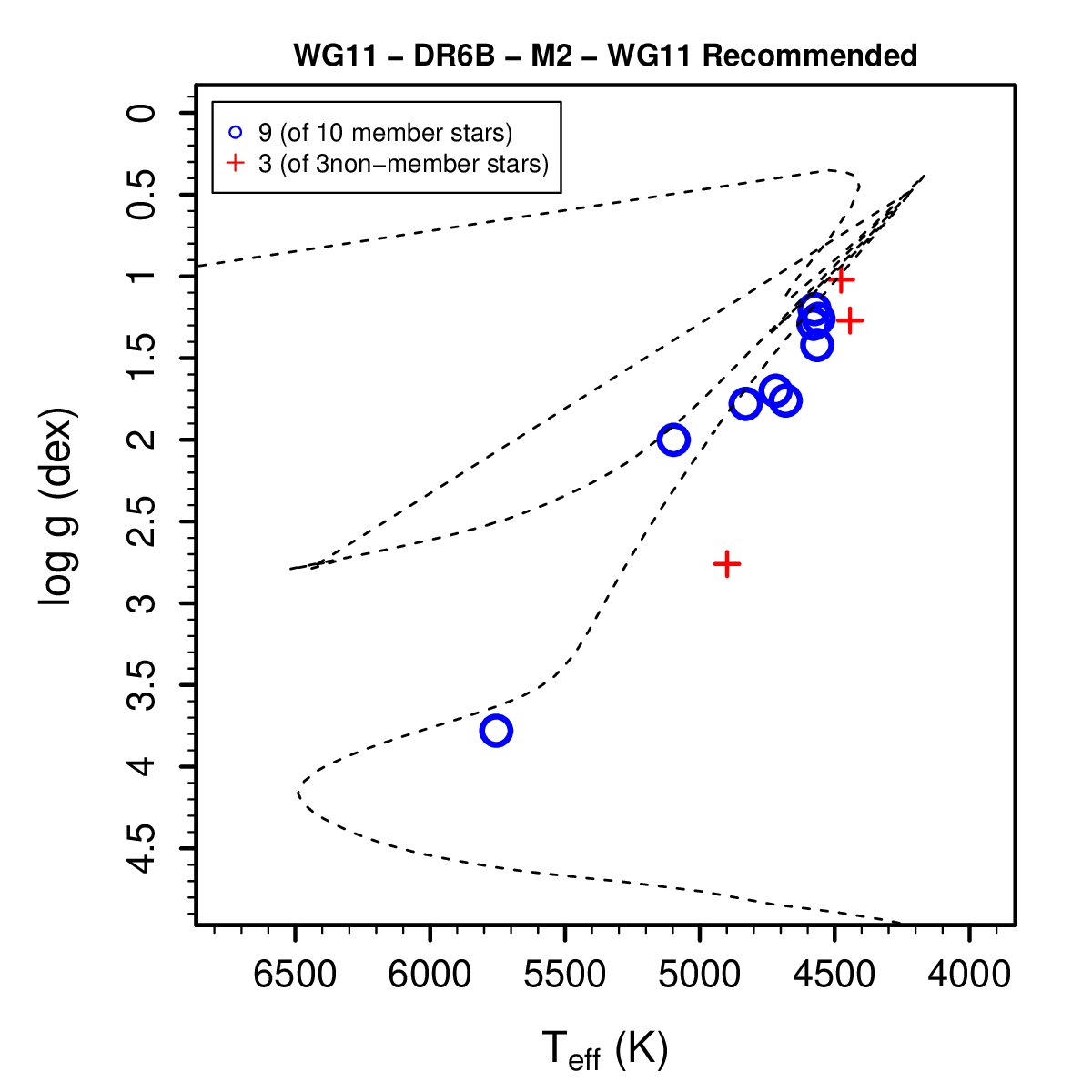}
\includegraphics[width=4.5cm]{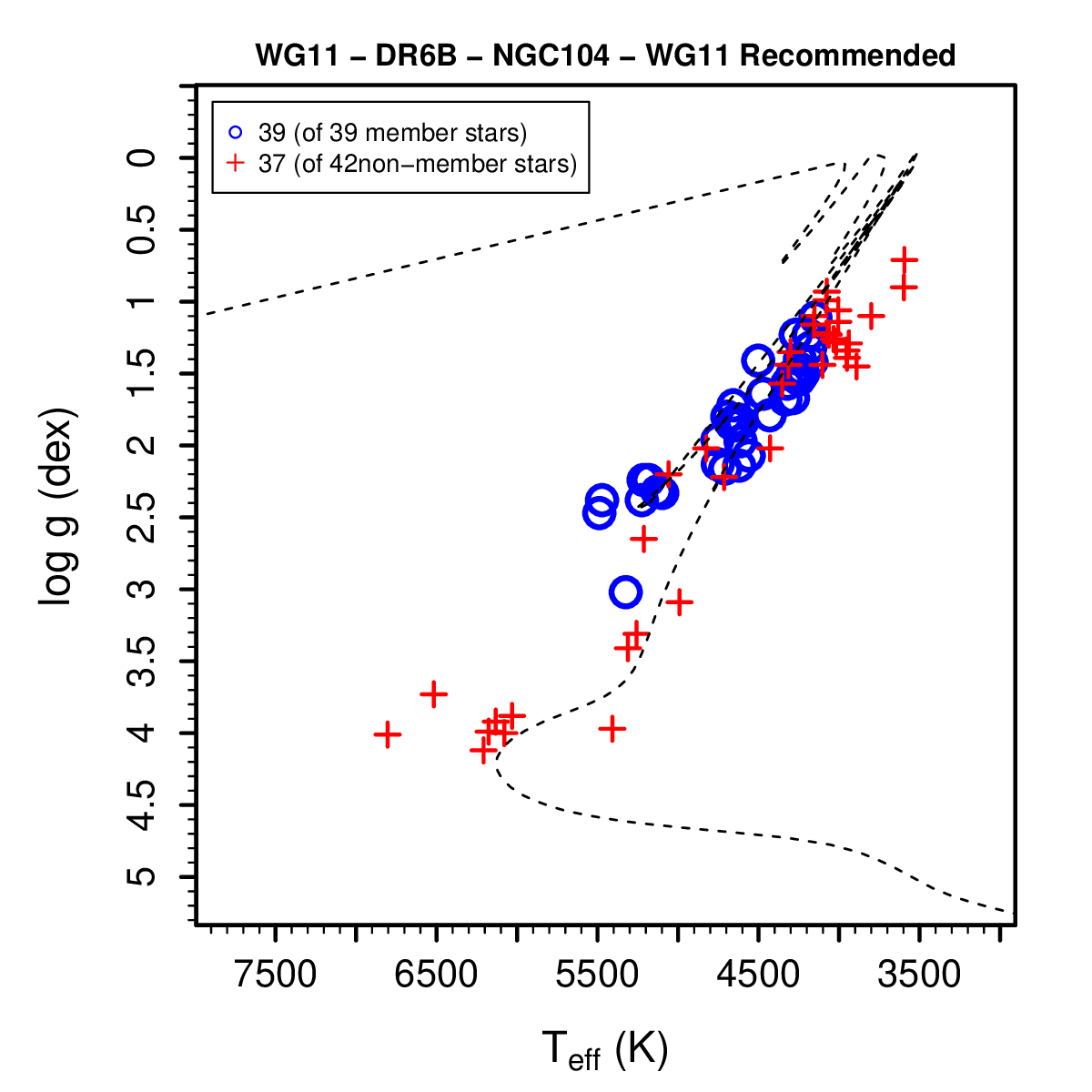}
\includegraphics[width=4.5cm]{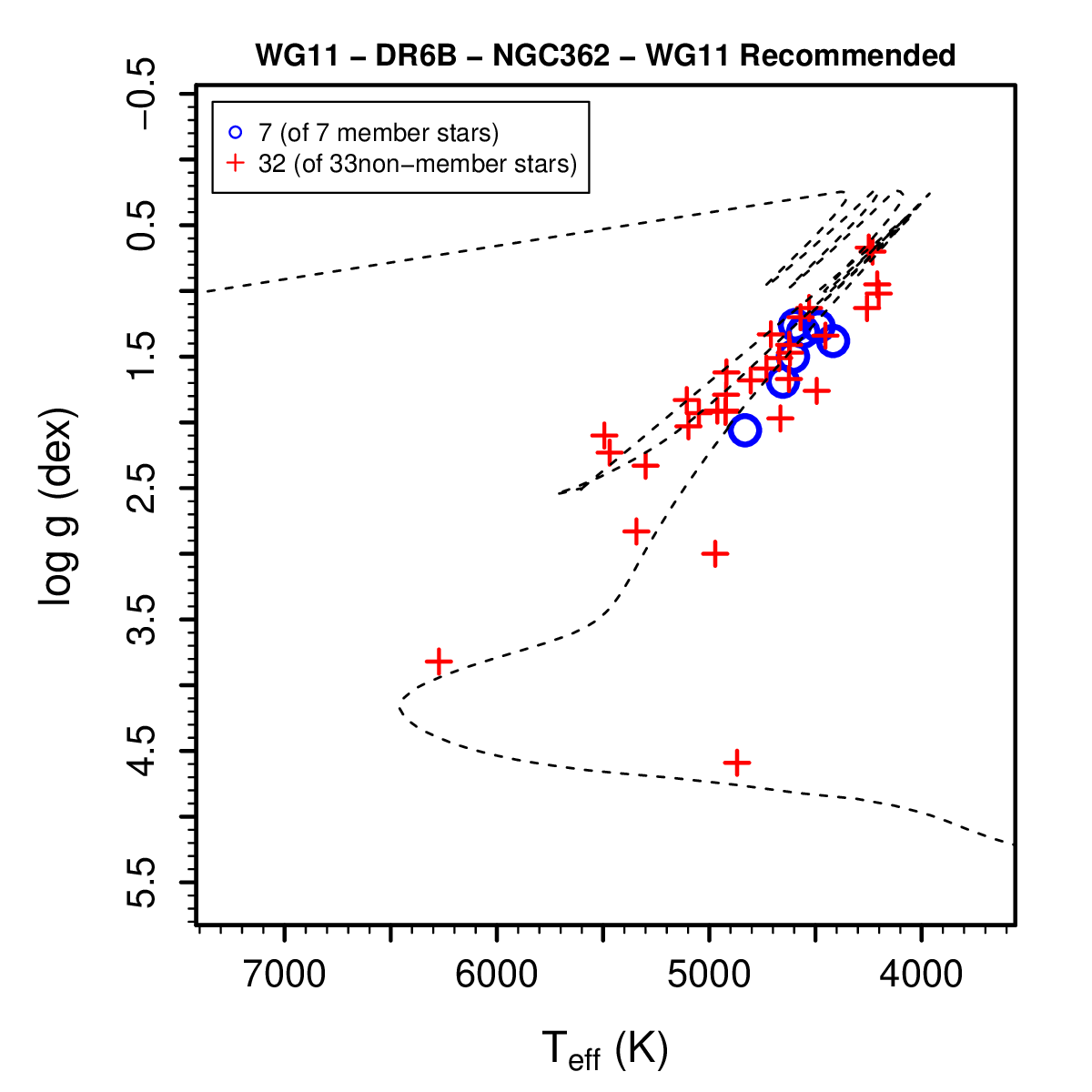}
\includegraphics[width=4.5cm]{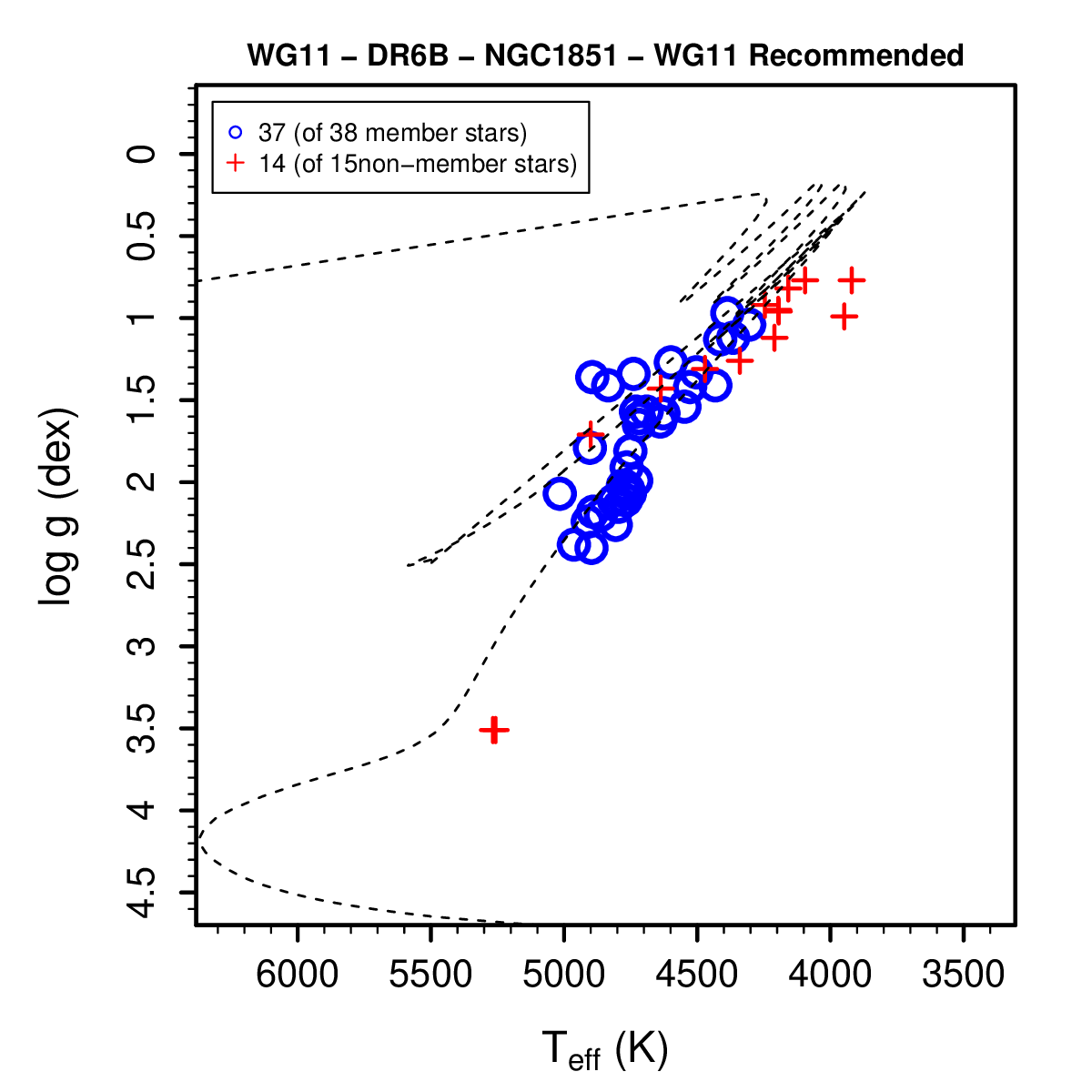}
\caption{\small{\teff-\logg{} diagrams for the globular clusters M2, NGC 104, NGC 362, and NGC 1851.}}
\label{fig:gcs}
\end{figure*}

Figures \ref{fig:ocs} and \ref{fig:gcs} show \teff-\logg\ diagrams for a few open and globular clusters. When possible, membership information was obtained from previous Gaia-ESO papers \citep{Spina2014,Magrini2017,Pancino2017b,Randich2018}. If that was not possible, a simple two-sigma cut in radial velocity was used as a first estimate of membership. (We note that since this analysis was carried out \cite{jackson2022} has provided key cluster membership lists. These are used in the verification of the final Gaia-ESO dataset in \cite{hourihane2023}). 

As can be seen, the agreement for the open clusters is excellent. The member stars tend to follow the isochrones, in particular for the case of red giants in older open clusters. In the case of young clusters, it happens that the main sequence stars are usually sitting slightly above the isochrone. We did find a few issues, however. For M67, the subgiants seem to have a homogenised $\log~g$ that is too high for their temperatures. In addition, the analysis of the Pleiades spectra did not return good atmospheric parameters.  Although in this last case, the spectra are not from UVES, and we believe there is something different in the data creating some kind of systematic problem in the analysis.

The agreement for the globular clusters is good in many cases, but there are cases of disagreement. In particular we mention NGC 1904, NGC 4833, NGC 5927, NGC 4372, and M 15, nearly all of which, except NGC 5927, are metal poor. In these cases, the stars tend to have temperatures that are cooler than expected from the position of the isochrones. In general, the stars in the globular clusters are bright giants ($\log~$ $<$ 1.5). For such stars, the WG11 analysis does not seem to be very robust. We recommend care when using results for these stars.

We also checked for trends between metallicity values and \teff{} or \logg{} for the same selection of open and globular clusters. In most cases, trends were not seen, are very small, or driven by one outlier whose membership could be questioned. However, there are cases (e.g. NGC 2243 and Trumpler 20) where correlations were detected. In other cases, large scatter can be present, as is particularly seen in younger clusters (see e.g. NGC 2516 or IC 4665). We point out that some of the trends and large scatter are not errors induced by the homogenisation, but are effects that appear from limitations in our methodology. The recent work by \citet{Baratella2020} suggests that traditional methods of analysis that rely on the various equilibria of Fe lines fail for young stars because the microturbulence is overestimated. Another example is the work of \citet{Semenova2020}, which indicates that abundance trends in the 2~Gyr open cluster NGC 2420 can be explained by neglected 3D non-LTE effects.

\subsection{Working Group 11: Homogenisation of chemical abundances}\label{sec:wg11.abundances}

Overall, the WG11 nodes attempted to derive abundances for 38 atomic species and two molecules. The atomic and molecular data that were used in the analysis are those described in \citet{heiterLL}. Abundances of \ion{O}{i} using the forbidden line at 6300 \AA, of carbon from molecular C$_2$, and nitrogen from CN bands were derived using spectrum synthesis by the Vilnius node only \citep[see][for details]{Taut2015}. Abundances and upper limits of \ion{Li}{i} come from measurements by the Arcetri node \citep{franciosini2022}. None of these abundances have gone through a homogenisation process.

For all other atomic species, we used individual line abundances for the homogenisation. The measurements come from a mix of equivalent width (by the CAUP, EPINARBO, and Vilnius nodes) and spectrum synthesis analyses (by LUMBA and, in the case of \ion{Mg}{i}, \ion{Ba}{ii}, \ion{Ce}{ii}, \ion{La}{ii}, \ion{Pr}{ii}, \ion{Y}{ii}, \ion{Zr}{i}, \ion{Zr}{ii}, and \ion{Nd}{ii}, by the Vilnius node). The final list of species from which abundances were estimated, in addition to the metallicity itself, is: \ion{Li}{i}, C (from C$_2$), \ion{C}{i}, N (from CN), \ion{O}{i}, \ion{Na}{i}, \ion{Mg}{i}, \ion{Al}{i}, \ion{Si}{i}, \ion{Si}{ii}, \ion{S}{i}, \ion{Ca}{i}, \ion{Ca}{ii}, \ion{Sc}{i}, \ion{Sc}{ii}, \ion{Ti}{i}, \ion{Ti}{ii}, \ion{V}{i}, \ion{Cr}{i}, \ion{Cr}{ii}, \ion{Mn}{i}, \ion{Co}{i}, \ion{Ni}{i}, \ion{Cu}{i}, \ion{Zn}{i}, \ion{Y}{ii}, \ion{Zr}{i}, \ion{Zr}{ii}, \ion{Nb}{i}, \ion{Mo}{i}, \ion{Ba}{ii}, \ion{La}{ii}, \ion{Ce}{ii}, \ion{Pr}{ii}, \ion{Nd}{ii}, \ion{Sm}{ii}, and \ion{Eu}{ii} (i.e., 30 different chemical elements). 

We advise particular care when using the abundances from \ion{S}{i}, \ion{Ca}{ii}, \ion{Sc}{i}, \ion{Mo}{i}, and \ion{Nb}{i}. These abundances come from a single (or a few) weak and/or blended lines. They were measured with equivalent widths, but precise results probably require spectrum synthesis. Quality control led us to reject all abundances from \ion{Sr}{i}, \ion{Ru}{i}, and \ion{Dy}{ii}, and therefore they are not part of the release.

Before homogenisation, we ran quality checks on the individual line abundances. For each line, we produced three plots of abundance as a function of \teff, \logg, and [Fe/H]. In these plots, we visually checked for trends, excessive scatter, and offsets among the nodes, and we removed anything that appeared suspicious. We also excluded lines that had been measured only in a small number of stars or by only one or two nodes (if there were other lines that had been measured by several nodes).

Homogenisation was also performed using a Bayesian modelling, with an adapted version of Eq.\ \ref{eq:multi} for a given chemical species:
\vspace{-0.15cm}
\begin{equation}
\overrightarrow{\mathrm{node.abundances}_n} \sim dmnorm( \overrightarrow{\mathrm{abun.}\mu_n},\, \Sigma_\mathrm{species}),\label{eq:multi.abun}
\end{equation}

where $\overrightarrow{\mathrm{node.abundances}_n}$ combines all line abundances measured by every node for star `$n$'. The covariance matrix $\Sigma_\mathrm{species}$ has dimensions equal to the sum of the number of lines used by all nodes, and, by itself, it introduced a large number of free parameters in the model. The mean vector $\overrightarrow{\mathrm{node.abundances}_n}$, expressed in a similar manner to Eq.\ \ref{eq:mu}, combines the true abundance, {\bf true.abun$_n$}, of that species in star `$n$' with the line biases {\bf line.bias$_{j}$}. 

The line bias was not introduced as a property of the node but of the spectral line $j$. This was meant to take into account a possible bias coming from uncertainties in the $\log~gf$ value of the lines. In principle, variation of this line bias in the parameter space could be introduced in order to model the changing importance of blends in different types of stars. However, this was not implemented, as it would introduce too many additional free parameters in the model. Distinct priors for the line biases were introduced depending on their quality flags in the form
\vspace{-0.15cm}
\begin{equation}
\overrightarrow{\mathrm{line.bias}_j} \sim dnorm(0.0,\, \mathrm{sigma.bias}),\label{eq:line.bias}
\end{equation}

where {\bf sigma.bias} is equal to 0.01, 0.02, 0.05, or 0.1, for lines with (SYNFLAG,LOGGFFLAG) = (Y,Y); (Y,U) or (U,Y); (U,U); (N,?) or (?,N), respectively\footnote{The SYNFLAG and LOGGFFLAG take values yes (Y), uncertain (U), or no (N), as explained in \citet{heiterLL}. SYNFLAG comes from tests of the blending of the lines in the spectra of the Sun and Arcturus. LOGGFFLAG is related to the confidence on the $\log~gf$ value of the line.}. To avoid that this line bias diverges when only one or two lines are measured, we found it necessary to change {\bf sigma.bias} to 0.002, 0.01, 0.02, and 0.05. 

Apart from the solar abundances, there are no other fundamental reference values that can be used to constrain the covariance matrix and the line biases. Because of that, the homogenisation of abundances was run as a single step, similar to what was done for $\xi$. Priors were used for the {\bf true.abun$_n$} of each star. For the Sun, the abundances from \citet{Grevesse07} were used as a strong Gaussian prior with $\sigma$ = 0.001. For the abundances of Mg, Ti, Ni, Mn, and V, we found it helpful to introduce the abundances of the benchmark stars as additional priors \citep{jofre2015}. For the other stars, we used a Gaussian distribution as the prior, with the mean at the metallicity-scaled solar abundance and $\sigma$ = 0.4. For \ion{Ba}{ii}, \ion{Cr}{i}, \ion{Cr}{ii}, \ion{Ca}{ii}, \ion{Ni}{i}, \ion{Y}{ii}, \ion{Mn}{i}, \ion{Zn}{i}, \ion{Si}{ii}, \ion{Sc}{i}, and \ion{V}{ii}, this had to be changed to $\sigma$ = 0.1 in order to decrease the final scatter of the abundances.  

Essentially, although it looks more complicated, the method can be considered as a sophisticated way to define a weighted mean. The sophistication lies within estimating the random errors of each node (i.e. the weights) directly from the data and in allowing for the line biases.

%--------------------------------------------------------------------
%--------------------------------------------------------------------
\section{Working Group 10: Homogenisation of stellar parameters}\label{sec:param_homog}
In this section, we describe the homogenisation of the WG10 stellar parameters. As shown in Table~\ref{tab:nodes_setups}, for WG10 there were five nodes that provided parameters across the four GIRAFFE SETUPs. The specific parameters and number of spectra analysed per SETUP per node is shown in Table~\ref{tab:nodes_summary}. For each SETUP there were two to four sets of node results available with which to perform the homogenisation.

%% Table nodes_summary
% Table generated by IDL routine gen_param_nodecoldata_plot.pro
\begin{table}[htbp]
\setlength{\tabcolsep}{3pt}
\centering
\small{
\caption{Number of parameters provided per NODE per SETUP. }
\begin{tabular}{lccccc}
Parameter & EP & LM & OT & IC & MP \\
\hline
\hline
\multicolumn{5}{l}{HR10} & \\
\hline
\teff{}                   & --           & --           & --           & 45679        & 35305        \\
$\sigma$ \teff{}          & --           & --           & --           & 45678        & 58959        \\
\logg{}                   & --           & --           & --           & 45679        & 35305        \\
$\sigma$ \logg{}          & --           & --           & --           & 45678        & 58959        \\
\feh{}                    & --           & --           & --           & 45679        & 35305        \\
$\sigma$ \feh{}           & --           & --           & --           & 45678        & 58959        \\
$\xi$                     & --           & --           & --           & --           & 35305        \\
$\sigma \xi$              & --           & --           & --           & --           & 58959        \\
TECH                      & --           & --           & --           & --           & 37623        \\
\hline
\multicolumn{5}{l}{HR10|HR21} & \\
\hline
\teff{}                   & --           & 56164        & --           & 50262        & --           \\
$\sigma$ \teff{}          & --           & 54020        & --           & 50262        & --           \\
\logg{}                   & --           & 56164        & --           & 50262        & --           \\
$\sigma$ \logg{}          & --           & 54020        & --           & 50243        & --           \\
\feh{}                    & --           & 56164        & --           & 50262        & --           \\
$\sigma$ \feh{}           & --           & 54020        & --           & 50261        & --           \\
$\xi$                     & --           & 56164        & --           & --           & --           \\
$\sigma \xi$              & --           & 59083        & --           & --           & --           \\
TECH                      & --           & 33948        & --           & --           & --           \\
\hline
\multicolumn{5}{l}{HR21: Bulge Fields \& Standards} & \\
\hline
\teff{}                   & --           & 11714        & --           & 11099        & 10819        \\
$\sigma$ \teff{}          & --           & 11515        & --           & 11098        & 11566        \\
\logg{}                   & --           & 11714        & --           & 11099        & 10819        \\
$\sigma$ \logg{}          & --           & 11515        & --           & 11097        & 11566        \\
\feh{}                    & --           & 11714        & --           & 11099        & 10819        \\
$\sigma$ \feh{}           & --           & 11515        & --           & 11099        & 11566        \\
$\xi$                     & --           & 11714        & --           & --           & 10819        \\
$\sigma \xi$              & --           & 11742        & --           & --           & 11566        \\
TECH                      & --           & 1922         & --           & --           & 2181         \\
\hline
\multicolumn{5}{l}{HR15N} & \\
\hline
\teff{}                   & 24969        & 26538        & 20022        & --           & --           \\
$\sigma$ \teff{}          & 24985        & 25295        & 20022        & --           & --           \\
\logg{}                   & 22271        & 26538        & 20022        & --           & --           \\
$\sigma$ \logg{}          & 22251        & 25295        & 20022        & --           & --           \\
\feh{}                    & 22584        & 26538        & 20022        & --           & --           \\
$\sigma$ \feh{}           & 22557        & 25295        & 20022        & --           & --           \\
$\xi$                     & --           & 26538        & --           & --           & --           \\
$\sigma \xi$              & --           & 26550        & --           & --           & --           \\
PECULI                    & 14           & --           & 466          & --           & --           \\
TECH                      & 6551         & 7147         & 7052         & --           & --           \\
\hline
\multicolumn{5}{l}{HR9} & \\
\hline
\teff{}                   & 960          & --           & 2495         & --           & --           \\
$\sigma$ \teff{}          & 960          & --           & 2495         & --           & --           \\
\logg{}                   & 960          & --           & 2495         & --           & --           \\
$\sigma$ \logg{}          & 960          & --           & 2495         & --           & --           \\
\feh{}                    & 960          & --           & 2494         & --           & --           \\
$\sigma$ \feh{}           & 960          & --           & 2494         & --           & --           \\
$\xi$                     & 960          & --           & --           & --           & --           \\
$\sigma \xi$              & 960          & --           & --           & --           & --           \\
PECULI                    & 114          & --           & 84           & --           & --           \\
TECH                      & 3510         & --           & 2136         & --           & --           \\
\hline\hline
\end{tabular}
\tablefoot{EP=EPINARBO, LM=Lumba, OT=OACT, IC=IAC, MP=MaxPlanck}
\label{tab:nodes_summary}
}
\end{table}

For the MW observing programme (not including the BL fields) two SETUPs were observed, HR10 and HR21. These were selected, as they contain key lines that have a different sensitivity to surface gravity depending on whether the star is a giant or a dwarf, thus breaking the dwarf-giant degeneracy due to effective temperature. See Section~\ref{sec:abun_homog} for details on the wavelength regions.

For this reason, for the MW fields, it was recommended that the nodes analysing the MW SETUPs combine HR10 and HR21 in a single analysis. However, nodes were free to analyse the data as suited their method, and all the data they provided were used in the homogenisation.

In particular, IAC provided two sets of analysis for the MW fields, the analysis of HR10 combined with HR21 (the HR10$|$HR21 SETUP) and the analysis of HR10-only. MaxPlanck provided results for the HR10-only SETUP. During the quality control phase, MaxPlanck investigated combining the HR10 and HR21 spectra (SETUP=HR10$|$HR21) as a single analysis but concluded that these results were not reliable for their process and therefore did not provide them.

As the HR10-only and HR10$|$HR21 analyses effectively covered the same sample, the Lumba HR10$|$HR21 results, the IAC HR10$|$HR21 and HR10-only results and the MaxPlanck HR10-only results were all used for the homogenisation of the MW fields. For the remainder of this work, the HR10$|$HR21 homogenisation refers to these four sets of node results.

The MW BL fields were observed at a higher S/N in the HR21 SETUP than the main MW fields and it was decided to not observe the same fields in HR10. (See \cite{gilmore2022} for more details on the observing strategy.) The nodes analysed the HR21-only BL fields and the standard fields that had also been observed in HR21. These samples were used in the homogenisation of the HR21-only SETUP. 

The OC SETUPs, HR15N and HR9B, covered different samples of open cluster stars, so the spectra of these SETUPs were analysed separately. (See \cite{randich2022, bragaglia2022} for further details on the observing strategy.) For HR15N three nodes provided results, while for HR9B, two nodes provided results.

\begin{figure*}
\centering
\includegraphics[width=0.99\linewidth]{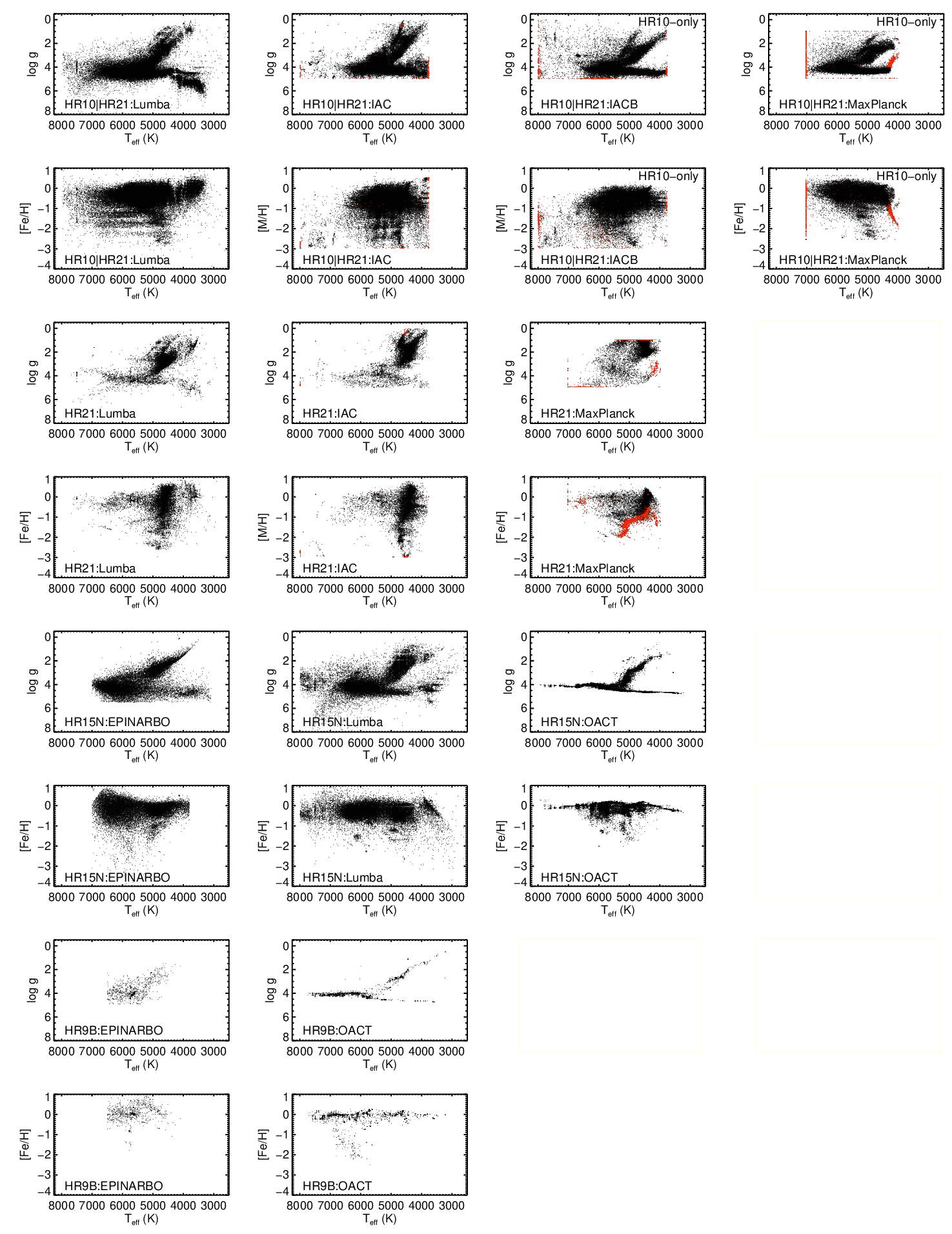}
\caption{Distribution of \teff{} with \logg{} and \teff{} with \feh{} for each node for each SETUP. Unflagged but rejected IAC and MaxPlanck results are shown in red. }\label{fig:nodes_hrd_feh}
\end{figure*}

Figure~\ref{fig:nodes_hrd_feh} shows the distribution of \teff{}, \logg{} and \feh{} provided by each node for each SETUP. The node results and associated reports were reviewed before homogenisation. 

The flag information (TECH, PECULI, REMARK) was inspected, and the WG14 flags that were used by each node were assessed to determine whether the associated results should be used in the homogenisation. The flags were assigned by each node based on the definitions in the WG14 Flag Dictionary.

There were 20 flags that were determined at the WG10 homogenisation level to mean that the associated results, if present, should not be used in the homogenisation. Which flags were reported as well as whether or not the results were provided varied between nodes so if the flag was present, the result (null or otherwise) was not used. The flag prefixes, the WG14 flag descriptions, and the number of spectra per node for which they were used are listed in Table~\ref{tab:nodes_paramflags}.

Inspection of the resulting node datasets showed that both the IAC and MaxPlanck analyses had results lying at the parameter grid limits that were not flagged.  The MaxPlanck analysis also showed a non-physical feature at \teff{}=4000~K, \logg{}=4~dex that is not present in the other node results. These are indicated as red points in Figure~\ref{fig:nodes_hrd_feh} and these results were removed prior to the homogenisation.

%% Table nodes_summary
% Table generated by IDL routine gen_flag_table.pro
\begin{table*}[ht]
\setlength{\tabcolsep}{3pt}
\centering
\small{
\caption{WG14 Flags used by WG10 NODES.}
\begin{tabular}{lp{11cm}ccccc}
Prefix    & WG14 Flag Prefix Description  &  EP  &  LM   &   OT  &    IC  &  MP  \\
\hline
\hline
10106 & Incomplete spectrum (missing wavelengths) & -- & -- & 2 & -- & -- \\
10302 & Code convergence issue: one or more convergence criteria (node-specific) could not be fulfilled & 330 & 247 & -- & 22520 & -- \\
10303 & Code convergence issue: temperature (\teff{}) is out of the node's grid & 4996 & -- & 1531 & -- & 2701 \\
10304 & Code convergence issue: gravity (\logg{}) is out of the node's grid & 1039 & -- & -- & -- & -- \\
10305 & Code convergence issue: metallicity ([M/H] or \feh{}) is out of the node's grid & 30 & -- & -- & -- & 285 \\
10308 & One or more parameter (which could not be identified) outside the node's grid & -- & 6739 & -- & -- & -- \\
10399 & No parameters provided because of lack of time & -- & -- & -- & -- & 29854 \\
11020 & $\nu sin(i)$ too high, preventing the determination of some/all parameters: $\nu sin(i)$ > 20 km/s & 55 & -- & -- & -- & -- \\
11100 & $\nu sin(i)$ too high, preventing the determination of some/all parameters: $\nu sin(i)$ > 100 km/s & 856 & -- & 3 & -- & -- \\
11150 & $\nu sin(i)$ too high, preventing the determination of some/all parameters: $\nu sin(i)$ > 150 km/s & -- & -- & 8 & -- & -- \\
11200 & $\nu sin(i)$ too high, preventing the determination of some/all parameters: $\nu sin(i)$ > 200 km/s & 227 & -- & 9 & -- & -- \\
11250 & $\nu sin(i)$ too high, preventing the determination of some/all parameters: $\nu sin(i)$ > 250 km/s & 78 & -- & 8 & -- & -- \\
13020 & Suspicious stellar parameters because temperature (\teff{}) is on the node's grid edge & -- & 23449 & 966 & -- & -- \\
13021 & Suspicious stellar parameters because gravity (\logg{}) is on the node's grid edge & -- & 3155 & -- & -- & -- \\
13022 & Suspicious stellar parameters because metallicity ([M/H] or \feh{}) is on the node's grid edge & 46 & 140 & 276 & -- & -- \\
13027 & Suspicious stellar parameters: multiple system & 4 & -- & -- & -- & -- \\
13028 & Suspicious stellar parameters because $\nu sin(i)$ is too high & -- & -- & 1 & -- & -- \\
20020 & Spectroscopic Binary: SBn, n >= 2 & 123 & -- & 474 & -- & -- \\
20030 & Spectroscopic Binary: SBn, n >= 3 & -- & -- & 20 & -- & -- \\
20070 & Composite spectrum & -- & -- & 38 & -- & -- \\
\hline\hline
\end{tabular}
\tablefoot{WG14 flags and associated descriptions as per the WG14 Dictionary used by the WG10 nodes in the parameter determination per spectrum. EP=EPINARBO, LM=Lumba, OT=OACT, IC=IAC, MP=MaxPlanck}
\label{tab:nodes_paramflags}
}
\end{table*}

\subsection{Parameter reference set}
Constructing a reference set using stars in common with WG11 was explored for each of the WG10 SETUPs for both the parameter phase and the abundance phase. Table~\ref{tab:wg11xsetups} gives the cross-match of each WG10 SETUP to WG11 and between WG10 SETUPs. 

%Table WG11 and SETUP cross-match
\begin{table*}[!ht]
  \centering
  \small{
  \caption{Cross-match of Stars (N$_{CN}$) and Spectra (N$_{SP}$) between WG11 and WG10 SETUPs.}
    \begin{tabular}{l|cc|cc|cc|cc|cc}
    \hline
    \hline
\multicolumn{1}{l|}{SETUP}      &      \multicolumn{2}{c|}{WG11}     &       \multicolumn{2}{c|}{HR15N}     &      \multicolumn{2}{c|}{HR10}      &      \multicolumn{2}{c|}{HR21}     &       \multicolumn{2}{c}{HR9B} \\
       &       N$_{CN}$  &    N$_{SP}$   &   N$_{CN}$  &    N$_{SP}$  &    N$_{CN}$   &   N$_{SP}$  &    N$_{CN}$  &    N$_{SP}$   &   N$_{CN}$  &    N$_{SP}$  \\
    \hline
HR15N    &     533 &    1295 &      -- &      -- &    3622 &    4387 &    3622 &    4387 &     794 &    1475 \\
HR10     &     170 &    1790 &    3622 &    8700 &      -- &      -- &   58354 &  118167 &     253 &    1796 \\
HR21     &     170 &    1797 &    3622 &    8767 &   58354 &  118342 &      -- &      -- &     253 &    1796 \\
HR21\_BL &      -- &      -- &      -- &      -- &     114 &     114 &    5935 &    5935 &      -- &      -- \\
HR9B     &     216 &     895 &     794 &    1448 &     253 &     912 &     253 &     912 &      -- &      -- \\
    \hline
    \end{tabular}%
    \tablefoot{Number of stars (CN=CNAME) cross-matched between each key SETUP sample and WG11, HR15N, HR10, HR21, HR9B, and number of associated spectra (SP) in the key SETUP.}
  \label{tab:wg11xsetups}%
  }
\end{table*}%

The decision to use the WG11 results as the source of the reference set against which to derive the WG10 results was driven by the reasoning that this would immediately put the WG10 results onto the WG11 scale.  The cross-match between WG11 and WG10 then comprised a larger more comprehensive set of stars in common than the process of using the reference sets that have more sparse coverage.

While the WG11 and WG10 observing programmes were not designed with an overlap in the parameter space for calibration purposes, the cross-match of each WG10 SETUP to WG11 was reasonably well sampled, in particular for \teff{} and \logg{}. The cross-match between WG10 SETUPs was also explored for use in the parameter phase as another way to expand the reference set for each SETUP.

As shown in Table~\ref{tab:wg11xsetups}, the SETUP with the largest per star (CNAME) cross-match to WG11 is the HR15N dataset. The cross-match of the HR10 dataset to WG11 is almost three times less than the HR15N dataset cross-match to WG11. However, the cross-match of the HR10 dataset to the HR15N dataset is over 21 times greater than the HR10 dataset cross-match to WG11. This is particularly due to the CoRoT sample for which all of the CoRoT fields were observed in all three SETUPS: HR10, HR21, and HR15N. The cross-match between HR21 and HR15N reflects that of HR10, as the HR21 targets were observed either in combination with HR10 or specifically for the BL fields. Therefore, the cross-match of HR21 to HR10 is particularly good. Targets in HR9B were observed to complement HR15N, so the cross-match with the HR15N dataset is almost four times greater than that with WG11 targets.

From this assessment a bootstrapping approach was taken to ensure all the SETUPs were homogenised onto a common scale. However, the last gap that needed to be covered was the lack of metal-poor reference stars, particularly for the HR10$|$HR21 (MW) and HR21 (BL) SETUPs in which metal-poor stars were most likely to be found,  rather than in HR15N and HR9B (OC).

\subsubsection{Filling in the metal poor tail with globular clusters}
Galactic globular clusters (GCs)were included as part of the Gaia-ESO calibration strategy to provide a reference set across the metal-poor regime.  (See \cite{Pancino17} for more details. ) Inspection of the cross-match between WG11 and the WG10 SETUPs revealed that there were not many globular cluster stars in common and what was in common was not sufficient to well sample the metal-poor end.  However, there were globular cluster stars present that could be used as \feh{} reference stars, assuming they could be confirmed as globular cluster members and thus could be assigned an associated \feh{} value.

To that end, a detailed globular cluster membership analysis was carried out using Gaia DR2 data in order to identify which stars in both WG11 and WG10 were globular cluster members \citep[][in prep]{worley2024}. The WG11 members per cluster were used to provide an average \feh{} and dispersion for that cluster. Then, for each WG10 cluster, if a cluster did not include a WG11 star that was a cluster member (none in the cross-match sample), the three most probable WG10 cluster members present in that SETUP were assigned the WG11 average \feh{} and dispersion as the value and its uncertainty. These then became the reference values against which those stars per node were compared. There are no globular cluster stars in the HR9B dataset, and the HR21 to HR10$|$HR21 overlap was already sufficient in globular cluster stars, so that this fill-in procedure was not implemented for these two SETUPs, and it was used only for HR10$|$HR21 and HR15N. While HR15N was used primarily to observe open cluster stars,  which are mainly solar metallicity and thus metal-poor stars were not expected, archival data of globular clusters were available in HR15N for some of the globular clusters that were observed in HR10 and HR21. This expanded the globular cluster star sample and provided another inter-SETUP calibration sample.

\subsubsection{Construction of per SETUP reference sets for parameter homogenisation}
For the parameter homogenisation phase, to maximise the size of the reference set for each SETUP, a bootstrapping procedure was implemented as listed in Table~\ref{tab:bootstrap_order}.

\begin{table}[htbp]
\setlength{\tabcolsep}{1.5pt}
\centering
\footnotesize{
\caption{Bootstrapping SETUP order and reference set content.}
\begin{tabular}{ccp{6cm}}
Order   &  SETUP & Reference set content \\
\hline
\hline
1 & HR15N &  FGK benchmark stars \\
  &             & Cross-match to WG11 \\
  &             & GC members assigned WG11 GC \feh{} \\
\hline
2 & HR9B &  FGK benchmark stars \\
  &             & Cross-match to HR15N \\
\hline
3 & HR10|HR21 & FGK benchmark stars \\
  &             & Cross-match to WG11 \\
  &             & Cross-match to HR15N \\
  &             & GC members assigned WG11 GC \feh{} \\
\hline
4 & HR21 BL & FGK benchmark stars  \\
  &             & Cross-match to HR10|HR21  \\
\hline\hline
\end{tabular}
\label{tab:bootstrap_order}
}
\end{table}

\begin{figure*}
\centering
\includegraphics[width=0.865\linewidth]{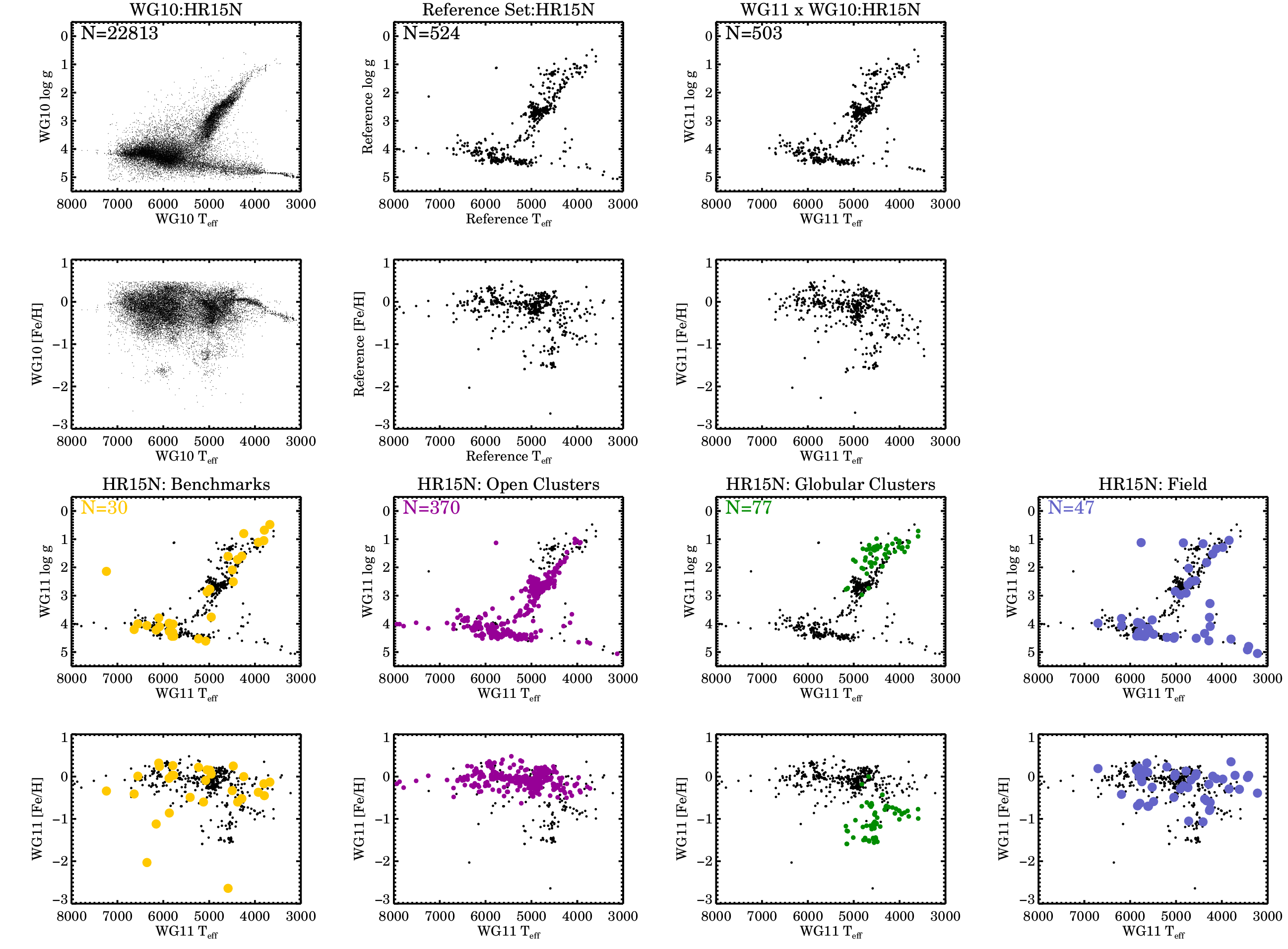}
%\label{fig:hr15n_refset}
\includegraphics[width=0.865\linewidth]{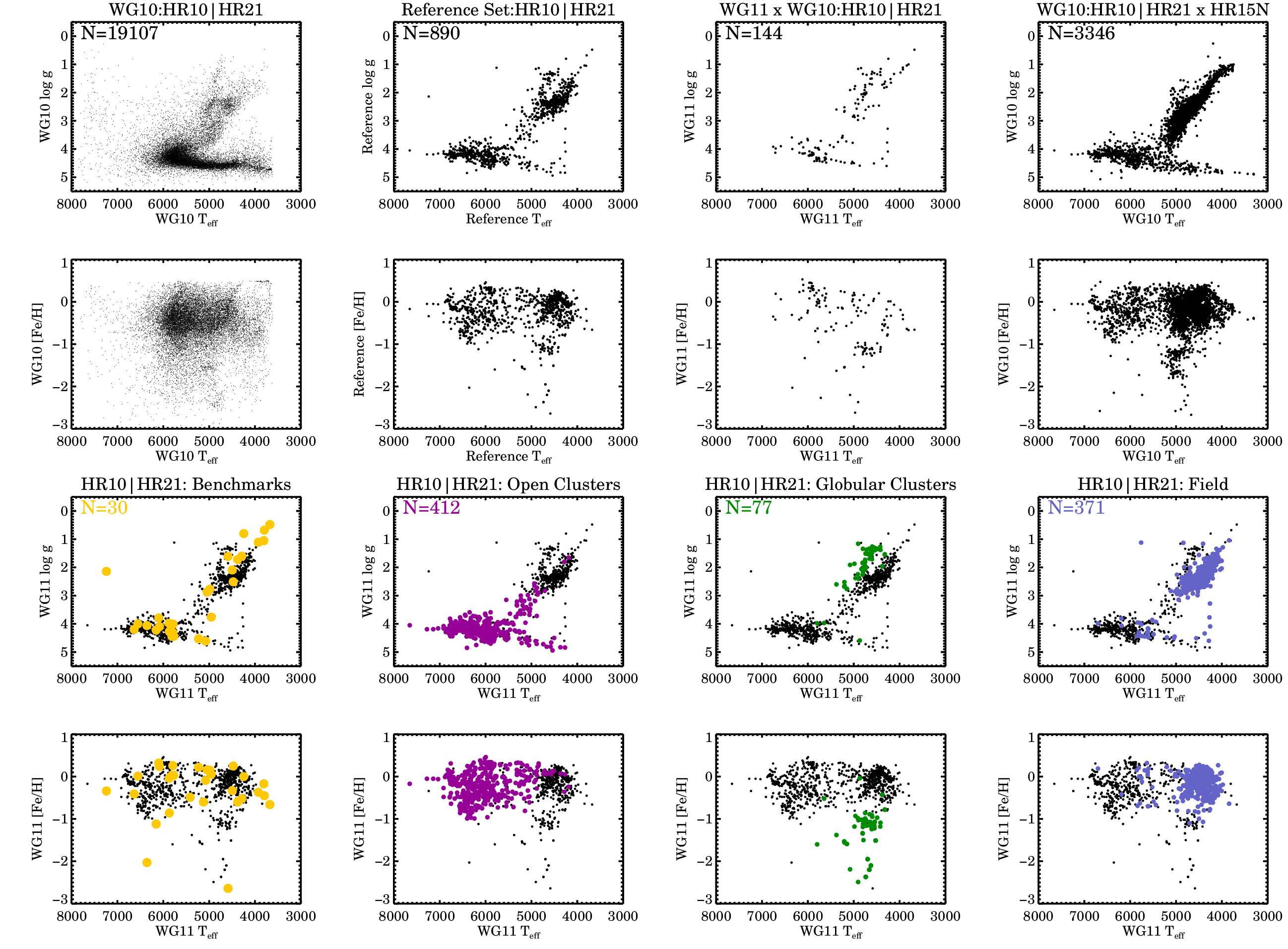}
\caption{Reference sets for HR15N and HR10$|$HR21 setups. Top row: Kiel diagrams for full HR15N sample, final HR15N reference set and WG11 cross-match with HR15N. Second row: Same as for the top row but for \feh{} versus \teff{}. Third row: Kiel diagrams of the four main samples: benchmarks (yellow), OC (magenta), GCs (green), and MW (blue) within the HR15N reference set overlaid on the full reference set. Fourth row: Same as for the third row but for \feh{} versus \teff{}.  Fifth to eighth rows: Same but for HR10$|$HR21.}
\label{fig:hr1021_refset}
\end{figure*}

\begin{figure*}
\centering
\includegraphics[width=0.865\linewidth]{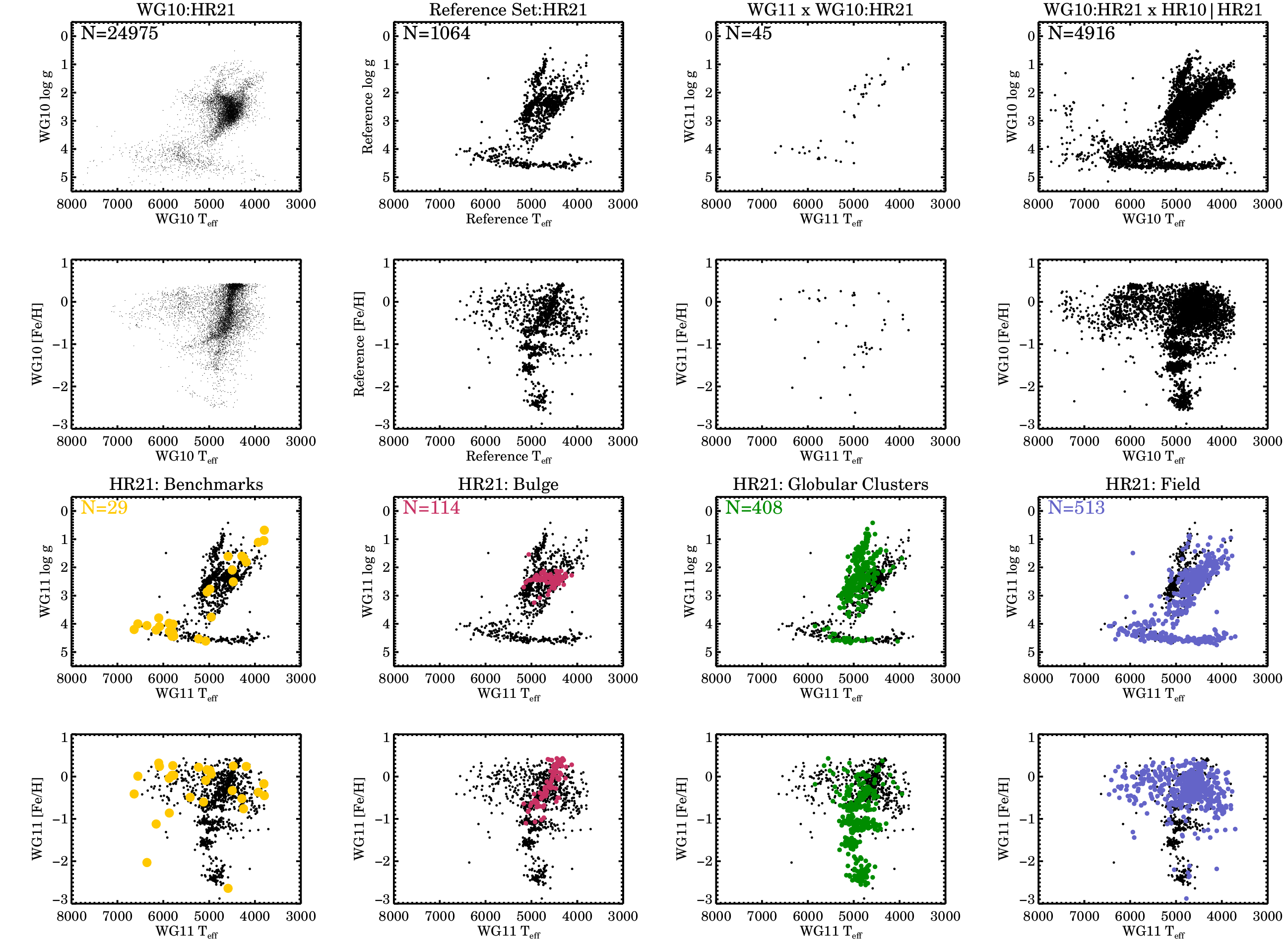}
%\label{fig:hr21_refset}
\includegraphics[width=0.865\linewidth]{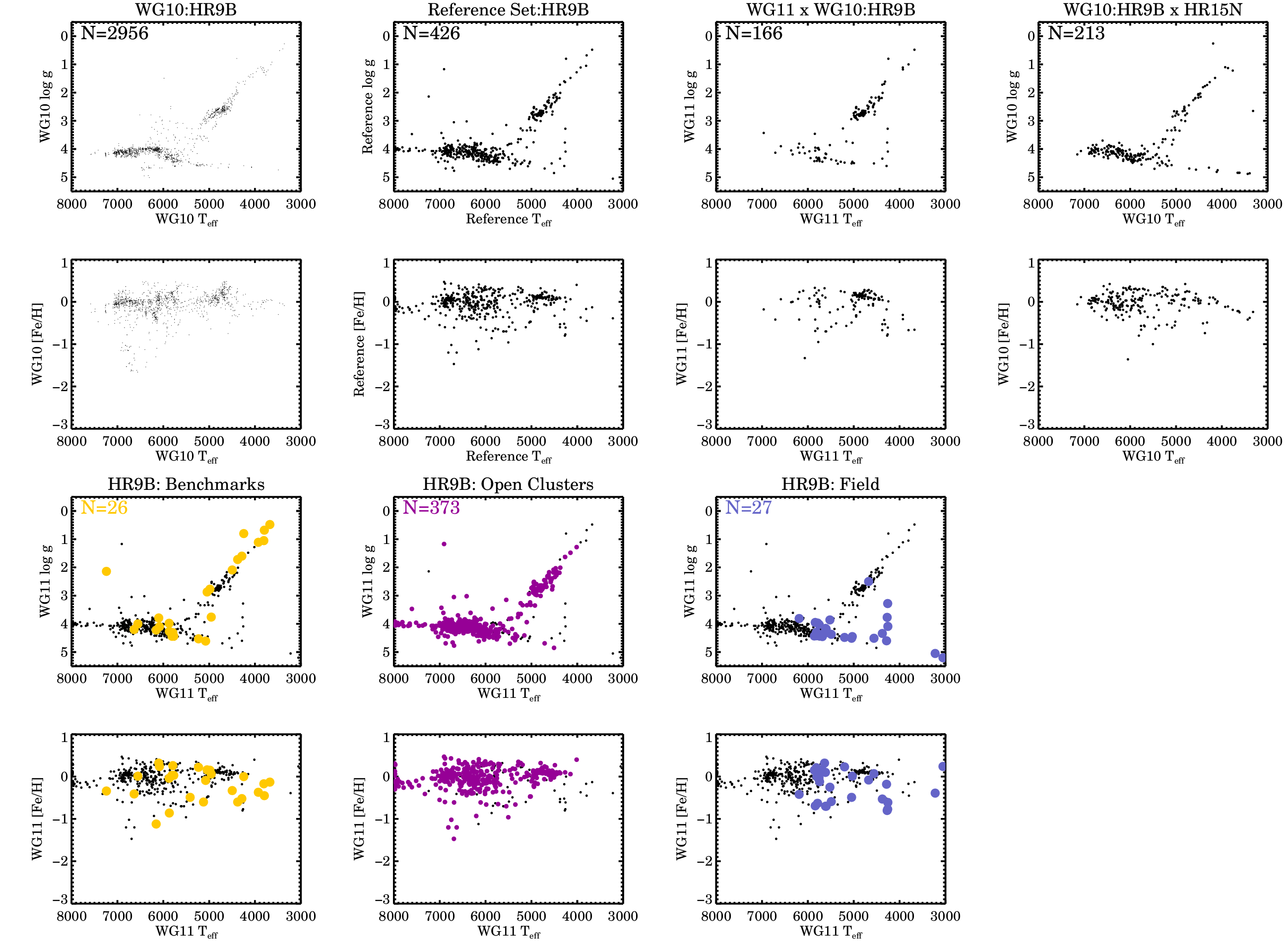}
\caption{Reference sets for HR21-only and HR9B SETUPs. Top to fourth rows: Same as in Figure~\ref{fig:hr1021_refset} but for HR21-only. The reference samples are benchmarks (yellow), BL (red), GCs (green) and MW (blue). Fifth to eighth rows: Same as in Figure~\ref{fig:hr1021_refset} but for HR9B.The reference samples are benchmarks (yellow), OC (magenta), and MW (blue).}
\label{fig:hr9b_refset}
\end{figure*}

Figures~\ref{fig:hr1021_refset} and \ref{fig:hr9b_refset} show the construction of the per SETUP reference sets as Kiel diagrams and \feh{} against \teff{}. The Kiel diagram of the full sample for each SETUP is shown, then the final reference set and the cross-match with WG11 are shown for comparison. The key samples as described above that are present in each reference set are also shown in order to show what part of the parameter space each key sample occupies.

\subsection{Corrections to node parameters based on reference sets}

\begin{figure*}
\centering
\includegraphics[width=0.99\linewidth]{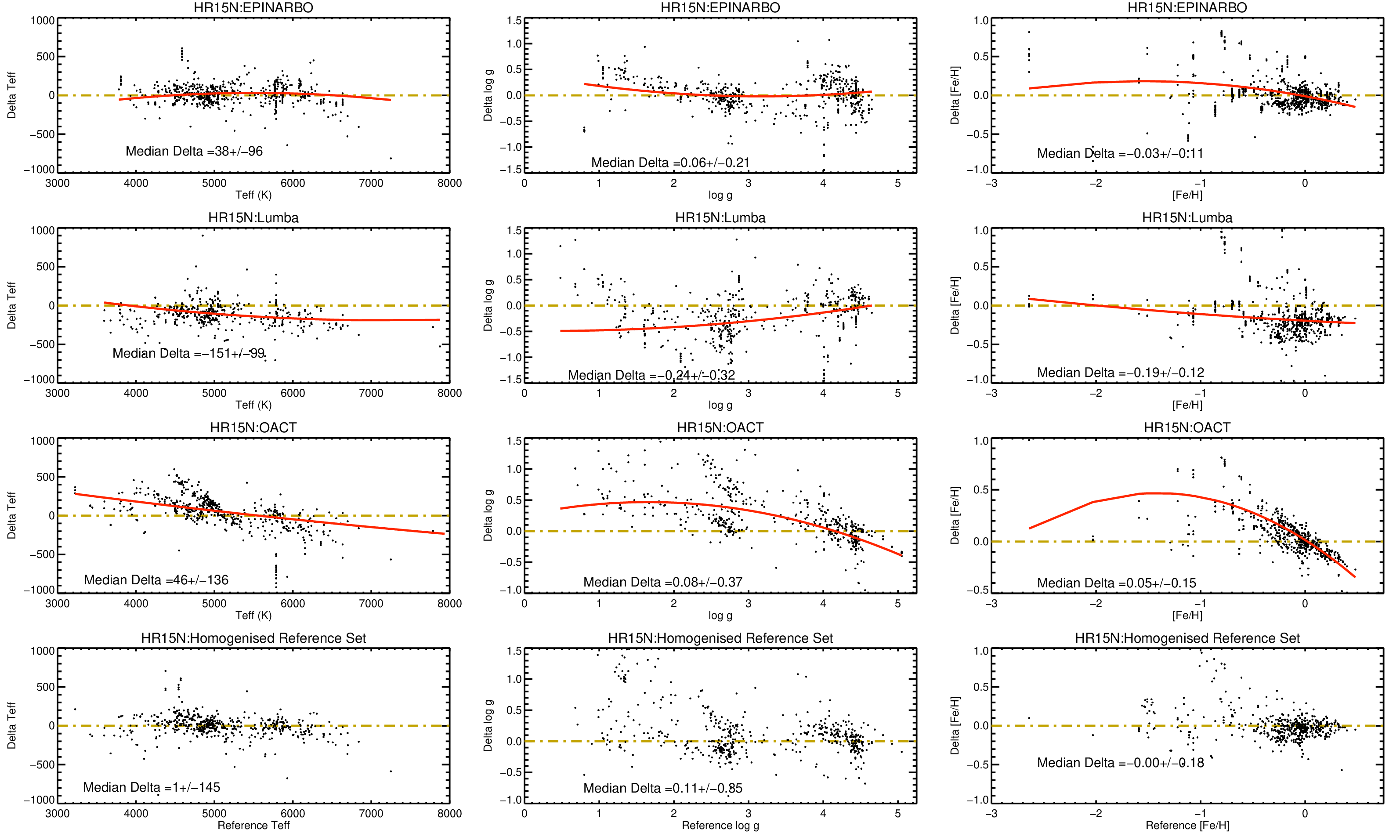}
\caption{Difference of node (EPINARBO, Lumba, OACT) values to reference set values for \teff{}, \logg{} and \feh{} for HR15N. The difference of the final homogenised reference set values from the reference set values per parameter for HR15N are shown in the bottom row..}
\label{fig:persu_hmg_hr15n}
\end{figure*}

As described in Section~\ref{sec:referencesets} the node results per SETUP were first assessed against the reference set to determine a bias correction. An example of the difference between the reference set values and the calculated bias corrections for the nodes and parameters for HR15N are shown in Figure~\ref{fig:persu_hmg_hr15n}. 
% Can produce whole set, but produce table of number of reference points (with values), offset and spread, bias correction coeeficients, scaling numbers, final setup median offset and spread.

Table~\ref{tab:nodes_paramcorcoefs} gives the coefficients for each bias correction per SETUP per node per parameter as well as the independent parameter against which each correction was calculated. The mean and standard deviation of the reference values per parameter are given per node. These values vary between the nodes as each node did not necessarily provide values for the complete set of reference stars. These values were used to normalise the node and reference values before the correction function was determined as described in Section~\ref{sec:wg11bayes}. The median and standard deviation of the difference between the node and reference values are also given per parameter per SETUP in Table~\ref{tab:nodes_paramcorcoefs}.

Each dataset was investigated in great detail using the relevant reference set in order to identify the polynomial function and independent parameter that provided the optimal correction. A variety of quality criteria were used to assess the agreement of the homogenised values to the reference values, such as difference measures (median and standard deviation) on the whole sample and sub-samples. In an extensive quality control process, results from different combinations of reference sets, polynomial fits, and independent parameters were compared to finally converge on the corrections provided in Table~\ref{tab:nodes_paramcorcoefs} using the WG11 Bayesian implementation (see Section~\ref{sec:wg11bayes})
.
%Despite efforts to enlarge the metal-poor part of the reference set it was still sparsely sampled and visual inspection was a necessary quality criteria. 

In the majority of cases, using \feh{} as the independent parameter provided the optimal correction. For HR10$|$HR21, the investigations showed that for all the nodes, there was a different trend with \feh{} between dwarfs and giants. Using \logg{} or \teff{} as the independent parameter did not capture the correction sufficiently either. Ultimately a two-parameter correction against both \feh{} and \logg{} was used in those cases, as listed in Table~\ref{tab:nodes_paramcorcoefs}.

%% Table per node/parameter  bias coefficients etc
% Table generated by IDL routine gen_node_bias_coefs_table.pro
\begin{sidewaystable*}
\setlength{\tabcolsep}{1pt}
\centering
\footnotesize{
\caption{Coefficients for the WG10 bias corrections.}
\begin{tabular}{llllcccccccccc}
SETUP   &  Node  &  Parameter  &  IP  &  a0   &  a1   &   a2  &  a3  &  a4  &  a5  & Mean & $\sigma$ & Median$\Delta$ & $\sigma\Delta$ \\
\hline
\hline
HR15N & Lumba & \teff{} & \feh{} & -0.167697384953 & -0.075824126601 & 0.011537441984 & -- & -- & -- & 5054 & 655 & -151 & 99 \\
HR15N & Lumba & \logg{} & \feh{} & -0.287599861622 & 0.140403673053 & 0.027343371883 & -- & -- & -- & 3.00 & 1.05 & -0.24 & 0.32 \\
HR15N & Lumba & \feh{} & \feh{} & -0.500064492226 & -0.076166518033 & 0.004766216502 & -- & -- & -- & -0.17 & 0.37 & -0.19 & 0.12 \\
HR15N & EPINARBO & \teff{} & \feh{} & 0.045188520104 & 0.021193103865 & -0.018201008439 & -- & -- & -- & 5134 & 618 & 38 & 96 \\
HR15N & EPINARBO & \logg{} & \feh{} & -0.018503241241 & -0.002362024272 & 0.043501205742 & -- & -- & -- & 3.14 & 1.02 & 0.06 & 0.21 \\
HR15N & EPINARBO & \feh{} & \feh{} & 0.066947847605 & -0.224765673280 & -0.029125435278 & -- & -- & -- & -0.15 & 0.37 & -0.03 & 0.11 \\
HR15N & OACT & \teff{} & \feh{} & 0.071406871080 & -0.113306954503 & 0.003251708345 & -- & -- & -- & 5107 & 711 & 46 & 136 \\
HR15N & OACT & \logg{} & \feh{} & 0.287826091051 & -0.217529743910 & -0.080652959645 & -- & -- & -- & 3.11 & 1.08 & 0.08 & 0.37 \\
HR15N & OACT & \feh{} & \feh{} & 0.291496694088 & -0.573199748993 & -0.082124628127 & -- & -- & -- & -0.15 & 0.36 & 0.05 & 0.15 \\
HR10|HR21 & Lumba & \teff{} & \feh{} & -0.039016053081 & 0.032239124179 & 0.021030869335 & -- & -- & -- & 4962 & 766 & 9 & 128 \\
HR10|HR21 & Lumba & \logg{} & \feh{} & -0.034867286682 & 0.066667184234 & -0.058418788016 & -- & -- & -- & 2.85 & 0.92 & -0.10 & 0.21 \\
HR10|HR21 & Lumba & \feh{} & \feh{}, \logg{} & -0.283124715090 & -0.306031435728 & 0.032216746360 & 0.284090071917 & 0.050201013684 & -0.124197311699 & -0.34 & 0.48 & -0.13 & 0.18 \\
HR10|HR21 & IAC & \teff{} & \feh{} & -0.123532809317 & -0.104557998478 & 0.042872212827 & -- & -- & -- & 4929 & 753 & -99 & 91 \\
HR10|HR21 & IAC & \logg{} & \feh{} & -0.569403111935 & 0.095033206046 & 0.056023135781 & -- & -- & -- & 2.80 & 0.90 & -0.32 & 0.31 \\
HR10|HR21 & IAC & \feh{} & \feh{}, \logg{} & -0.591382920742 & -0.113700605929 & -0.037862878293 & 0.105327464640 & 0.048286218196 & -0.083991259336 & -0.28 & 0.43 & -0.31 & 0.11 \\
HR10 & IAC & \teff{} & \feh{} & -0.086688518524 & -0.088158629835 & 0.048037074506 & -- & -- & -- & 4918 & 736 & -57 & 92 \\
HR10 & IAC & \logg{} & \feh{} & 0.055968381464 & -0.081914037466 & -0.028329251334 & -- & -- & -- & 2.79 & 0.90 & -0.06 & 0.27 \\
HR10 & IAC & \feh{} & \feh{}, \logg{} & -0.540781557560 & -0.212909042835 & -0.043102603406 & -0.006950300653 & -0.009692860767 & -0.011048068292 & -0.32 & 0.50 & -0.34 & 0.12 \\
HR10 & MaxPlanck & \teff{} & \feh{} & 0.090736053884 & -0.033964030445 & 0.045537993312 & -- & -- & -- & 4865 & 690 & 104 & 74 \\
HR10 & MaxPlanck & \logg{} & \feh{} & 0.010179239325 & -0.247776672244 & 0.079962849617 & -- & -- & -- & 2.73 & 0.89 & 0.08 & 0.19 \\
HR10 & MaxPlanck & \feh{} & \feh{}, \logg{} & 0.109406463802 & -0.192167446017 & -0.026684174314 & 0.117145642638 & -0.010831248015 & 0.012366904877 & -0.32 & 0.53 & 0.02 & 0.12 \\
HR21 & Lumba & \teff{} & \feh{} & -0.075124010444 & -0.131862834096 & 0.044369023293 & -- & -- & -- & 4739 & 525 & -89 & 108 \\
HR21 & Lumba & \logg{} & \feh{} & 0.178803175688 & -0.007759041619 & -0.016072347760 & -- & -- & -- & 2.63 & 0.67 & 0.02 & 0.15 \\
HR21 & Lumba & \feh{} & \feh{} & -0.101629003882 & 0.019427971914 & 0.006261380389 & -- & -- & -- & -0.63 & 0.73 & -0.01 & 0.00 \\
HR21 & IAC & \teff{} & \feh{} & -0.400498867035 & -0.305235207081 & 0.065208077431 & -- & -- & -- & 4719 & 484 & -297 & 120 \\
HR21 & IAC & \logg{} & \feh{} & -0.926806926727 & 0.118088871241 & -0.025646008551 & -- & -- & -- & 2.60 & 0.60 & -0.57 & 0.17 \\
HR21 & IAC & \feh{} & \feh{} & -0.380962938070 & 0.217933356762 & -0.017874522135 & -- & -- & -- & -0.54 & 0.63 & -0.50 & 0.00 \\
HR21 & MaxPlanck & \teff{} & \feh{} & -0.155645251274 & -0.184893101454 & 0.037887506187 & -- & -- & -- & 4722 & 492 & -61 & 163 \\
HR21 & MaxPlanck & \logg{} & \feh{} & -0.997872114182 & -0.109319239855 & 0.097923599184 & -- & -- & -- & 2.59 & 0.63 & -0.78 & 0.39 \\
HR21 & MaxPlanck & \feh{} & \feh{} & -0.087564639747 & -0.152285099030 & 0.025769094005 & -- & -- & -- & -0.60 & 0.71 & 0.10 & 0.00 \\
HR9B & EPINARBO & \teff{} & \feh{} & -0.015919817612 & -0.005959096830 & 0.037611626089 & -- & -- & -- & 5490 & 715 & -87 & 121 \\
HR9B & EPINARBO & \logg{} & \feh{} & -0.111597761512 & -0.097315594554 & -0.008452756330 & -- & -- & -- & 3.57 & 0.80 & -0.30 & 0.23 \\
HR9B & EPINARBO & \feh{} & \feh{} & 0.116780869663 & 0.078230343759 & -0.069311171770 & -- & -- & -- & -0.03 & 0.27 & 0.00 & 0.09 \\
HR9B & OACT & \teff{} & \feh{} & -0.001508530462 & -0.014328612015 & 0.013083978556 & -- & -- & -- & 5946 & 1078 & -11 & 93 \\
HR9B & OACT & \logg{} & \feh{} & 0.004861100577 & -0.046551045030 & 0.029044212773 & -- & -- & -- & 3.61 & 0.85 & -0.02 & 0.13 \\
HR9B & OACT & \feh{} & \feh{} & -0.059049095958 & -0.573390781879 & -0.052157264203 & -- & -- & -- & -0.02 & 0.25 & -0.04 & 0.13 \\
\hline\hline
\end{tabular}
\tablefoot{Coefficients for the bias corrections per SETUP (HR10|HR21, HR21, HR15N, HR9B), per Node (EPINARBO, Lumba, OACT, IAC, MaxPlanck), per Parameter (\teff{}, \logg{}, \feh{}). The independent parameter(s) (IP) against which the bias is calculated is listed. The mean and standard deviation ($\sigma$) of the parameter for the reference set per node used for scaling are provided. The mean and standard deviation of the offset ($\Delta$) between the node and reference values of the reference set for the node are given.}
\label{tab:nodes_paramcorcoefs}
}
\end{sidewaystable*}

%Parameter corrections are:
%HR15N_1ParamFEH_WG11xmat_BM_GGC
%HR10HR21_1ParamFEH_WG11xmat_BM_HR15N_CRT_GGC
%HR9B_1ParamFEH_WG11xmat_BM_HR15N
%HR21_1ParamFEH_HR10HR21xmat_BM_GGC

%FEH corrections
%HR15N_1ParamFEH_WG11xmat_Clean_BM_GGCmem3
%HR10HR21_2ParamFEHLOGG_WG11xmat_Clean_BM_GGCmem3_HR15N
%HR9B_1ParamFEH_WG11xmat_Clean_BM_HR15N
%HR21_1ParamFEH_HR10HR21xmat_Clean_BM_SD_CRT_GGCmem

\subsection{Combining SETUPs for final Working Group 10 parameter homogenisation}
The final step in the WG10 homogenisation process was to combine the per SETUP homogenisation into the per CNAME homogenisation, which is the single star catalogue for all CNAMEs analysed within WG10. Due to the bootstrapping procedure used to construct the reference sets, each homogenisation per SETUP was ultimately bootstrapped onto the WG11 scale.

Table~\ref{tab:setups_paramdeltas} lists the mean and standard deviation of the difference between the homogenised values and reference values per SETUP. In all cases, the offsets are close to zero and within the spread of the differences ($\Delta$) given by the standard deviation, which indicates very good agreement between the per SETUP homogenised values and the reference values. The dispersion of the difference (standard deviation) is generally two or three times higher than the typical uncertainties of the homogenised stellar parameters. Therefore, due to the bootstrapping procedure, the homogenised values per SETUP were all assumed to be on the WG11 parameter scale and can thus be combined without further correction.

%% Table per SETUP offsets : homogenised to reference
% Table generated by IDL routine plot_node_ref_biases.pro
\begin{table}[htbp]
\setlength{\tabcolsep}{2.5pt}
\centering
\footnotesize{
\caption{Difference ($\Delta$) of WG10 SETUP homogenisations to reference.}
\begin{tabular}{lccccc}
SETUP   &  Parameter & Mean $\Delta$ & $\sigma \Delta$ & Mean Error & $\sigma$ Error \\
\hline
\hline
HR15N & \teff{} & 1 & 145 & 70 & 15 \\
HR15N & \logg{} & 0.11 & 0.35 & 0.19 & 0.05 \\
HR15N & \feh{} & -0.00 & 0.18 & 0.06 & 0.02 \\
HR10|HR21 & \teff{} & 10 & 149 & 64 & 15 \\
HR10|HR21 & \logg{} & 0.03 & 0.31 & 0.16 & 0.03 \\
HR10|HR21 & \feh{} & -0.03 & 0.15 & 0.08 & 0.03 \\
HR21 & \teff{} & -28 & 126 & 27 & 14 \\
HR21 & \logg{} & -0.09 & 0.31 & 0.16 & 0.04 \\
HR21 & \feh{} & 0.04 & 0.15 & 0.11 & 0.02 \\
HR9B & \teff{} & 9 & 183 & 76 & 46 \\
HR9B & \logg{} & 0.05 & 0.24 & 0.09 & 0.08 \\
HR9B & \feh{} & -0.02 & 0.18 & 0.02 & 0.02 \\
\hline\hline
\end{tabular}
\tablefoot{Mean and standard deviation of difference ($\Delta$) between per SETUP (HR10|HR21, HR21, HR15N, HR9B) homogenised values and reference values per Parameter (\teff{}, \logg{}, \feh{}). The mean and standard deviation of the error for each parameter are also given.}
\label{tab:setups_paramdeltas}
}
\end{table}

The results per SETUP were combined into the final WG10 per CNAME catalogue. The majority of CNAMEs were observed using only one SETUP. However, for the cases in which results from multiple SETUPs were available (e.g. the reference sets) an order of priority was implemented reflecting the science programmes and calibration samples as specified by GES\_TYPE (see Table~\ref{tab:setup_info} for definitions). The priority order depending on the GES\_TYPE are given in Table~\ref{tab:priority_setups}.

\begin{table}[htbp]
\setlength{\tabcolsep}{1.0pt}
\centering
\footnotesize{
\caption{Priority order for selecting final result in cases of results from multiple SETUPs.}
\begin{tabular}{c|cc}
GES\_TYPE(s)   &  Priority & SETUP \\
\hline
\hline
MW (excluding BL), BM, & 1 & HR10$|$HR21 \\
BW, CR, K2, GC,  & 2 & HR21 \\
RV, TL, PC, MC & 3 & HR15N \\
 & 4 & HR9B \\
 \hline
CL, OC, BC & 1 & HR10$|$HR21 \\
 & 2 & HR15N \\
 & 3 & HR9B \\
 & 4 & HR21 \\
 \hline
BL & 1 & HR21 \\
 \hline\hline
\end{tabular}
\label{tab:priority_setups}
}
\end{table}
%'BM','GC','MW','BW','CR','K2','MC',PC,RV,TL -> prior1
%'_CL','SD_OC','BC' -> prior 2
%'_BL' -> prior3
%prior1 = ['HR10$|$HR21','HR21','HR15N','HR9B']
%prior2 = ['HR10$|$HR21','HR15N','HR9B','HR21']
%prior3 = ['HR21','HR10$|$HR21','HR15N','HR9B']

\begin{figure*}
\centering
\includegraphics[width=0.99\linewidth]{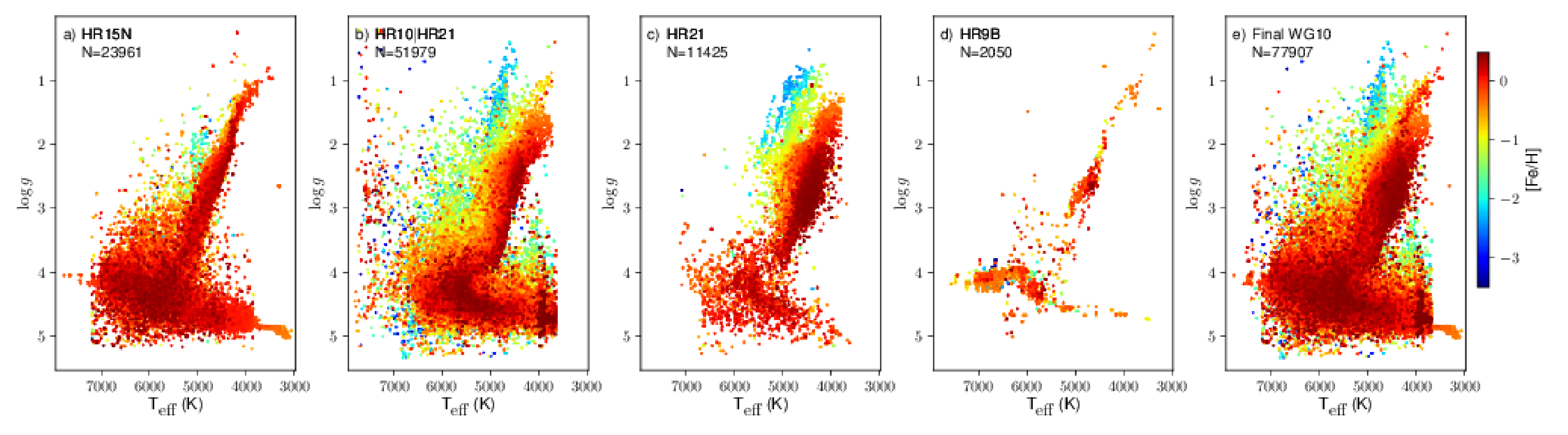}
\caption{Kiel diagram with a metallicity colour map for the per SETUP homogenised parameters: Panels from left to right: a) HR15N, b) HR10$|$HR21, c) HR21, d) HR9B, and e) the final WG10 homogenised parameters. All panels are on the same colour map scale.}
\label{fig:hrd_setups_fin}
\end{figure*}

Figure~\ref{fig:hrd_setups_fin} shows the homogenised parameters per SETUP and the final WG10 homogenised parameters as Kiel diagrams with a metallicity colour map. As no combining of values was performed, features in each per SETUP Kiel diagram can be identified within the final Kiel diagram, which itself coherently displays the morphology of the branches of stellar evolution. 

\begin{figure*}
\centering
\includegraphics[width=0.99\linewidth]{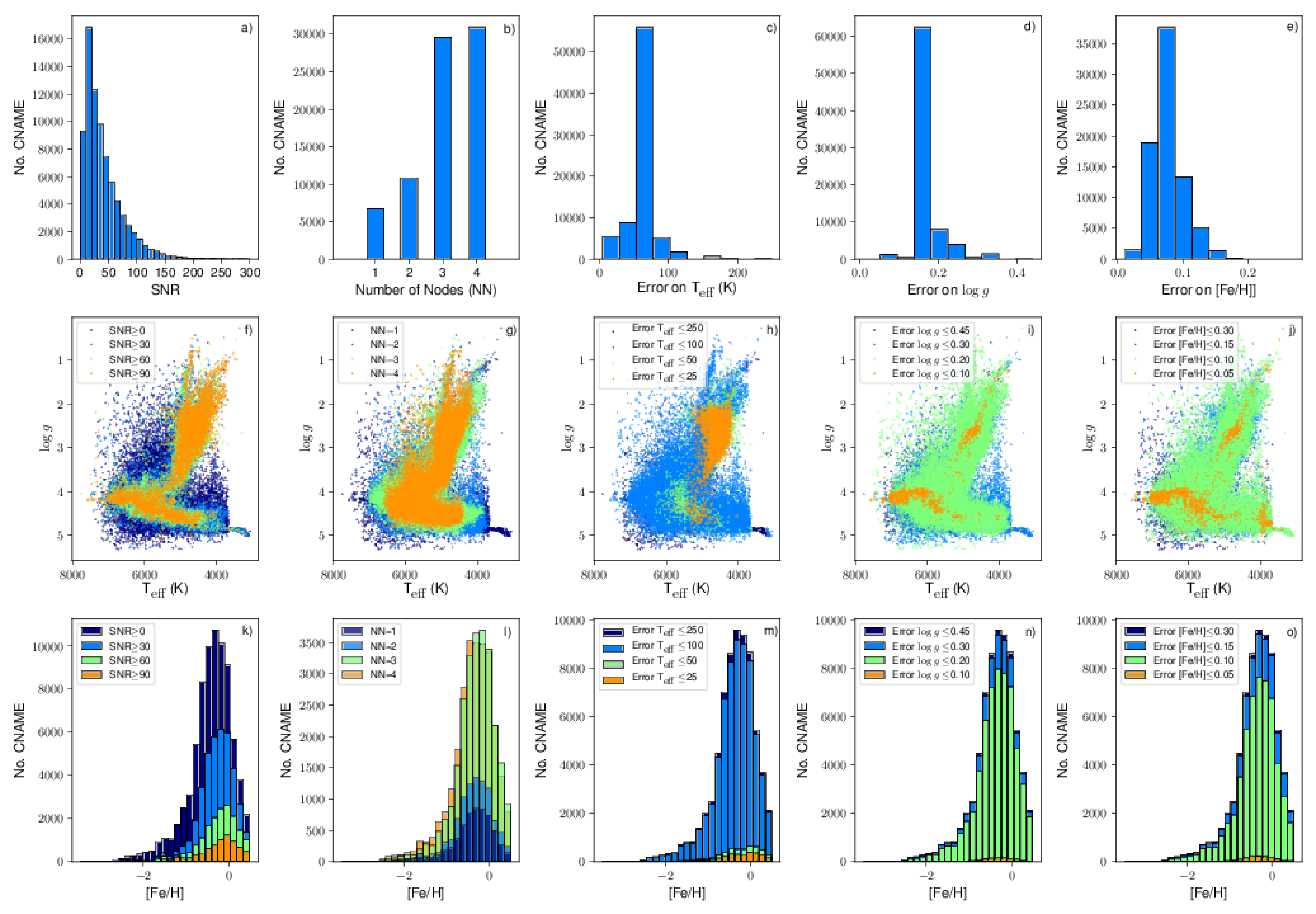}
\caption{Characterisation of the final WG10 stellar parameters with histograms (top row), bins in Kiel diagrams (middle row) and bins in metallicity distribution (bottom row). Specific panel content are: a,f,k) S/N; b,g,l) Number of Nodes (NN); c,h,m) Error on \teff{}; d,i,n) Error on \logg{}; e,j,o) Error on \feh{}}
\label{fig:hrd_fin_qc}
\end{figure*}

Figure~\ref{fig:hrd_fin_qc} characterises the final WG10 stellar parameters with respect to the key quality measures of S/N, number of nodes (NN) contributing to the final result, error on \teff{}, error on \logg{}, and error on \feh{}. The top row shows the histogram of each of these quantities, the middle row shows the Kiel diagrams of the final stellar parameters binned with respect to each quantity, and the bottom row shows the metallicity distribution as a histogram also binned with respect to each quantity. 

The S/N and NN both show a decrease in scatter and a more refined stellar evolution morphology in the Kiel diagrams with better quality results (e.g. more signal and more node results contributing, respectively). For \teff{}, \logg{}, and \feh{}, the errors have a significant peak around a particular value ($\sim$65~K, $\sim$0.17, $\sim$0.08, respectively), which is reflected in the binning of those quantities in the Kiel diagrams. However, the scatter is reduced with bins of decreasing error.

The metallicity distributions are less informative on this aspect. For the error quantities shown in Figure~\ref{fig:hrd_fin_qc}~(m, n, and o), the bulk of the values lie about a single value and thus fall mainly in a single bin.

However, Figure~\ref{fig:hrd_fin_qc}~k shows the peak of the metallicity distribution moving towards solar with bins of increasing S/N. This reflects the sampling of the medium-resolution data, as the fainter targets are typically more distant, and so the peak reflects the more metal-poor populations of the thick disk and the halo. The brighter targets are typically closer, sampling the thin disk and the solar neighbourhood, which are typically solar metallicity. Hence the metallicity distribution of the medium-resolution data reflects the expected trend of metallicity with stellar populations.

Figure~\ref{fig:hrd_fin_qc}~l shows the metallicity distribution binned with NN. The peak does not shift between bins and the shape of the distributions are similar, indicating that the bins with fewer than the maximum NN contain a similar sample in general, and that NN does not necessarily track with S/N. 

This discussion illustrates how the quality measures can, and indeed should, be used to refine the WG10 dataset for any study in Galactic Archeaology.  To be most effective these quality measures should be considered both individually and together.

\subsection{Verification of the homogenised Working Group 10 stellar parameters}
For the parameter homogenisation, as shown in Figures~\ref{fig:hr1021_refset} and \ref{fig:hr9b_refset}, key sub-samples were included within the parameter reference set. Verification of the WG10 homogenisation as part of the greater homogenisation of Gaia-ESO is explored in detail in \cite{hourihane2023} with particular attention to these sub-samples. As such, only the FGK benchmark stellar parameters, the WG11 cross-match stellar parameters, and the GC metallicities are reviewed in this section.

\begin{figure*}[ht]
\centering
\includegraphics[width=0.90\linewidth]{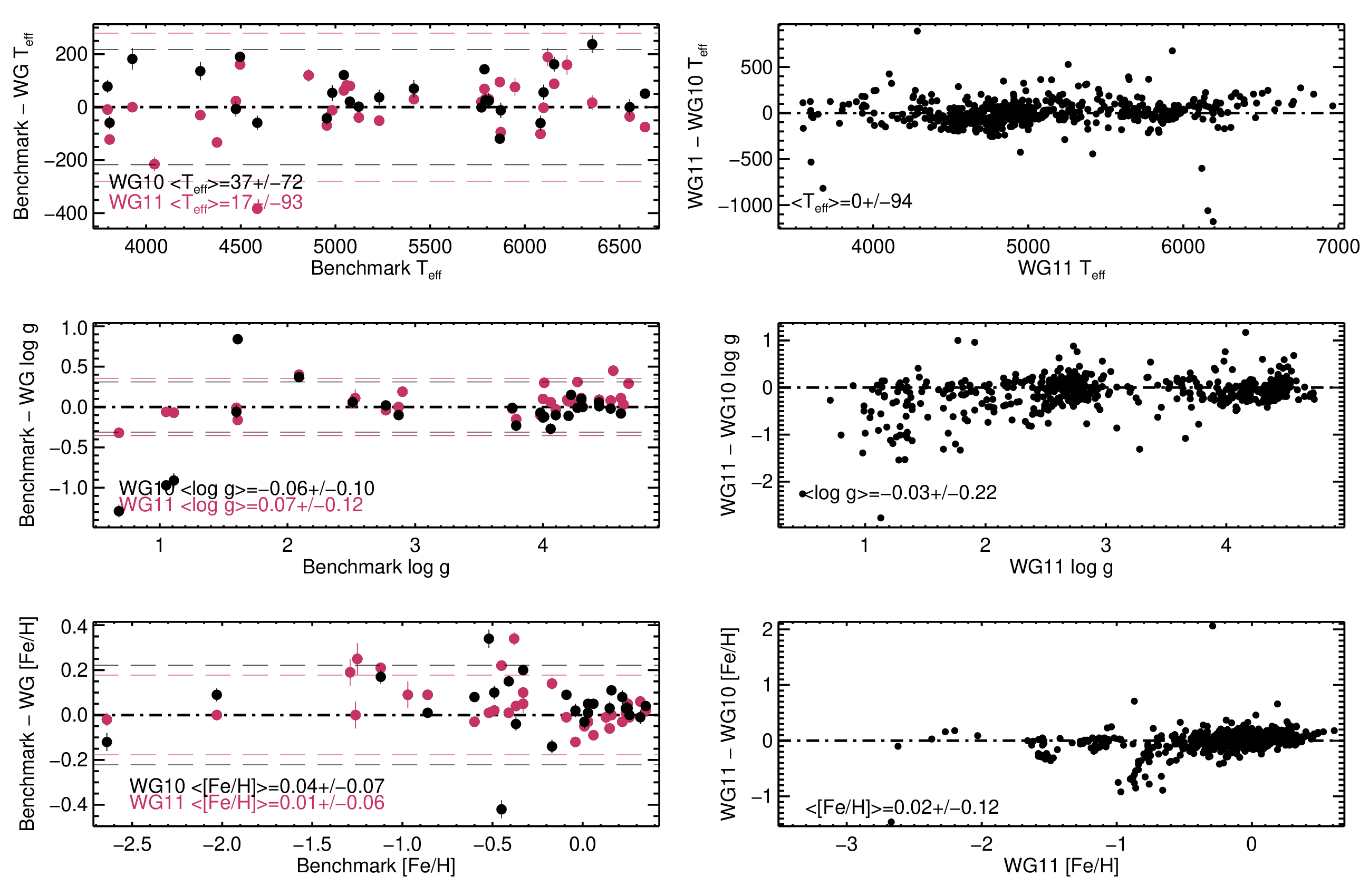}
\caption{Comparison of WG10 (black) and WG11 (red) stellar parameters for: Left column) the FGK benchmarks stars against the reference values.  The mean difference and standard deviation are given. Three sigma limits are shown as dashed lines. Right column) the cross-match between WG10 and WG11 against the WG11 values.  The mean difference and standard deviation are given.}
\label{fig:wg10_bm_wg11_ver}
\end{figure*}

Figure~\ref{fig:wg10_bm_wg11_ver} shows a comparison of the reference stellar parameters for the FGK benchmark stars with the values determined in both WG11 and WG10, and a comparison of the WG11 stellar parameters with the final WG10 stellar parameters for the cross-match between WG11 and WG10. The median and standard deviation of the differences are also given. Overall, the agreement between these reference sets and the final parameters is good, with relatively small offsets and small spread in differences within the typical errors of the stellar parameters.

Inspecting further the stars with large differences ($>3\sigma$) to the benchmark parameters, for WG11, 61\_Cyg\_B shows a notable disagreement in both \teff{} (-215 K) and \feh{} (0.34 dex). It is a close-to-solar-metallicity K dwarf which the nodes analyses should have dealt with quite well. However, the spectrum analysed was in fact non-UVES archive spectra from the benchmark spectral library made to be UVES-like for the WG11 node analyses in an expansion of the calibration effort.  Making the spectrum UVES-like may have caused an issue with the archived data, although this star is not in common with WG10 and thus was not used in the WG10 homogenisation.

HD122563, the very metal-poor (\feh{} = -2.64) luminous giant (\logg{} = 1.61), represents a difficult combination of parameters. The difficulty shows up as a significant difference compared to the benchmark parameters in \teff{} for the WG11 result (-383 K), and in \logg{} for the WG10 result (0.84). HD84937 is also a metal-poor (\feh{} = -2.03), albeit dwarf, star for which there was a significant difference in \teff{} for the WG10 result compared to the FGK benchmark result.

The WG10 node analyses also struggled with luminous giants, as shown by the trio of low \logg{} stars with large differences compared to the benchmark \logg{}.

Finally there is a difference in \feh{} of 0.34 dex for the WG10 results for the K giant, Arcturus, placing it as more metal poor than the FGK benchmark accepted value.

Overall, these discrepancies indicate that the WG10 and WG11 results in the parameter space of metal-poor stars and luminous giants are not as robust as in the parameter space of more metal-rich, high gravity stars within the survey dataset. This is not unexpected, as metal-poor stars and luminous giants were not the primary FGK science targets of the Gaia-ESO survey \citep{gilmore2022, randich2022},thus ensuring robust parameters for these types of stars was not the main focus of the node analyses.

However, the metal-poor stars, whilst few and only comprising two benchmarks, end up in quite good agreement with the reference \feh{} values for the WG10 homogenisation. As described above, to supplement the very few metal-poor benchmarks, the mean WG11 \feh{} value per GC was used to try to anchor the metal-poor end in the WG10 analysis by imposing that value on the respective highly probable cluster members in the WG10 and including them in the parameter reference set.

\begin{figure}[ht]
\centering
\includegraphics[width=0.95\linewidth]{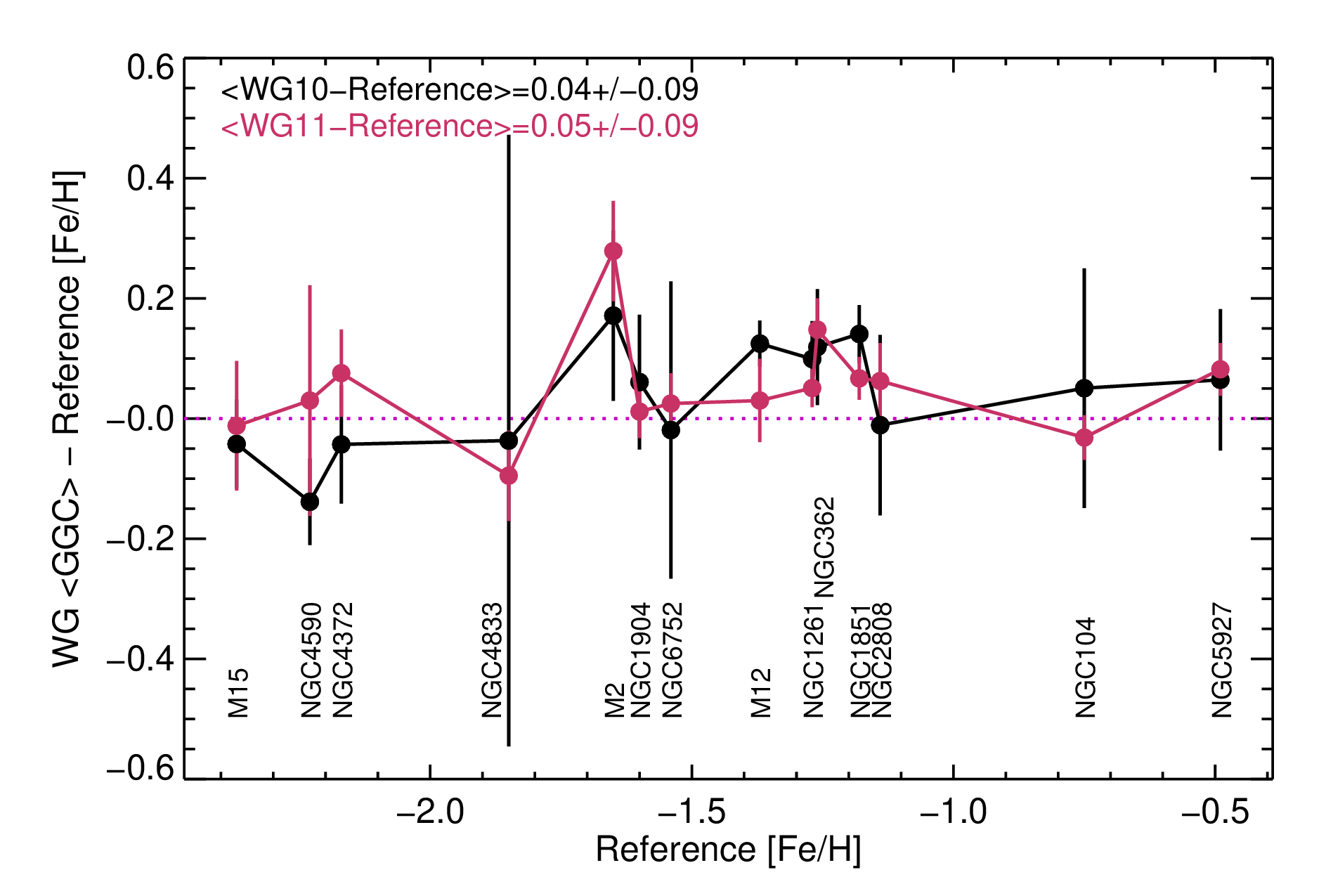}
\caption{Comparison of mean \feh{} per GC for WG10 (black) and WG11 (red) against the reference values. The mean difference and standard deviation are given.}
\label{fig:wg10_wg11_ggc_ver}
\end{figure}

Figure~\ref{fig:wg10_wg11_ggc_ver} shows the outcome of this effort, by comparing the mean \feh{} values of the GC members in WG11 and WG10 to the reference values \citep[][(2010 edition)]{harris1996}, where the WG11 values are those that were imposed in the reference set if needed. The mean and standard deviation of each WG sample to the reference values are also given.

The majority of the mean GC values for both WG10 and WG11 are within 0.1 dex of the reference values, with M2 being the main outlier. There is a large spread in \feh{} values for the WG10 stars defined here as members of NGC4833, although the mean value agrees well with the reference value. Otherwise, the spread in \feh{} per GC, particularly at the metal-poor end, are reasonable, and the mean \feh{} of each GC for WG10 generally track with WG11, indicating that the attempt to anchor the metal-poor end of the WG10 dataset with the WG11 GC mean values was relatively successful.

%\clearpage\afterpage
\section{Working Group 10 homogenisation of chemical abundances }\label{sec:abun_homog}
\begin{table*}[htbp]
   \setlength{\tabcolsep}{3pt}
   \centering
   \scriptsize{
   \caption{Summary of element abundance detections (D) and lines measured (L).}
   \begin{tabular}{|l|ccc|ccc|ccc|c|c|p{5cm}|}
   \hline
    \multicolumn{1}{|c}{} & \multicolumn{3}{c|}{HR10 (D,L)} & \multicolumn{3}{c|}{HR21 (D,L)}  & \multicolumn{3}{c|}{HR15N (D,L)} & HR9B (D,L) & WG11 & \multicolumn{1}{c|}{}\\ 
    \multicolumn{1}{|c}{Species} & Lumba & CAUP & Vilnius & Lumba & CAUP & Vilnius & EPINARBO & Lumba & CAUP & EPINARBO & X-Match & \multicolumn{1}{l|}{References} \\ 
\hline
\hline
AL1 & -,-,- & -,-,- & 10173,2 & 27068,1 & 23983,2 & 32238,2 & 9057,2 & 6021,2 & -,-,- & -,-,- & 555 & \citet{K75, WSM} \\
{\bf B}A2 & -,-,- & -,-,- & -,-,- & -,-,- & -,-,- & -,-,- & 1008,1 & 11946,1 & -,-,- & -,-,- & 553 &  \citet{MW} \\
C1 & 634,1 & -,-,- & -,-,- & -,-,- & -,-,- & -,-,- & 5536,1 & 1779,1 & -,-,- & -,-,- & 460 & \citet{K10, NIST10} \\
CA1 & -,-,- & 3065,1 & 33719,6 & -,-,- & -,-,- & -,-,- & 6561,2 & 12667,3 & 4758,1 & 887,2 & 583 & \citet{Sm, S0, SR} \\
CA2 & -,-,- & -,-,- & -,-,- & -,-,- & -,-,- & 40787,3 & -,-,- & -,-,- & -,-,- & -,-,- & 318 & \citet{T, TB} \\
CE2 & -,-,- & -,-,- & -,-,- & -,-,- & -,-,- & -,-,- & -,-,- & -,-,- & -,-,- & 686,2 & 387 & \citet{LSCI} \\
CO1 & 2883,1 & 300,1 & 21164,6 & -,-,- & -,-,- & -,-,- & 9139,2 & 4854,2 & -,-,- & 814,2 & 575 & \citet{K08} \\
CR1 & 17040,2 & 11275,2 & 39180,6 & -,-,- & -,-,- & -,-,- & -,-,- & -,-,- & -,-,- & 888,5 & 582 & \citet{WLHK} \\
CR2 & -,-,- & -,-,- & -,-,- & -,-,- & -,-,- & -,-,- & -,-,- & -,-,- & -,-,- & 205,1 & 540 & \citet{PGBH} \\
CU1 & -,-,- & -,-,- & -,-,- & -,-,- & -,-,- & -,-,- & -,-,- & -,-,- & -,-,- & 368,1 & 572 & \citet{K12} \\
DY2 & -,-,- & -,-,- & -,-,- & -,-,- & -,-,- & -,-,- & -,-,- & -,-,- & -,-,- & 75,1 & -  & \citet{WLN} \\
EU2 & -,-,- & -,-,- & -,-,- & -,-,- & -,-,- & -,-,- & 8019,1 & 2070,1 & -,-,- & -,-,- & 474 & \citet{LWHS} \\
FE1 & 35301,16 & 22137,3 & 40969,28 & 28165,7 & 20606,2 & 44674,7 & 9263,18 & -,-,- & 7117,4 & 888,17 & 580  & \citet{FMW, BK0, K07, BWL} \\
FE2 & -,-,- & -,-,- & 24823,2 & -,-,- & -,-,- & -,-,- & 8918,2 & -,-,- & -,-,- & 887,4 & 563  & \citet{K13}\\
LA2 & -,-,- & -,-,- & -,-,- & -,-,- & -,-,- & -,-,- & -,-,- & -,-,- & -,-,- & 465,1 & 316  &  \citet{LBS} \\
LI1 & -,-,- & -,-,- & -,-,- & -,-,- & -,-,- & -,-,- & 4905,1 & 319,1 & -,-,- & -,-,- & 548  &  \citet{LN} \\
MG1 & 17542,1 & -,-,- & 41162,2 & 24751,2 & 40834,2 & 44250,3 & -,-,- & -,-,- & -,-,- & 2,2 & 569  & \citet{KP, NIST10} \\
MN1 & 12838,2 & 2745,1 & 33859,4 & -,-,- & -,-,- & -,-,- & 6,1 & -,-,- & -,-,- & -,-,- & 550  &  \citet{K07} \\
MO2 & 198,1 & -,-,- & -,-,- & -,-,- & -,-,- & -,-,- & -,-,- & -,-,- & -,-,- & -,-,- & 222  &  \citet{WBb}\\
ND2 & 736,1 & -,-,- & -,-,- & -,-,- & -,-,- & -,-,- & 3493,1 & -,-,- & -,-,- & 601,1 & 567  & \citet{HLSC, MC} \\
NI1 & 4265,2 & 570,1 & 34672,5 & -,-,- & -,-,- & -,-,- & 8978,3 & 9961,4 & 4417,2 & 255,1 & 579  & \citet{K08, WLa}\\
PR2 & -,-,- & -,-,- & -,-,- & -,-,- & -,-,- & -,-,- & -,-,- & -,-,- & -,-,- & 274,1 & 234  & \citet{ILW, BLQS}\\
S1 & -,-,- & -,-,- & -,-,- & -,-,- & -,-,- & -,-,- & 7733,2 & -,-,- & -,-,- & -,-,- & 218  &  \citet{BQZ, K04} \\
SC2 & 14039,1 & 616,1 & -,-,- & -,-,- & -,-,- & -,-,- & 9063,1 & -,-,- & 2937,1 & -,-,- & 575  &  \citet{LD} \\
SI1 & -,-,- & -,-,- & -,-,- & 9219,1 & -,-,- & 19448,1 & 9230,2 & 5416,2 & 4074,1 & -,-,- & 575  &  \citet{K07, GARZ} \\
SR1 & -,-,- & -,-,- & -,-,- & -,-,- & -,-,- & -,-,- & 3516,1 & -,-,- & -,-,- & -,-,- & -  & \citet{GC} \\
TI1 & -,-,- & -,-,- & 31799,7 & -,-,- & -,-,- & 18235,2 & 9027,2 & 7878,2 & -,-,- & 888,5 & 584  & \citet{K10, NWL, LGWSC} \\
TI2 & -,-,- & 511,1 & 35712,6 & -,-,- & -,-,- & -,-,- & -,-,- & -,-,- & -,-,- & 884,2 & 575  & \citet{WLSC} \\
V1 & 2209,1 & -,-,- & -,-,- & -,-,- & -,-,- & -,-,- & 7321,1 & -,-,- & -,-,- & -,-,- & 575  &  \citet{K09} \\
Y2 & 406,2 & -,-,- & 5647,2 & -,-,- & -,-,- & -,-,- & 4117,1 & -,-,- & -,-,- & 861,2 & 571  &  \citet{PN, K11, BBEHL, HLGBW} \\
ZR1 & 243,1 & -,-,- & -,-,- & -,-,- & -,-,- & -,-,- & 84,1 & -,-,- & -,-,- & -,-,- & 399  &  \citet{BGHL} \\
    \hline
    \end{tabular}
    \tablefoot{Summary per node per SETUP of the number of CNAMEs that have abundance detections (D) and average number of spectral lines measured per species (L). The number of CNAMEs in the cross-match with all WG10 SETUPs for which each abundance is measured by WG11.}
  \label{tab:node_abun_det}
  }
\end{table*}

The strategy of the chemical abundance homogenisation was to combine in a single step the per spectral line element abundances derived by each node for each SETUP per CNAME. We refer to these element abundances as the line-by-line (LbL) abundances. Hence, all SETUPs were combined at once per CNAME rather than homogenising the results for one CNAME one SETUP at a time and then combining the per SETUP results.

\begin{figure*}
\centering
\includegraphics[width=0.77\linewidth]{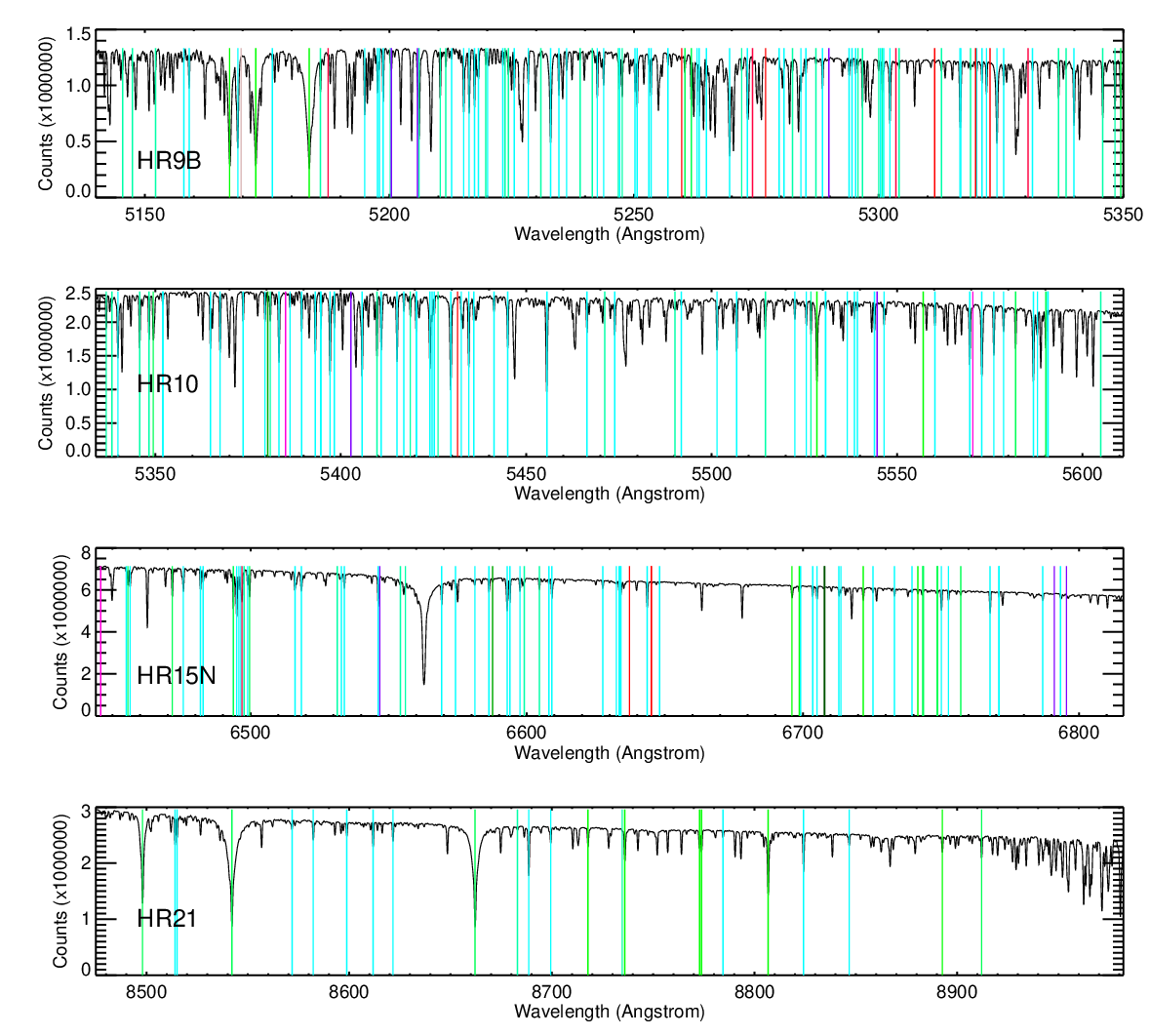}
%\hfill
\includegraphics[width=0.1275\linewidth,trim={2cm 0cm 0cm 0cm},clip]{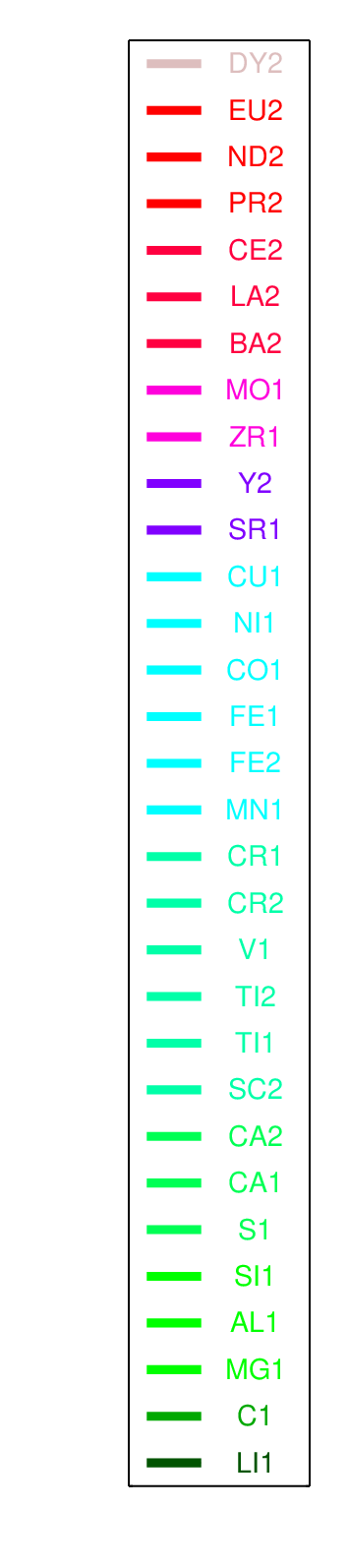}
\caption{ESO Solar spectrum (\url{https://www.eso.org/observing/dfo/quality/UVES/pipeline/FLAMES_solar_spectrum.html}) reduced with the Gaia-ESO GIRAFFE reduction pipeline for the WG10 SETUPs. Spectral lines analysed by the WG10 nodes are indicated by vertical lines coloured by groupings of elements.}
\label{fig:wg10_spectra}
%\end{figure*}
\includegraphics[width=0.95\linewidth]{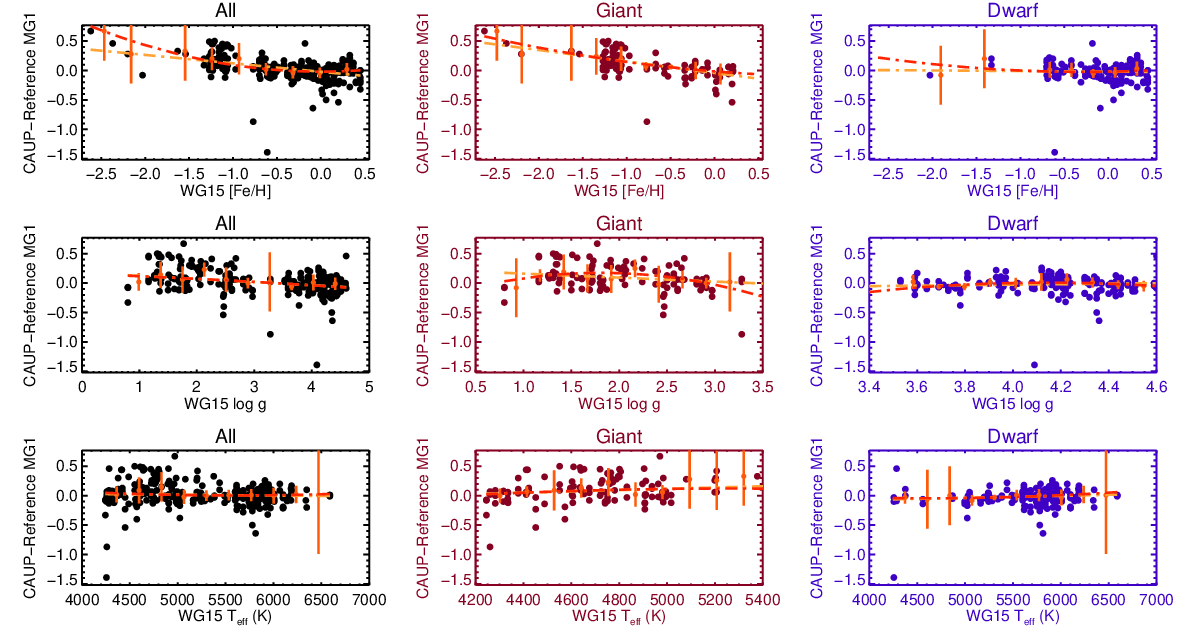}
%/Users/charlotteworley/Documents/GES/WG10/iDR6AbundanceHomog/Analysis
\caption{Difference between node values and reference values in example case of the CAUP analysis of Mg\,\textsc{i} measured in the HR21 spectrum. Top row: Differences against \feh{} for the full sample (left-black),  the giants (middle-red),  and the dwarfs (right-blue). Middle row: Same but against \logg{}.  Bottom row: Same but against \teff{}. The differences as median differences calculated per bins are shown as orange points with error bars. The linear fit (orange dot-dash) and quadratic fit (red dot-dash) are shown in each case.}
\label{fig:abun_corrections}
\end{figure*}

The wavelength ranges across the solar spectrum for each of the four WG10 SETUPs, and the location of the spectral lines used by the nodes to measure abundances are shown in Figure~\ref{fig:wg10_spectra}. The list of all lines measured is provided in Table~\ref{tab:spectralines}. These are taken from the Gaia-ESO line list \citep{heiterLL}. In the following tables and figures, a capitalised format for designating the elements is used in which the final digit indicates the ionisation state (1=neutral, 2=singly ionised). This matches the data model used within the survey.

Table~\ref{tab:node_abun_det} gives the number of CNAMEs analysed by each node per element species per SETUP as well as the specific line list references. Two numbers are provided: `D' is the number of detections, and `L' is the average number of spectral lines measured per species. The number of CNAMEs in the WG11 cross-match to all the WG10 SETUPs with WG11 abundances per element species is also given. There was no requirement on the nodes to measure the abundance of every possible element in all four SETUPs. Hence, as can be seen in Table~\ref{tab:node_abun_det}, the node results are a complex dataset with varied coverage of the chemical abundance space.

% Table generated manuall
\begin{table}[htbp]
\setlength{\tabcolsep}{3pt}
\centering
\footnotesize{
\caption{Element measurements rejection criteria.}
\begin{tabular}{lcc}
SETUP & ELEMENT & Crit.  \\
\hline
\hline
%EPINARBO HR15N V1 ERR >0.8
HR15N &  AL1 &  >10 \\
HR15N &  CO1 & >7 \\
HR15N &  EU2 &  >4 \\
HR15N &  NI1 &  >8 \\
HR15N &  NI1 &  <2 \\
HR15N &  TI1  & >9 \\
HR15N &  TI1  & <2 \\
HR10 &  C1 &  >10 \\
HR10 &  MO1 &  >4 \\
HR10 &  V1 &  >5 \\
HR10 &  Y2 &  >5 \\
HR21 &  MG1 &  >15 \\
HR21 &  SI1 &   >20 \\
\hline\hline
\end{tabular}
\tablefoot{Elements for specific SETUPs and Node analyses for which extrema in absolute abundance were rejected if above (>) or below (<) the rejection criteria (Crit.) threshold.}
\label{tab:elements_extrej}
}
\end{table}

\subsection{Removing extrema from node results}
Table~\ref{tab:elements_extrej} gives some broad rejection criteria that were applied to specific element datasets, as extreme values (in absolute abundance) were identified that were not reasonable compared to the bulk of the distribution. Further cleaning was of course possible, but the goal was to take the node analyses as provided and to try to use as much information as possible. Quality measures such as S/N, NN, and errors should be used to refine the dataset as needed. This allows for differences between scientific studies regarding the tolerable level of uncertainty in the data.

\subsection{Homogenisation procedure for Working Group 10 chemical abundances}

It was important to follow a methodical procedure to obtain the optimal homogenisation of the WG10  LbL chemical abundances. The Bayesian inference method used in the parameter phase was not used here due to the range of incompleteness in the measurements, as the gaps made it difficult to apply the method consistently. 

A simple procedure was therefore employed, of which the key steps are as follows. (1) Calculate the correction to the WG11 element abundance scale for each element based on LbL abundances for each SETUP for each node, using the set of cross-matched stars of the SETUP to WG11. (2) On a per star per element basis, apply the correction for each node for each setup. (3) Reject LbL abundances following rules set by WG11. (4) Take the median of the corrected LbL abundances across all nodes and SETUPs per element per star to calculate the final abundance for that star. (5)Take the standard deviation of the corrected LbL abundances for the element for the star as the error on the element abundance. (6)Take the number of NODE+SETUP analysed as the NN contributions to the abundance determination.

The homogenised abundances were then assessed using the quality control samples that are described in the following section. As the full distribution could then be inspected, this revealed issues with the corrections that could not be detected on the much smaller cross-match with WG11. Each element distribution was inspected, and adjustments to the correction were made when warranted such that the homogenisation was run again. This iterative process from correction to homogenisation to quality control to correction was repeated several times to home in on the optimal homogenisation.

\subsection{Correction to Working Group 11 cross-match reference set}\label{sec:wg10_abund_xmat_wg11}
The strategy used in the parameter homogenisation (i.e. bootstrap each SETUP onto a reference set based on the previous SETUP plus other reference stars) could not be employed for the abundance analysis due to the decision to homogenise all spectral lines for an element across all SETUPs at once. Thus, for each CNAME, all the spectral line abundances from all the possible setups from all the possible nodes were combined to derive the final abundance.  There was no homogenisation of each setup in turn.  Thus the only reference set available was the cross-match to WG11.  No bootstrapping between SETUPs was possible.

The WG11 cross-match was not complete in the parameter space; in particular, there were gaps in the \feh{} space.  When the cross-match was deemed insufficient, alternate procedures were adopted, which are explained below. The number of stars (CNAMEs) in the cross-match between WG11 and each of the WG10 SETUPs is given in Table~\ref{tab:wg11xsetups}.  However, depending on the node and the SETUP, there were not necessarily abundances for the full set of stars in each cross-match.  In some cases, there were no abundances available for the WG11 cross-match stars, or there were no WG11 abundances available at all.

In the general case, for each ELEMENT+SETUP+NODE combination a set of corrections were calculated between the node values and the reference set values for each parameter as given in Table~\ref{tab:set_abuncor}.

\begin{table}[h!]
\setlength{\tabcolsep}{3pt}
\centering
\footnotesize{
\caption{Set of corrections calculated for each parameter for each ELEMENT+SETUP+NODE combination.}
\begin{tabular}{cp{7cm}}
\hline
\hline
1 & Median and standard deviation of the difference \\
\hline
2 & Linear fit to the difference against: \\
  & a) FEH \\
  & b) TEFF \\
  & c) LOGG \\
\hline
3 & Quadratic fit to the difference against: \\
  & a) FEH \\
  & b) TEFF \\
  & c) LOGG \\
\hline\hline
\end{tabular}
\label{tab:set_abuncor}
}
\end{table}

The corrections were calculated on the binned difference between the node values and the WG11 values in the associated SETUP cross-match. The corrections were calculated for the whole sample as well as separately for the dwarf (\logg $> 3.4$) and the giant (\logg $\le 3.4$) samples.  

An example of the set of corrections that was calculated for a particular element for a particular node for a particular SETUP is shown in Figure~\ref{fig:abun_corrections}. This example (CAUP+HR21+MG1) shows how the difference between the node values and the reference values can behave differently depending on the independent variable that is used and how the sample is or is not separated. In this case, the median offset was applied for the dwarf sample, while the quadratic fit against FEH was applied for the giant sample, see Table~\ref{tab:abun_corrections}.

For each NODE+SETUP+ELEMENT combination,  the difference was calculated between the LbL abundances and the reference abundance value for the WG11 cross-match (black points in Figure~\ref{fig:abun_corrections}). The set was then divided into ten evenly distributed bins spanning the range of the reference values for the respective parameter (\teff{}, \logg{}, or \feh{}). The median and standard deviation of the differences in each bin were then calculated (orange points with error bars in Figure~\ref{fig:abun_corrections}). The median difference and standard deviation, linear fit, and quadratic fit to the binned data points were then calculated (shown as orange dot-dashed line and red dot-dashed line, respectively in Figure~\ref{fig:abun_corrections}).  The coefficients and goodness of fit for the range of corrections were returned and examined.  

Table~\ref{tab:abun_corrections} gives the coefficients of the fit and the parameter range for the final set of corrections. While useful numbers with which to derive a correction were returned for the majority of SETUP+NODE+ELEMENT combinations (indicated as WG11xmat in the `Calibration' column of the table), there were nonetheless cases for which there were not enough data points with which to work. 

There were three exceptions to the general case: (1) Insufficient element abundances in the WG11 cross-match sample but a reasonably useful number in the rest of the WG11 dataset (WG11-full). (2) No WG11 abundances at all for that element  (Scaled Solar). (3) Super-solar trend in HR21 compared to HR10  (HR21toHR10-WG11).

In the first case, there were nine SETUP+NODE+ELEMENT combinations for which the full WG11 element abundance distribution was used to estimate a correction. These combinations are listed in Table~\ref{tab:wg11full_element}.

% Table generated by hand
\begin{table}[htbp]
\setlength{\tabcolsep}{3pt}
\centering
\footnotesize{
\caption{Abundance corrections derived using the full WG11 dataset.}
\begin{tabular}{lll}
SETUP  &  NODE & ELEMENT \\
\hline
\hline
HR10  &  Lumba  &  MO1 \\
HR10  &  Lumba  &  ZR1 \\
HR15N  &  EPINARBO  &  BA2 \\
HR15N  &  EPINARBO  &  ND2 \\
HR15N  &  EPINARBO  &  ZR1 \\
HR9B  &  EPINARBO  &  CR2 \\
HR9B  &  EPINARBO  &  CU1 \\
HR9B  &  EPINARBO  &  LA2 \\
HR9B  &  EPINARBO  &  PR2 \\
\hline\hline
\end{tabular}
\tablefoot{List of SETUP+NODE+ELEMENT combinations for which the sample was compared to the full W11 dataset for deriving the correction relation, WG11-full.}
\label{tab:wg11full_element}
}
\end{table}

In the second case, there were two combinations for which the only option was to scale to the solar abundance. These two combinations are listed in Table~\ref{tab:elecor_scalesol}.
% Table generated by hand
\begin{table}[htbp]
\setlength{\tabcolsep}{3pt}
\centering
\footnotesize{
\caption{Abundance corrections derived using the Solar Chemical Composition.}
\begin{tabular}{lll}
SETUP  &  NODE & ELEMENT \\
\hline
\hline
HR15N  &  EPINARBO  &  SR1 \\
HR9B  &  EPINARBO  &  DY2 \\
\hline\hline
\end{tabular}
\tablefoot{List of SETUP+NODE+ELEMENT combinations for which the sample was compared to the solar chemical abundance for deriving the correction offset, Scaled Solar.}
\label{tab:elecor_scalesol}
}
\end{table}

In the third case, comparison of the HR21 BL to HR10 BL abundances revealed an exaggerated upturning to enhanced abundances at the metal-rich end. Further exploration showed that this was a difference between the giants and dwarfs. The dwarf sample did not show this in neither HR21 nor HR10. This upturning seemed extreme for an astrophysical effect, but it could not be compared to the reference sample as there were no stars in common between HR21 and WG11 for the bulge sample.

However, there was the cross-match sample between HR21 and HR10 to examine. The equivalent set in HR10 did not show such an extreme upturning at the metal-rich end, though for some abundances it was slightly present which could indicate an astrophysical effect. The goal was to put giants in HR21 onto the same scale as giants in HR10 but not to remove the feature completely if present in both sets of results.

Thus HR21 giants cross-matched to the HR10 giants sample were used to remove any systematic without erasing a potential astrophysical signature.  However,  just because the giants in HR21 behaved differently compared to the dwarfs in HR21, this did not necessarily mean that the dwarfs in HR21 behaved the same as those in HR10. It was necessary to investigate a correction to HR10 for the dwarf targets in HR21 to also ensure all targets were put onto the HR10 scale. As HR10 was corrected onto the WG11 scale separately, this was carried out first. Then HR21 was corrected onto HR10, which had already been corrected onto the WG11 scale. The SETUP+NODE+ELEMENT combinations for which the corrections needed to be calculated are listed in Table~\ref{tab:hr21hr10supersol}.

% Table generated by hand
\begin{table}[htbp]
\setlength{\tabcolsep}{3pt}
\centering
\footnotesize{
\caption{Extreme abundance enhancement present at Super-solar metallicity.}
\begin{tabular}{lll}
SETUP  &  NODE & ELEMENT \\
\hline
\hline
HR21  &  CAUP  &  FE1 \\
HR21  &  Lumba  &  FE1 \\
HR21  &  Vilnius  &  CA2 \\
HR21  &  Vilnius  &  FE1 \\
HR21  &  Vilnius  &  MG1 \\
HR21  &  Vilnius  &  TI1 \\
\hline\hline
\end{tabular}
\tablefoot{List of SETUP+NODE+ELEMENT combinations for which the sample showed an upturn at super-solar \feh{} in HR21 but not in HR10, HR21toHR10-WG11.}
\label{tab:hr21hr10supersol}
}
\end{table}

\begin{figure*}
\centering
\includegraphics[width=1.0\linewidth,trim={2cm 0cm 0cm 0cm}, clip]{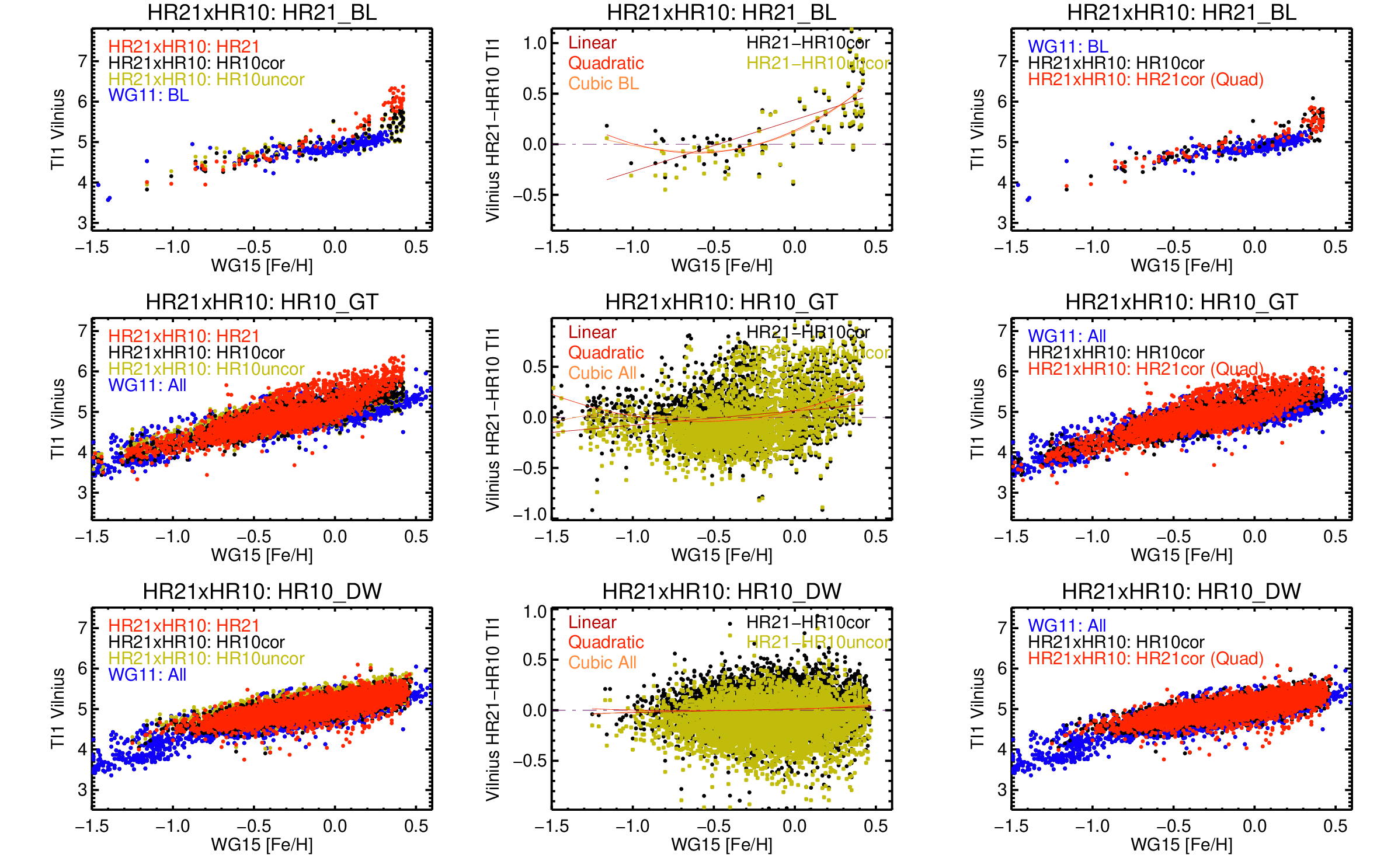}
\caption{Comparing HR21 (by cross-match) with HR10 uncorrected (HR10uncor), HR10 corrected to WG11 (HR10cor), and WG11. Top row: the bulge sample (BL); Middle row: the giant sample (GT); Bottom row: the dwarf sample (DW). First column: Ti\,\textsc{i} abundances against [Fe/H]; Second column: Difference between HR21 and HR10 abundances (HR10uncor and HR10cor) with linear, quadratic and cubic fits; Third column: Application of correction from quadratic fit to HR21 (HR21cor).}
\label{fig:vilti1hr21}
\end{figure*}

Figure~\ref{fig:vilti1hr21} illustrates the process of determining and applying the correction using the Vilnius and Ti\,\textsc{i} results as an example. The panels show the Ti\,\textsc{i} abundances against [Fe/H], comparing HR21 with the uncorrected HR10 (HR10uncor), HR10 corrected to WG11 (HR10cor), and WG11. The first row shows the cross-match to HR10 for the bulge sample (BL), the second row shows the giant sample (GT) and the third row shows the dwarf sample (DW). The first column shows HR21 uncorrected. The upturn at super solar is clear in the HR21 giant sample (we note that the bulge stars are also giants) when compared to the HR10 giant sample, the dwarfs in both HR21 and HR10, and WG11. The second column shows the linear, quadratic and cubic fit to the difference in Ti\,\textsc{i} values of the cross-match between HR21 and HR10uncor and HR10cor for the bulge, giants, and dwarfs. The third column applies the correction from the quadratic fit to the HR21 values in each case. The procedure successfully scales HR21 and HR10 to WG11 while retaining any subtle potentially astrophysical effects.

\subsection{Errors in the homogenised abundances}
\begin{figure*}
\centering
\includegraphics[width=0.99\linewidth]{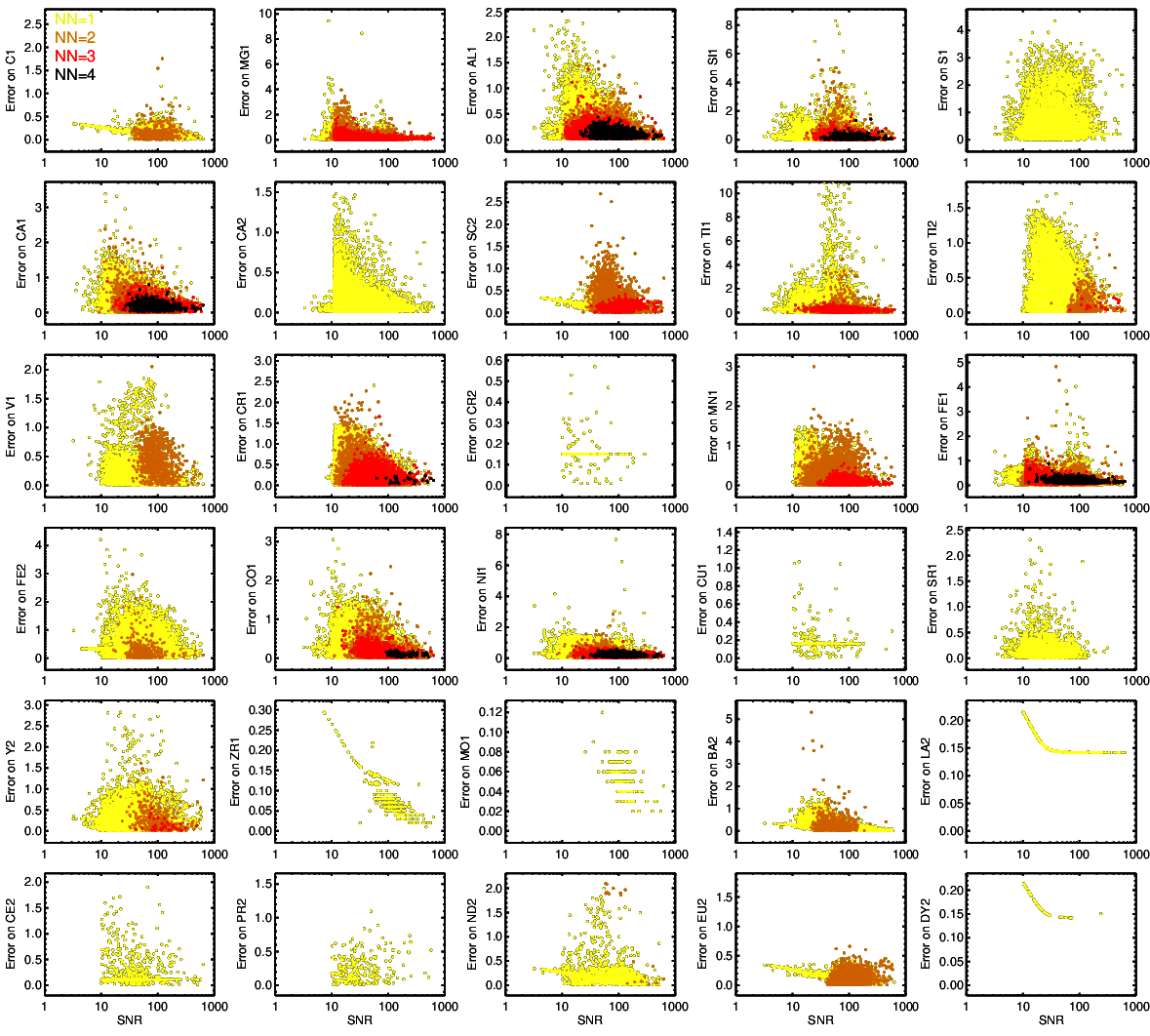}
\caption{Errors on WG10-recommended abundances (in absolute abundance) against S/N on a log scale. Sub-samples per NN are shown as specified in the top-left panel.} \label{fig:rec_abunerrs}
\end{figure*}

Figure~\ref{fig:rec_abunerrs} shows the error distributions against S/N for the homogenised WG10 abundances. For the LbL abundances, the error in the abundances was calculated as the standard deviation of the set of node LbL abundances used per target in the homogenisation. In some cases if a single line abundance from a single node was the only abundance available, then the error provided by that node was reported as the error. 

There were three situations in which errors would potentially end up missing from the final homogenisation: (1) No error was provided with the single line abundance measurement although errors for other measurements for the same element for that node were provided. (2) No errors were provided at all by the node for the LbL abundances of that element. (3) No errors were provided at all by the node for the LbL abundances of any element.

Three relations were derived to complete the final errors. (1) An error relation with S/N was generated for each NODE+SETUP+ELEMENT combination. (2) An error relation with S/N was generated by combining all reported errors for all abundances provided by the node for the SETUP. (3) For each CNAME and SETUP, the spread in each abundance that had more than one line was calculated across the node values and was used to calculate an uncertainty relation with S/N per abundance per SETUP.

Thus, in the homogenisation procedure, if the final abundance was based on a single node value that did not have an associated error, the appropriate relation was used to provide an estimate of the uncertainty, which was then reported as the final error on that abundance. In this way, all values in the WG10 abundance homogenisation have associated error values as shown in Figure~\ref{fig:rec_abunerrs}.

However, we observed some extreme outliers and large spread in error values for some elements. Figure~\ref{fig:rec_abunerrs} shows the errors by sub-sample of NN, in which the lowest errors and highest S/N are typically represented by the maximum NN sub-sample. The highest errors typically occur when less than three node results are combined. In particular, when (and despite the fact that) the S/N is high ($\sim100$), the homogenised error is also high ($>2$), such as for MG1, SI1, TI1, and FE1. This may not just be attributed to a better result by combining more nodes, but also to differences in the level of data quality for which nodes reported results. A consistent error model imposed across the nodes would have improved the resulting dataset.

%Might need some numbers on how often this was the case.
%For cases of eabun < 0.01 then eabun was set to 0.01 as we were providing abundnaces nominally to 2s.f. or 2d.p.

\subsection{Verification of the Working Group 10 homogenised chemical abundances}

\begin{figure*}
\centering
\includegraphics[width=0.99\linewidth]{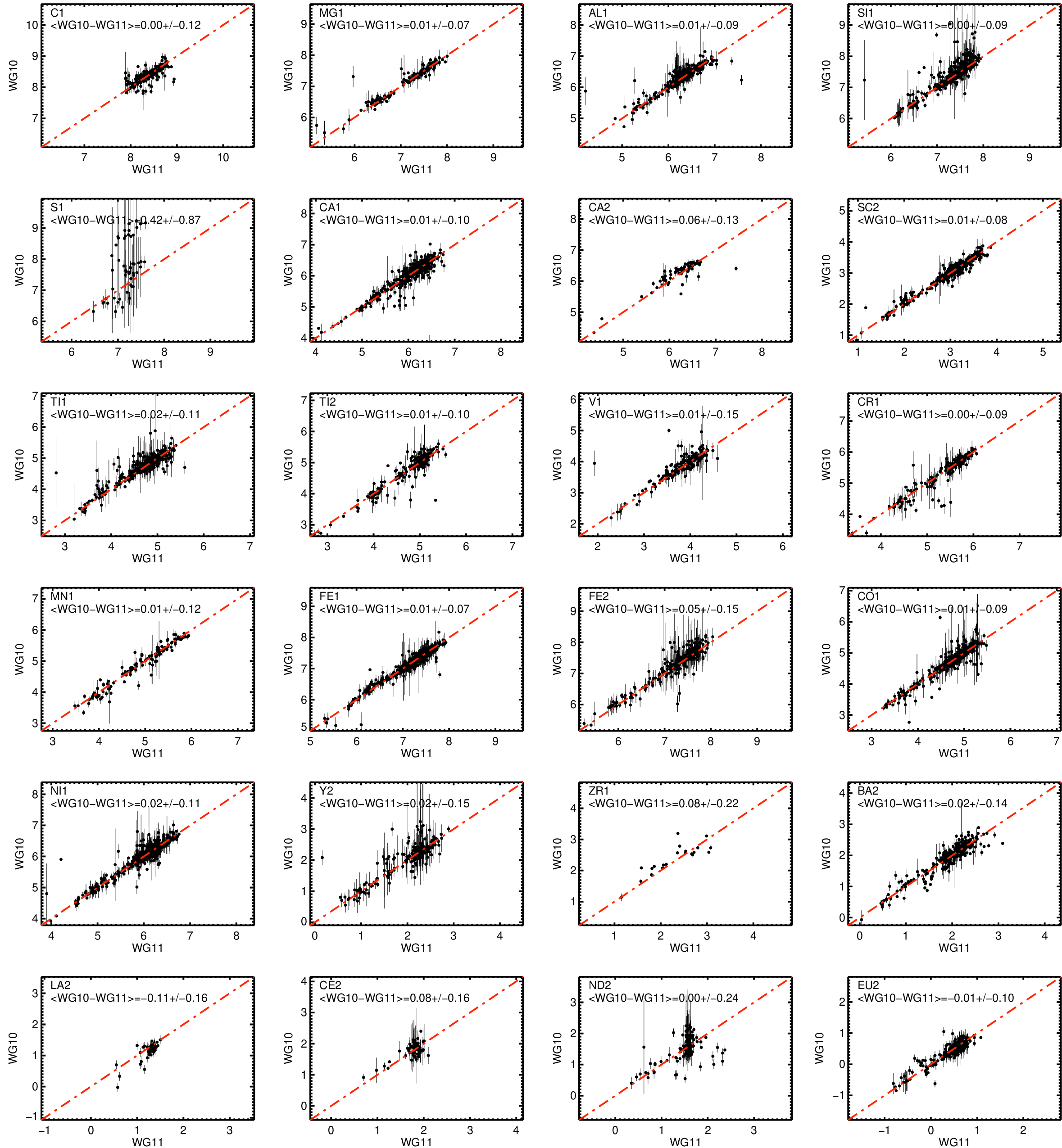}
\caption{Final WG10 abundances against WG11 abundances for the CNAMEs in common with WG11 for each WG10 homogenised element. The median difference and spread are provided for each element.}
\label{fig:finabuncomp_wg10_wg11}
\end{figure*}

The reference set used for the calibration of the WG10 chemical abundances was the cross-match with WG11 as described in Section~\ref{sec:wg10_abund_xmat_wg11}. Detailed quality checks on the WG10 homogenisation regarding key sub-samples in the context of the full survey have been carried out in \cite{hourihane2023}. In this work, we only inspect the comparison to WG11 in Figure~\ref{fig:finabuncomp_wg10_wg11}.

The median difference and standard deviation are given for each element. In general, the agreement is very good across the elements, with a spread on the order of typical uncertainties in abundance measurements. With a subset of the cross-match S\,\textsc{i} shows an issue, though the bulk of the cross-match are in reasonable agreement. The error bars per CNAME are also large indicating greater uncertainty in the measurement of this element.

\section{Conclusions}\label{sec:conclusion}

The homogenisation of the WG10 results across four SETUPs with analyses from multiple nodes that covered, often sparsely, different ranges in stellar parameters and chemical abundances was a challenging process. The goal was to produce a robust and well-calibrated single star catalogue that could be homogenised with the rest of the survey results. This meant optimally combining the node results following the WG11 Bayesian inference method for the WG10 and WG11 stellar parameters as well as for the WG11 chemical abundances, while for the WG10 chemical abundance, a simple per analysis calibration to WG11 was carried out. Crucial to the robustness of the final WG10 catalogue is understanding the quality of the results. In particular, the S/N, NN, and errors are key to refining the sample for any scientific study.

\begin{figure*}[ht]
\centering
\includegraphics[width=0.48\linewidth,trim={1cm 0cm 2cm 1cm}, clip]{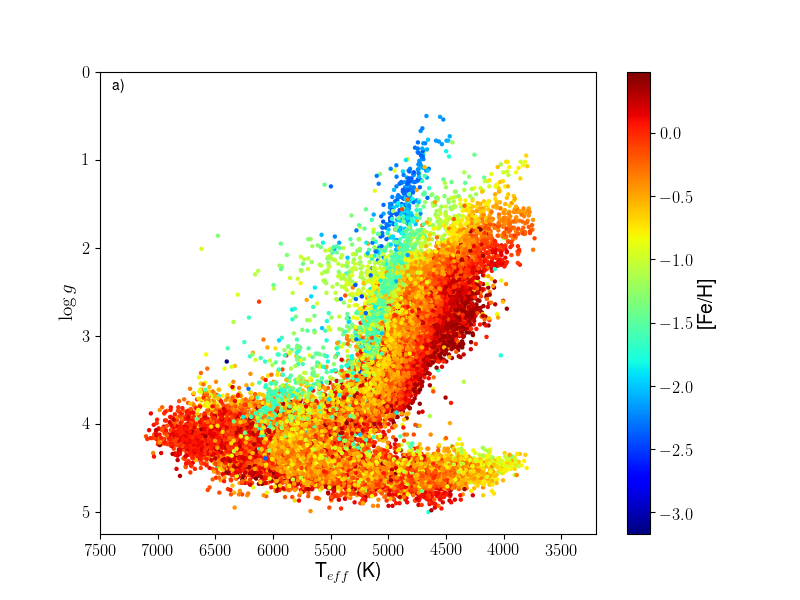}
\hfill
\includegraphics[width=0.48\linewidth,trim={0.5cm 0cm 2cm 1cm}, clip]{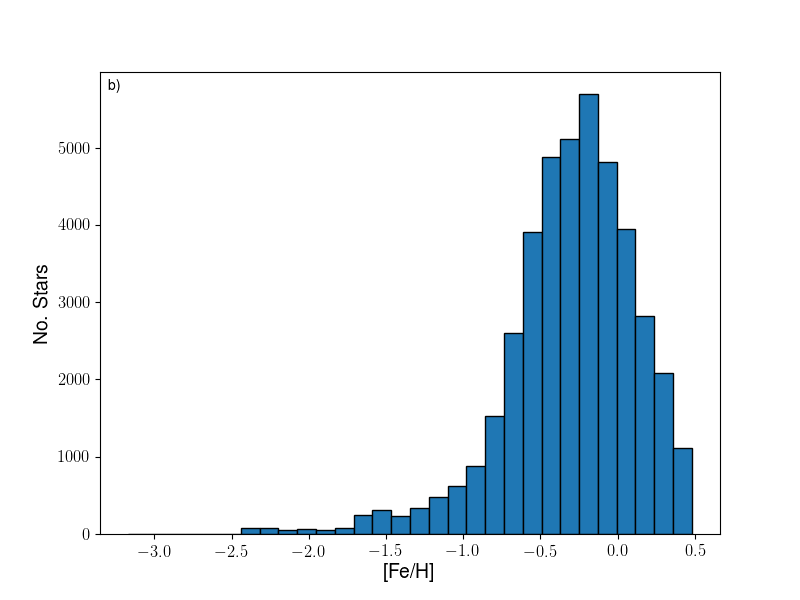}
\caption{Final WG10 stellar parameters with S/N$\ge$25 and NN$\ge$2 as a) Kiel diagram with metallicity colour map, and b) Metallicity distribution.}
\label{fig:hrd_fin}
\end{figure*}

The final stellar parameters as a Kiel diagram and metallicity distribution, with a simple cleaning of S/N$\ge$25 and NN$\ge$2, is shown in Figure~\ref{fig:hrd_fin}. The scatter is considerably reduced in the Kiel diagram with respect to the full sample. The shift of the RGB with metallicity is clearly discernible. The metallicity distribution shows a left-handed asymmetry indicative of the metal-poor contribution from the thick disk. Small peaks at -1.5 and -2.4 coincide with globular cluster samples.

\begin{figure*}[ht]
\centering
\includegraphics[width=0.95\linewidth]{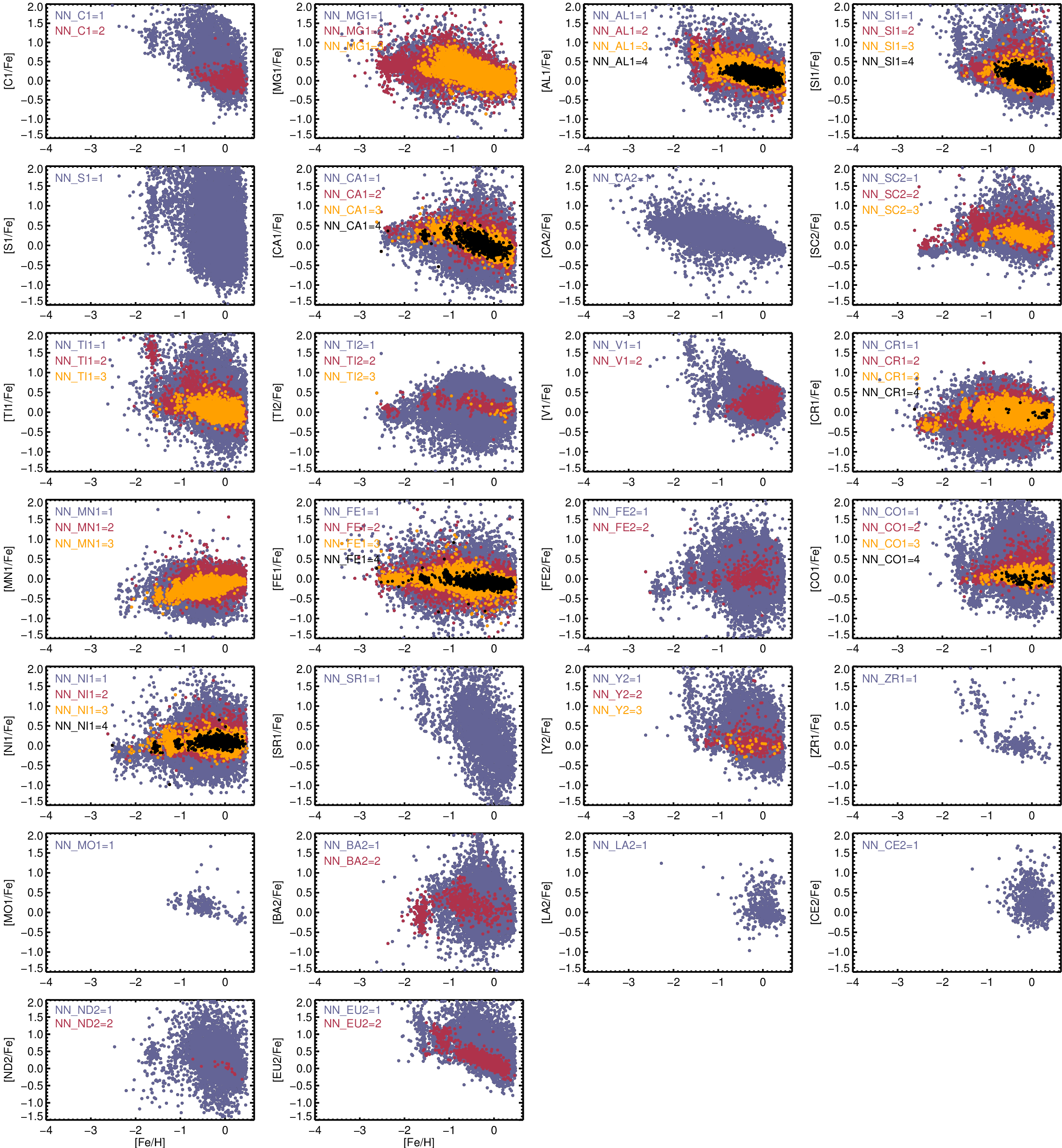}
\caption{Final WG10 chemical abundances as [X/Fe] against [Fe/H] illustrating the process of selecting by the NN contributing to the final abundance.}
\label{fig:chemabun_fin}
\end{figure*}

The final WG10 chemical abundances are shown in Figure~\ref{fig:chemabun_fin} as [X/Fe] against [Fe/H]. The abundances are binned by the NN that contributed to each abundance. The greater the NN, the clearer the morphology of the distribution, illustrating how these quality measures can be used to interpret this complex and intriguing dataset.

\begin{acknowledgements}
Based on data products from observations made with ESO Telescopes at the La Silla Paranal Observatory under programme ID 188.B-3002. These data products have been processed by the Cambridge Astronomy Survey Unit (CASU) at the Institute of Astronomy, University of Cambridge (supported by UKRI-STFC grants: ST/N005805/1, ST/T003081/1 and ST/X001857/1), and by the FLAMES/UVES reduction team at INAF/Osservatorio Astrofisico di Arcetri.

These data have been obtained from the Gaia-ESO Survey Data Archive, prepared and hosted by the Wide Field Astronomy Unit, Institute for Astronomy, University of Edinburgh, which is funded by the UK Science and Technology Facilities Council.

This work was partly supported by the European Union FP7 programme through ERC grant number 320360 and by the Leverhulme Trust through grant RPG-2012-541. We acknowledge the support from INAF and Ministero dell' Istruzione, dell' Universit{\`a}' e della Ricerca (MIUR) in the form of the grant ``Premiale VLT 2012". The results presented here benefit from discussions held during the Gaia-ESO workshops and conferences supported by the ESF (European Science Foundation) through the GREAT Research Network Programme.

D.M. acknowledges financial support from the Agencia Estatal de Investigaci{\'o}n 10.13039/501100011033 of the Ministerio de Ciencia e Innovaci{\'o}n and the ERDF ``A way of making Europe” through project PID2019-109522GBC54.

F.J.E acknowledges support from ESA through the Faculty of the European Space Astronomy Centre (ESAC) - Funding reference 4000139151/22/ES/CM.

H.M.T. acknowledges financial support from the Agencia Estatal de Investigaci{\'o}n (AEI/10.13039/501100011033) of the Ministerio de Ciencia e Innovaci{\'o}n and the ERDF ``A way of making Europe'' through project PID2019-109522GB-C51.

S.V. gratefully acknowledges the support provided by Fondecyt reg. 1220264 and by the ANID BASAL projects ACE210002 and FB210003.

\v{S}.M. thanks the COST Action CA18104: MW-Gaia.

U.H. acknowledges support from the Swedish National Space Agency (SNSA/Rymdstyrelsen).

E.J.A. acknowledges financial support from the State Agency for Research of the Spanish MCIU through the ``Center of Excellence Severo Ochoa" award to the Instituto de Astrof{\'i}sica de Andaluc{\'i}a (CEX2021- 001131-S).

T.B. was funded by the project grant No. 2018-04857 from the Swedish Research Council.

G.G. acknowledges support by the Collaborative Research Centre SFB 881 (projects A5, A10), Heidelberg University, of the Deutsche Forschungsgemeinschaft (DFG, German Research Foundation) and by the European Research Council (ERC) under the European Union’s Horizon 2020 research and innovation programme (grant agreement 949173).

P.J. acknowledges support from Fondecyt Regular Ladder Number 1231057, Millenium Nucleus ERIS NCN2021\_017, Centros ANID Iniciativa Milenio.

J.I.G.H. acknowledges financial support from the Spanish Ministry of Science and Innovation (MICINN) project PID2020-117493GB-I00.

E. M. acknowledges financial support through a ``Margarita Salas" postdoctoral fellowship from Universidad Complutense de Madrid (CT18/22), funded by the Spanish Ministerio de Universidades with NextGeneration EU funds.

\end{acknowledgements}

\bibliographystyle{aa}
\bibliography{worley_wg10homog}

\onecolumn
\begin{appendix}

% Lines used in WG10
\section{Working Group 10 spectral line information}
% Code: LbL/gen_finlinelist.pro
\begin{table*}[h!]
   \setlength{\tabcolsep}{2pt}
   \centering
   \scriptsize{
   \caption{Spectral lines used by WG10 Nodes.}
   \begin{tabular}{lp{17cm}}
   \hline
Species & Wavelength (\AA) \\ 
\hline
\hline
Al\,\textsc{i} & 5557.06, 6696.02, 6698.67, 8772.87, 8773.90 \\
Ba\,\textsc{ii} & 6496.90 \\
C\,\textsc{i} & 5380.33, 6587.61 \\
Ca\,\textsc{i} & 5260.39, 5261.70, 5349.46, 5581.96, 5590.11, 6455.60, 6471.66, 6493.78, 6499.65 \\
Ca\,\textsc{ii} & 8498.02, 8542.09, 8662.14, 8912.07 \\
Ce\,\textsc{ii} & 5187.46, 5274.23, 5330.56 \\
Co\,\textsc{i} & 5176.08, 5212.69, 5280.63, 5301.04, 5331.45, 5352.04, 5352.07, 5530.73, 5530.78, 5590.66, 5590.74, 6454.99, 6632.44, 6771.02, 6771.03 \\
Cr\,\textsc{i} & 5200.17, 5206.02, 5238.96, 5241.46, 5247.56, 5272.00, 5287.18, 5296.69, 5300.75, 5304.18, 5312.86, 5318.77, 5345.80, 5348.31, 5409.78 \\
Cr\,\textsc{ii} & 5246.77 \\
Cu\,\textsc{i} & 5218.20, 5220.07 \\
Dy\,\textsc{ii} & 5169.69 \\
Eu\,\textsc{ii} & 6645.06, 6645.10 \\
Fe\,\textsc{i} & 5159.06, 5194.94, 5197.94, 5198.71, 5215.18, 5216.27, 5217.39, 5223.18, 5225.53, 5228.38, 5232.94, 5236.20, 5242.49, 5243.78, 5247.05, 5250.21, 5250.65, 5253.02, 5253.46, 5262.88, 5263.31, 5269.54, 5273.16, 5279.65, 5285.13, 5288.52, 5293.96, 5294.55, 5295.31, 5300.40, 5302.30, 5320.04, 5321.11, 5322.04, 5324.18, 5339.93, 5364.87, 5367.47, 5373.71, 5379.57, 5383.37, 5386.33, 5389.48, 5393.17, 5397.13, 5398.28, 5405.77, 5410.91, 5415.20, 5417.03, 5424.07, 5429.70, 5434.52, 5441.34, 5445.04, 5455.61, 5466.40, 5473.90, 5491.83, 5501.46, 5506.78, 5522.45, 5525.54, 5536.58, 5538.52, 5539.28, 5543.94, 5546.51, 5560.21, 5569.62, 5572.84, 5576.09, 5586.76, 6475.62, 6481.87, 6494.98, 6495.74, 6496.47, 6498.94, 6518.37, 6533.93, 6546.24, 6569.21, 6574.23, 6581.21, 6592.91, 6593.87, 6597.56, 6608.02, 6609.11, 6627.54, 6633.41, 6633.75, 6634.11, 6648.08, 6699.14, 6703.57, 6705.10, 6713.05, 6713.74, 6725.36, 6733.15, 6739.52, 6750.15, 6752.71, 6786.86, 6793.26, 8514.07, 8515.11, 8571.80, 8582.26, 8598.83, 8611.80, 8621.60, 8688.62, 8699.45, 8784.44, 8824.22, 8846.74 \\
Fe\,\textsc{ii} & 5169.03, 5197.57, 5234.62, 5256.93, 5264.80, 5284.10, 5316.61, 5316.78, 5325.55, 5425.25, 6442.96, 6456.38, 6516.08 \\
La\,\textsc{ii} & 5303.53 \\
Li\,\textsc{i} & 6707.76, 6707.91 \\
Mg\,\textsc{i} & 5167.32, 5172.68, 5183.60, 5528.40, 8717.83, 8736.02, 8806.76 \\
Mn\,\textsc{i} & 5394.62, 5394.67, 5420.35, 5420.42, 5432.50, 6440.93 \\
Mo\,\textsc{i} & 5570.44 \\
Nd\,\textsc{ii} & 5276.87, 5311.45, 5319.81, 5431.52, 6637.19 \\
Ni\,\textsc{i} & 5157.98, 5424.65, 5435.86, 5578.72, 5587.86, 6482.80, 6532.87, 6586.31, 6643.63, 6767.77 \\
Pr\,\textsc{ii} & 5259.73, 5322.77 \\
S\,\textsc{i} & 6743.48, 6743.54, 6743.58, 6748.57, 6748.58, 6748.79, 6757.15 \\
Sc\,\textsc{ii} & 5526.79, 6604.60 \\
Si\,\textsc{i} & 6721.85, 6741.63, 8892.72 \\
Sr\,\textsc{i} & 6546.78, 6791.02 \\
Ti\,\textsc{i} & 5145.46, 5147.48, 5152.18, 5210.38, 5219.70, 5223.62, 5224.30, 5230.97, 5282.38, 5295.78, 5300.01, 5338.31, 5426.25, 5471.19, 5490.15, 5514.53, 6497.68, 6554.22, 6556.06, 6599.10, 8682.98, 8734.71 \\
Ti\,\textsc{ii} & 5185.90, 5211.53, 5336.79, 5381.02, 5418.77 \\
V\,\textsc{i} & 5604.90, 6531.42 \\
Y\,\textsc{ii} & 5200.41, 5205.72, 5289.81, 5402.77, 5544.61, 6795.41 \\
Zr\,\textsc{i} & 5385.14, 6445.74 \\
    \hline
    \end{tabular}
    \tablefoot{Species and wavelength of spectral lines for which abundances were measured by the WG10 nodes for the WG10 setups. See the Gaia-ESO linelist \citep{heiterLL} for the complete information per line.}
  \label{tab:spectralines}
  }
\end{table*}

%\begin{table*}
%\input{velocities_giraffe.tex}
%\input{velocities_uves.tex}
%\caption{}\label{6xGaia DR2}
%\end{table*}

\section{Working Group 10 element abundance corrections}
% Table generated by Excel2LaTeX from sheet 'tab-wg10_corrections'
%\onecolumn
%\longtab{
\begin{spacing}{.98}
{\scriptsize
\setlength\tabcolsep{1pt}
\setlength{\LTcapwidth}{\textwidth}
\begin{longtable}{ccccccccccccc}

\caption{Coefficients of WG10 bias corrections.}
\label{tab:abun_corrections}
\\
\hline\hline
SETUP & NODE & ELEMENT & IP & a0 & a1 & a2 & a3 & min(IP) & max(IP) & Median$\Delta$ & $\sigma\Delta$ & Calibration \\
\hline
\endfirsthead
\caption{continued.} \\
 \hline\hline
SETUP & NODE & ELEMENT & IP & a0 & a1 & a2 & a3 & min(IP) & max(IP) & Median$\Delta$ & $\sigma\Delta$ & Calibration \\
\hline 
\endhead
\hline
\endfoot
    HR10  & CAUP  & CA1   & FEH\_D & -0.0128635771 & -0.0853183007 & -     & -     & -2.52 & 0.32  & 0.0000 & 0.1878 & WG11xmat \\
    HR10  & CAUP  & CA1   & FEH\_G & 0.0671735328 & 0.0720870430 & -     & -     & -2.00 & 0.21  & -0.0100 & 0.1762 & WG11xmat \\
    HR10  & CAUP  & CO1   & FEH\_D & -0.1904706240 & 0.0279439500 & -     & -     & -0.30 & 0.41  & -0.1800 & 0.0909 & WG11xmat \\
    HR10  & CAUP  & CO1   & FEH\_G & -0.0299998950 & -     & -     & -     & -1.21 & -0.21 & -0.0300 & 0.0824 & WG11xmat \\
    HR10  & CAUP  & CR1   & LOGG\_D & -5.6450875166 & 2.8419813136 & -0.3506320046 & -     & 3.53  & 4.59  & 0.0800 & 0.1470 & WG11xmat \\
    HR10  & CAUP  & CR1   & FEH\_G & 0.0624965243 & -0.0223026185 & -0.0403518382 & -     & -2.62 & 0.21  & 0.0200 & 0.3286 & WG11xmat \\
    HR10  & CAUP  & FE1   & LOGG\_D & -0.0379078330 & -0.0468408820 & 0.0117258380 & -     & 0.95  & 4.68  & -0.0600 & 0.1338 & WG11xmat \\
    HR10  & CAUP  & FE1   & FEH\_G & -0.0388349210 & 0.0275701220 & -0.0078109530 & -     & -2.52 & 0.04  & -0.0500 & 0.1593 & WG11xmat \\
    HR10  & CAUP  & MN1   & FEH\_D & -0.2178134436 & -0.1944580752 & -     & -     & -0.25 & 0.45  & -0.2200 & 0.1458 & WG11xmat \\
    HR10  & CAUP  & MN1   & FEH\_G & -0.3584931025 & -0.0358944088 & -     & -     & -1.54 & 0.21  & -0.3100 & 0.1742 & WG11xmat \\
    HR10  & CAUP  & NI1   & FEH\_D & -0.0322487332 & -0.2028706838 & -     & -     & -0.70 & 0.45  & 0.0100 & 0.1130 & WG11xmat \\
    HR10  & CAUP  & NI1   & FEH\_G & -0.0539072155 & -0.0685051063 & -     & -     & -1.54 & 0.21  & 0.0200 & 0.1747 & WG11xmat \\
    HR10  & CAUP  & SC2   & FEH\_D & -0.1411620289 & -0.1317874827 & -     & -     & -1.10 & 0.45  & -0.1300 & 0.1385 & WG11xmat \\
    HR10  & CAUP  & SC2   & FEH\_G & -0.3906970169 & -0.5386362834 & -0.1670547663 & -     & -1.10 & 0.21  & -0.0400 & 0.1691 & WG11xmat \\
    HR10  & CAUP  & TI2   & FEH\_D & -0.1060732326 & -0.0850312307 & -     & -     & -2.52 & 0.45  & -0.1100 & 0.1078 & WG11xmat \\
    HR10  & CAUP  & TI2   & FEH\_G & -0.3173366324 & -0.3105625844 & -0.0567801729 & -     & -2.62 & 0.21  & -0.0800 & 0.2249 & WG11xmat \\
    HR10  & Lumba & C1    & FEH\_D & 0.0732768207 & -0.0491589552 & -     & -     & -0.70 & 0.45  & 0.0500 & 0.1135 & WG11xmat \\
    HR10  & Lumba & C1    & OFFSET\_G & 0.2500 & -     & -     & -     & -     & -     & 0.2500 & 0.0683 & WG11xmat \\
    HR10  & Lumba & CO1   & OFFSET\_D & -0.0500 & -     & -     & -     & -     & -     & -0.0500 & 0.1646 & WG11xmat \\
    HR10  & Lumba & CO1   & FEH\_G & -0.0313088310 & 0.0788521830 & 0.1184216920 & -     & -1.26 & 0.04  & 0.0200 & 0.1091 & WG11xmat \\
    HR10  & Lumba & CR1   & TEFF\_D & -16.7383167700 & 0.0061316870 & -0.0000005546 & -     & 4842  & 6483  & 0.1100 & 0.1228 & WG11xmat \\
    HR10  & Lumba & CR1   & TEFF\_G & 6.7606788150 & -0.0022318690 & 0.0000001747 & -     & 4423  & 6720  & 0.0300 & 0.2739 & WG11xmat \\
    HR10  & Lumba & FE1   & FEH\_G & 0.0976131640 & 0.0838707160 & 0.0123411550 & -     & -2.52 & 0.37  & 0.0700 & 0.2463 & WG11xmat \\
    HR10  & Lumba & FE1   & LOGG\_D & 4.3500420690 & -2.2274475100 & 0.2789431000 & -     & 3.45  & 4.68  & -0.0700 & 0.1282 & WG11xmat \\
    HR10  & Lumba & MG1   & FEH   & 0.0464074730 & -0.0687944546 & -     & -     & -1.10 & 0.45  & 0.0600 & 0.1221 & WG11xmat \\
    HR10  & Lumba & MN1   & FEH   & -0.0721551932 & -0.2434403206 & -0.2944977537 & -     & -1.10 & 0.30  & -0.1200 & 0.2009 & WG11xmat \\
    HR10  & Lumba & MO1   & FEH   & 0.4932369290 & 1.3317375180 & -0.6803415420 & -3.55216122 & -0.49 & 0.35  & 0.5000 & 3.8600 & WG11-full \\
    HR10  & Lumba & ND2   & OFFSET\_D & 0.7200 & -     & -     & -     & -     & -     & 0.7200 & 0.1259 & WG11xmat \\
    HR10  & Lumba & ND2   & FEH\_G & 0.5507316480 & 0.6873607070 & 0.2871853550 & -     & -1.23 & 0.04  & 0.2400 & 0.1836 & WG11xmat \\
    HR10  & Lumba & NI1   & FEH\_D & 0.1131103971 & 0.0412906286 & -     & -     & -0.40 & 0.45  & 0.1000 & 0.1028 & WG11xmat \\
    HR10  & Lumba & NI1   & OFFSET\_G & 0.1800 & -     & -     & -     & -     & -     & 0.1800 & 0.1969 & WG11xmat \\
    HR10  & Lumba & SC2   & OFFSET\_D & -0.1100 & -     & -     & -     & -     & -     & -0.1100 & 0.1490 & WG11xmat \\
    HR10  & Lumba & SC2   & OFFSET\_G & 0.0700 & -     & -     & -     & -     & -     & 0.0700 & 0.1842 & WG11xmat \\
    HR10  & Lumba & V1    & OFFSET\_D & -0.0800 & -     & -     & -     & -     & -     & -0.0800 & 0.0283 & WG11xmat \\
    HR10  & Lumba & V1    & OFFSET\_G & -0.0400 & -     & -     & -     & -     & -     & -0.0400 & 0.0752 & WG11xmat \\
    HR10  & Lumba & Y2    & OFFSET\_D & -0.2300 & -     & -     & -     & -0.42 & 0.45  & -0.2300 & 0.1245 & WG11xmat \\
    HR10  & Lumba & Y2    & OFFSET\_G & -0.0900 & -     & -     & -     & -1.27 & 0.06  & -0.0900 & 0.1608 & WG11xmat \\
    HR10  & Lumba & ZR1   & FEH   & 0.1523604990 & 0.5486256480 & 1.1509851220 & 0.65947002 & -1.25 & 0.37  & 0.1300 & 4.4100 & WG11-full \\
    HR10  & Vilnius & AL1   & FEH\_D & 0.0598238978 & 0.0004466455 & -     & -     & -0.70 & 0.45  & 0.0800 & 0.2045 & WG11xmat \\
    HR10  & Vilnius & AL1   & FEH\_G & 0.1159365538 & -0.0336255123 & -     & -     & -1.50 & 0.21  & 0.1800 & 0.1879 & WG11xmat \\
    HR10  & Vilnius & CA1   & FEH\_D & 0.1303592641 & 0.0529698156 & -     & -     & -0.70 & 0.45  & 0.1400 & 0.1557 & WG11xmat \\
    HR10  & Vilnius & CA1   & OFFSET\_G & 0.0400 & -     & -     & -     & -1.10 & 0.21  & 0.0400 & 0.1886 & WG11xmat \\
    HR10  & Vilnius & CO1   & FEH\_D & 0.1415773330 & 0.0838384140 & -0.0796160610 & -     & -0.71 & 0.41  & 0.1200 & 0.1454 & WG11xmat \\
    HR10  & Vilnius & CO1   & FEH\_G & 0.2384163850 & 0.2857059710 & 0.1121185940 & -     & -1.26 & 0.37  & 0.0600 & 0.2132 & WG11xmat \\
    HR10  & Vilnius & CR1   & LOGG\_D & 3.1411860370 & -1.5389358300 & 0.1933153520 & -     & 3.45  & 4.68  & 0.0900 & 0.1543 & WG11xmat \\
    HR10  & Vilnius & CR1   & LOGG\_G & 0.3495480500 & -0.4913832760 & 0.1403916450 & -     & 0.95  & 3.39  & -0.0200 & 0.2645 & WG11xmat \\
    HR10  & Vilnius & FE1   & LOGG\_D & 2.6557655300 & -1.3217719570 & 0.1695725040 & -     & 3.45  & 4.68  & 0.1000 & 0.1555 & WG11xmat \\
    HR10  & Vilnius & FE1   & FEH\_G & 0.2328753010 & 0.1948311820 & 0.0523083140 & -     & -2.52 & 0.37  & 0.0900 & 0.2349 & WG11xmat \\
    HR10  & Vilnius & FE2   & LOGG  & 0.6344379240 & -0.2852883300 & 0.0442309150 & -     & 0.95  & 4.68  & 0.2200 & 0.2018 & WG11xmat \\
    HR10  & Vilnius & MG1   & OFFSET\_D & 0.1600 & -     & -     & -     & -0.70 & 0.45  & 0.1600 & 0.1779 & WG11xmat \\
    HR10  & Vilnius & MG1   & FEH\_G & -0.0406796642 & -0.3352504432 & -     & -     & -1.10 & 0.21  & 0.2700 & 0.2601 & WG11xmat \\
    HR10  & Vilnius & MN1   & TEFF\_D & -1.4101857502 & 0.0004340610 & -0.0000000316 & -     & 4256  & 6756  & 0.0100 & 0.1615 & WG11xmat \\
    HR10  & Vilnius & MN1   & FEH\_G & 0.0285548670 & 0.1218215121 & -     & -     & -1.54 & 0.21  & -0.0900 & 0.1970 & WG11xmat \\
    HR10  & Vilnius & NI1   & LOGG\_D & 6.1505396041 & -3.1638494044 & 0.4061924846 & -     & 3.53  & 4.51  & 0.0500 & 0.1508 & WG11xmat \\
    HR10  & Vilnius & NI1   & TEFF\_G & 9.0197287067 & -0.0035566944 & 0.0000003496 & -     & 4189  & 5207  & 0.0000 & 0.2335 & WG11xmat \\
    HR10  & Vilnius & TI1   & FEH\_D & 0.1215496359 & 0.0156832702 & -     & -     & -0.70 & 0.45  & 0.1600 & 0.2127 & WG11xmat \\
    HR10  & Vilnius & TI1   & FEH\_G & -0.0248537984 & -0.2107111023 & -     & -     & -0.70 & 0.00  & 0.2000 & 0.2336 & WG11xmat \\
    HR10  & Vilnius & TI2   & FEH\_G & 0.3898855087 & 0.0237904194 & -     & -     & -2.62 & 0.21  & 0.3300 & 0.2273 & WG11xmat \\
    HR10  & Vilnius & TI2   & FEH\_D & 0.1096975581 & -0.0688373436 & -0.0399396966 & -     & -2.03 & 0.45  & 0.1100 & 0.2071 & WG11xmat \\
    HR10  & Vilnius & Y2    & OFFSET\_D & -0.0200 & -     & -     & -     & -0.70 & 0.45  & -0.0200 & 0.2140 & WG11xmat \\
    HR10  & Vilnius & Y2    & OFFSET\_G & 0.0700 & -     & -     & -     & -1.27 & 0.20  & 0.0700 & 0.2044 & WG11xmat \\
    HR15N & CAUP  & CA1   & FEH\_D & -0.0987282970 & -0.0744936740 & 0.0912719410 & -     & -1.01 & 0.32  & -0.0500 & 0.1928 & WG11xmat \\
    HR15N & CAUP  & CA1   & FEH\_G & -0.0558260260 & -0.0515666940 & -0.0056638010 & -     & -2.54 & 0.29  & -0.0400 & 0.1498 & WG11xmat \\
    HR15N & CAUP  & FE1   & TEFF\_D & 1.5395357310 & -0.0005183000 & 0.0000000444 & -     & 4092  & 6524  & 0.0400 & 0.1338 & WG11xmat \\
    HR15N & CAUP  & FE1   & TEFF\_G & 6.0155818810 & -0.0025553130 & 0.0000002721 & -     & 4171  & 5132  & 0.0400 & 0.1189 & WG11xmat \\
    HR15N & CAUP  & NI1   & TEFF\_D & -0.9724628770 & 0.0003342880 & -0.0000000277 & -     & 4092  & 6524  & 0.0200 & 0.1552 & WG11xmat \\
    HR15N & CAUP  & NI1   & TEFF\_G & 9.5562424000 & -0.0041799340 & 0.0000004513 & -     & 4171  & 5132  & -0.0800 & 0.1454 & WG11xmat \\
    HR15N & CAUP  & SC2   & FEH\_D & -0.1003669214 & -0.1982793021 & -     & -     & -1.00 & 0.00  & -0.0900 & 0.1188 & WG11xmat \\
    HR15N & CAUP  & SC2   & FEH\_G & -0.1855341097 & -0.1870818927 & -0.0423925519 & -     & -1.66 & 0.33  & -0.1300 & 0.1105 & WG11xmat \\
    HR15N & CAUP  & SI1   & FEH\_D & 0.0506171736 & 0.0450048868 & -     & -     & -0.80 & 0.49  & 0.0500 & 0.1275 & WG11xmat \\
    HR15N & CAUP  & SI1   & FEH\_G & 0.0823885330 & 0.2563482820 & 0.3457082370 & -     & -0.86 & 0.29  & 0.0600 & 0.1468 & WG11xmat \\
    HR15N & EPINARBO & AL1   & TEFF\_D & -0.9569700663 & 0.0005238327 & -0.0000000558 & -     & 5003  & 6710  & 0.2060 & 0.2588 & WG11xmat \\
    HR15N & EPINARBO & AL1   & FEH\_G & 0.3162735083 & 0.2239841044 & 0.0792369502 & -     & -1.66 & 0.33  & 0.2550 & 0.1659 & WG11xmat \\
    HR15N & EPINARBO & BA2   & FEH\_D & 0.0985018910 & -0.7698388700 & -0.5755655770 & -0.13481981 & -2.09 & 0.32  & 0.3130 & 2.9850 & WG11-full \\
    HR15N & EPINARBO & BA2   & FEH\_G & -0.4055081900 & 0.0141792960 & 1.4568006990 & 0.66517806 & -2.54 & 0.17  & 0.0560 & 2.6590 & WG11-full \\
    HR15N & EPINARBO & C1    & OFFSET\_D & -0.0800 & -     & -     & -     & -     & -     & -0.0800 & 0.1779 & WG11xmat \\
    HR15N & EPINARBO & C1    & FEH\_G & 0.0846544789 & -0.2307559980 & -     & -     & -0.42 & 0.33  & 0.0040 & 0.2960 & WG11xmat \\
    HR15N & EPINARBO & CA1   & OFFSET\_D & 0.1740 & -     & -     & -     & -0.56 & 0.49  & 0.1740 & 0.1957 & WG11xmat \\
    HR15N & EPINARBO & CA1   & FEH\_G & 0.1069150534 & 0.2156591825 & 0.1333802684 & -     & -1.66 & 0.29  & 0.0640 & 0.1319 & WG11xmat \\
    HR15N & EPINARBO & CO1   & FEH\_D & 0.1945363800 & -0.1662533510 & 0.3524026810 & -     & -2.20 & 0.36  & 0.1980 & 0.3100 & WG11xmat \\
    HR15N & EPINARBO & CO1   & FEH\_G & 0.2644097580 & -0.0484794770 & -0.1019780170 & -     & -1.66 & 0.29  & 0.2180 & 0.2778 & WG11xmat \\
    HR15N & EPINARBO & EU2   & OFFSET\_D & 0.0590 & -     & -     & -     & -     & -     & 0.0590 & 0.1505 & WG11xmat \\
    HR15N & EPINARBO & EU2   & FEH\_G & 0.1269190636 & 0.1666165730 & -     & -     & -1.58 & 0.33  & 0.0250 & 0.1917 & WG11xmat \\
    HR15N & EPINARBO & FE1   & FEH\_D & 0.0462124040 & -0.2231451800 & 0.4096522430 & -     & -0.75 & 0.36  & 0.0790 & 0.3726 & WG11xmat \\
    HR15N & EPINARBO & FE1   & FEH\_G & 0.0987415140 & 0.2073014350 & 0.1384381260 & -     & -2.20 & 0.36  & 0.0810 & 0.2919 & WG11xmat \\
    HR15N & EPINARBO & FE2   & FEH\_D & 0.2666048930 & 0.0828850200 & -     & -     & -2.20 & 0.36  & 0.2730 & 0.3504 & WG11xmat \\
    HR15N & EPINARBO & FE2   & FEH\_G & 0.3398705230 & 0.1513853770 & -     & -     & -1.66 & 0.29  & 0.2750 & 0.2321 & WG11xmat \\
    HR15N & EPINARBO & ND2   & FEH   & 1.5361205339 & -0.0615894571 & -0.2475007921 & -0.23830388 & -1.76 & 0.40  & 1.5890 & 2.4480 & WG11-full \\
    HR15N & EPINARBO & NI1   & FEH\_D & 0.0104620694 & -0.1388466011 & 0.4247540473 & -     & -0.50 & 0.49  & 0.0710 & 0.2495 & WG11xmat \\
    HR15N & EPINARBO & NI1   & LOGG\_G & 0.0377553625 & -0.0227948910 & 0.0089080360 & -     & 0.98  & 3.35  & 0.0420 & 0.1586 & WG11xmat \\
    HR15N & EPINARBO & S1    & TEFF  & 20.2390962443 & -0.0078025044 & 0.0000007638 & -     & 4651  & 6184  & 0.6020 & 0.7445 & WG11xmat \\
    HR15N & EPINARBO & SC2   & TEFF\_D & 0.8796531494 & -0.0001731191 & -     & -     & 5003  & 6300  & -0.1330 & 0.1535 & WG11xmat \\
    HR15N & EPINARBO & SC2   & TEFF\_G & 0.6089881206 & -0.0001382218 & -     & -     & 4269  & 5171  & -0.0530 & 0.1050 & WG11xmat \\
    HR15N & EPINARBO & SI1   & FEH\_D & 0.1413808311 & 0.0756314018 & 0.3421655219 & -     & -0.57 & 0.49  & 0.1300 & 0.2265 & WG11xmat \\
    HR15N & EPINARBO & SI1   & FEH\_G & 0.2668795320 & 0.2973452999 & 0.1215426521 & -     & -1.66 & 0.33  & 0.2270 & 0.1737 & WG11xmat \\
    HR15N & EPINARBO & SR1   & OFFSET & 1.6330 & -     & -     & -     & -     & -     & 1.6330 & -     & Scaled Solar \\
    HR15N & EPINARBO & TI1   & TEFF\_D & 17.1293465052 & -0.0064595794 & 0.0000006138 & -     & 5003  & 6300  & 0.3030 & 0.4986 & WG11xmat \\
    HR15N & EPINARBO & TI1   & TEFF\_G & -9.9478597542 & 0.0044616249 & -0.0000004934 & -     & 4269  & 5100  & 0.1020 & 0.1317 & WG11xmat \\
    HR15N & EPINARBO & V1    & FEH\_D & 0.0867662461 & -0.3446519198 & -     & -     & -0.51 & 0.49  & 0.1100 & 0.4483 & WG11xmat \\
    HR15N & EPINARBO & V1    & TEFF\_G & 0.2239432282 & -0.0000467122 & -     & -     & 4269  & 5095  & 0.0020 & 0.1956 & WG11xmat \\
    HR15N & EPINARBO & Y2    & LOGG\_D & 18.5786753232 & -10.6917378985 & 1.5217598504 & -     & 3.46  & 4.20  & -0.0660 & 0.5153 & WG11xmat \\
    HR15N & EPINARBO & Y2    & LOGG\_G & -0.1684343003 & 0.0395075169 & -     & -     & 0.98  & 2.97  & -0.0950 & 0.2638 & WG11xmat \\
    HR15N & EPINARBO & ZR1   & OFFSET & 1.1620 & -     & -     & -     & -     & -     & 1.1620 & 4.9400 & WG11-full \\
    HR15N & Lumba & AL1   & FEH\_D & 0.1258188370 & 0.0201138850 & 0.0890347210 & -     & -1.66 & 0.34  & 0.1400 & 0.1370 & WG11xmat \\
    HR15N & Lumba & AL1   & FEH\_G & 0.1615605640 & 0.0683883570 & 0.0985345720 & -     & -1.66 & 0.30  & 0.1700 & 0.1050 & WG11xmat \\
    HR15N & Lumba & BA2   & LOGG\_D & 6.2405706048 & -2.8519959072 & 0.3395505367 & -     & 3.46  & 4.58  & 0.2800 & 0.2128 & WG11xmat \\
    HR15N & Lumba & BA2   & LOGG\_G & 1.2262712441 & -0.2678149175 & -     & -     & 0.98  & 3.35  & 0.5900 & 0.2151 & WG11xmat \\
    HR15N & Lumba & C1    & OFFSET\_D & 0.1234004990 & 0.0507172730 & -     & -     & -0.43 & 0.36  & 0.1300 & 0.1383 & WG11xmat \\
    HR15N & Lumba & C1    & OFFSET\_G & 0.2393637190 & 0.3464880590 & -     & -     & -0.51 & 0.29  & 0.2200 & 0.1388 & WG11xmat \\
    HR15N & Lumba & CA1   & OFFSET\_D & 0.1400 & -     & -     & -     & -0.70 & 0.49  & 0.1400 & 0.1595 & WG11xmat \\
    HR15N & Lumba & CA1   & FEH\_G & 0.1455244593 & -0.0630665992 & -     & -     & -1.10 & 0.33  & 0.1800 & 0.1932 & WG11xmat \\
    HR15N & Lumba & CO1   & TEFF\_D & -0.7148061500 & 0.0001486460 & -     & -     & 4113  & 6452  & 0.1000 & 0.1431 & WG11xmat \\
    HR15N & Lumba & CO1   & FEH\_G & 0.0176560850 & -0.1413575310 & -0.0389629720 & -     & -1.30 & 0.30  & 0.0500 & 0.0842 & WG11xmat \\
    HR15N & Lumba & EU2   & OFFSET\_D & 0.3400 & -     & -     & -     & -     & -     & 0.3400 & 0.1436 & WG11xmat \\
    HR15N & Lumba & EU2   & FEH\_G & 0.1914570730 & 0.2859310470 & 0.2835082150 & -     & -1.30 & 0.29  & 0.1400 & 0.1288 & WG11xmat \\
    HR15N & Lumba & NI1   & FEH\_D & 0.1008589225 & 0.0544224804 & -     & -     & -1.33 & 0.49  & 0.0900 & 0.1769 & WG11xmat \\
    HR15N & Lumba & NI1   & FEH\_G & 0.2560589917 & 0.0509889047 & -     & -     & -1.66 & 0.33  & 0.2400 & 0.2184 & WG11xmat \\
    HR15N & Lumba & SI1   & LOGG\_D & 0.1048324560 & 0.0224807643 & -     & -     & 3.46  & 4.58  & 0.1900 & 0.1059 & WG11xmat \\
    HR15N & Lumba & SI1   & TEFF\_G & 24.1362275043 & -0.0093215520 & 0.0000009122 & -     & 4400  & 5095  & 0.4500 & 0.4346 & WG11xmat \\
    HR15N & Lumba & TI1   & FEH   & 0.0851134142 & -0.0407456089 & -     & -     & -2.03 & 0.49  & 0.1000 & 0.2972 & WG11xmat \\
    HR21  & CAUP  & AL1   & FEH\_D & -0.1330458850 & 0.0144045030 & 0.1180994800 & -     & -1.33 & 0.45  & -0.1200 & 0.1356 & WG11xmat \\
    HR21  & CAUP  & AL1   & FEH\_G & 0.0724923910 & 0.1917499230 & 0.1070101150 & -     & -1.63 & 0.21  & -0.0100 & 0.1311 & WG11xmat \\
    HR21  & CAUP  & FE1   & FEH\_D & 0.0308106900 & 0.2023866500 & 0.0894924000 & -     & -0.90 & 0.47  & -0.0060 & 0.1830 & HR21toHR10-WG11 \\
    HR21  & CAUP  & FE1   & FEH\_G & 0.2812608200 & 0.4460150900 & 0.0927387200 & -     & -0.80 & 0.42  & 0.1580 & 0.2360 & HR21toHR10-WG11 \\
    HR21  & CAUP  & MG1   & OFFSET\_D & -0.0100 & -     & -     & -     & -2.03 & 0.45  & -0.0100 & 0.1694 & WG11xmat \\
    HR21  & CAUP  & MG1   & FEH\_G & -0.0149402040 & -0.1185022070 & 0.0415395750 & -     & -1.10 & 0.21  & 0.0700 & 0.2260 & WG11xmat \\
    HR21  & Lumba & AL1   & FEH\_D & -0.2658367920 & 0.0818413460 & 0.3999028370 & -     & -0.70 & 0.45  & -0.2300 & 0.1332 & WG11xmat \\
    HR21  & Lumba & AL1   & FEH\_G & -0.2010914430 & 0.1827250330 & 0.1375634280 & -     & -1.63 & 0.21  & -0.2300 & 0.1854 & WG11xmat \\
    HR21  & Lumba & FE1   & FEH\_D & 0.1221244200 & -0.0673754000 & -0.0510338100 & -     & -2.52 & 0.45  & 0.1320 & 0.1570 & HR21toHR10-WG11 \\
    HR21  & Lumba & FE1   & FEH\_G & 0.2208933400 & 0.0717466300 & 0.0159230300 & -     & -3.11 & 0.42  & 0.1920 & 0.1700 & HR21toHR10-WG11 \\
    HR21  & Lumba & MG1   & OFFSET\_D & 0.1200 & -     & -     & -     & -2.03 & 0.45  & 0.1200 & 0.1850 & WG11xmat \\
    HR21  & Lumba & MG1   & FEH\_G & 0.0967134805 & -0.0665194907 & 0.0651249740 & -     & -1.10 & 0.21  & 0.2000 & 0.2860 & WG11xmat \\
    HR21  & Lumba & SI1   & LOGG\_D & -0.3013328157 & 0.1152338589 & -     & -     & 3.53  & 4.59  & 0.1800 & 0.1160 & WG11xmat \\
    HR21  & Lumba & SI1   & FEH\_G & 0.5240188932 & 0.9555999724 & 0.5280106112 & -     & -1.10 & 0.00  & 0.2000 & 0.4609 & WG11xmat \\
    HR21  & Vilnius & AL1   & FEH\_D & 0.0123871914 & 0.0024776950 & -     & -     & -1.33 & 0.45  & 0.0300 & 0.1644 & WG11xmat \\
    HR21  & Vilnius & AL1   & FEH\_G & 0.3107618178 & 0.1954399937 & -     & -     & -1.27 & 0.21  & 0.1600 & 0.1964 & WG11xmat \\
    HR21  & Vilnius & CA2   & FEH\_G & 0.1236401000 & 0.1339087600 & -     & -     & -2.53 & 0.42  & 0.0420 & 0.2710 & HR21toHR10-WG11 \\
    HR21  & Vilnius & CA2   & FEH\_D & -0.0241594400 & 0.2321338100 & 0.1195865000 & -     & -0.70 & 0.47  & -0.0690 & 0.1870 & HR21toHR10-WG11 \\
    HR21  & Vilnius & FE1   & FEH\_D & 0.1630460800 & 0.0028094900 & -0.0104126700 & -     & -2.57 & 0.47  & 0.1600 & 0.1330 & HR21toHR10-WG11 \\
    HR21  & Vilnius & FE1   & FEH\_G & 0.5288046000 & 0.6799116700 & 0.2248112100 & -     & -3.17 & 0.93  & 0.2890 & 0.2610 & HR21toHR10-WG11 \\
    HR21  & Vilnius & MG1   & OFFSET\_D & 0.1580 & -     & -     & -     & -2.57 & 0.47  & 0.1580 & 0.1320 & HR21toHR10-WG11 \\
    HR21  & Vilnius & MG1   & FEH\_G & 0.2158573700 & 0.4460271900 & 0.5062285700 & -     & -0.30 & 0.42  & 0.2920 & 0.2060 & HR21toHR10-WG11 \\
    HR21  & Vilnius & SI1   & LOGG\_D & 0.0299960908 & 0.0267704496 & -     & -     & 3.59  & 4.60  & 0.1200 & 0.1307 & WG11xmat \\
    HR21  & Vilnius & SI1   & FEH\_G & 0.2776062190 & 0.4798213770 & 0.2179009050 & -     & -1.27 & 0.21  & 0.0700 & 0.2174 & WG11xmat \\
    HR21  & Vilnius & TI1   & FEH\_D & 0.1499854500 & 0.0088722400 & -0.0068887000 & -     & -1.25 & 0.47  & 0.1480 & 0.1890 & HR21toHR10-WG11 \\
    HR21  & Vilnius & TI1   & FEH\_G & 0.3295874000 & 0.4103813500 & 0.3194012300 & -     & -0.50 & 0.42  & 0.2650 & 0.2430 & HR21toHR10-WG11 \\
    HR9B  & EPINARBO & CA1   & TEFF\_D & -5.6125887540 & 0.0017516880 & -0.0000001347 & -     & 5030  & 6184  & -0.0250 & 0.1675 & WG11xmat \\
    HR9B  & EPINARBO & CA1   & TEFF\_G & 31.6180845900 & -0.0134122070 & 0.0000014091 & -     & 4585  & 5375  & -0.2780 & 0.1321 & WG11xmat \\
    HR9B  & EPINARBO & CE2   & FEH\_G & -0.2340408971 & -0.6577443268 & -     & -     & -0.10 & 0.26  & -0.2980 & 0.2162 & WG11xmat \\
    HR9B  & EPINARBO & CE2   & FEH\_D & 0.0889999659 & -0.7001094103 & -     & -     & -0.57 & 0.20  & -0.0150 & 0.4717 & WG11xmat \\
    HR9B  & EPINARBO & CO1   & FEH\_D & -0.1717849057 & 0.0075862458 & 0.6403065816 & -     & -1.33 & 0.33  & -0.0970 & 0.3113 & WG11xmat \\
    HR9B  & EPINARBO & CO1   & TEFF\_G & 45.2018809500 & -0.0179687140 & 0.0000017819 & -     & 4585  & 5375  & -0.0230 & 0.1880 & WG11xmat \\
    HR9B  & EPINARBO & CR1   & LOGG\_D & -0.0278139970 & 0.0055499350 & -     & -     & 3.52  & 4.57  & 0.0100 & 0.2581 & WG11xmat \\
    HR9B  & EPINARBO & CR1   & LOGG\_G & -0.1809654460 & 0.0729570610 & -     & -     & 2.24  & 3.28  & 0.0000 & 0.1448 & WG11xmat \\
    HR9B  & EPINARBO & CR2   & OFFSET & 0.2400 & -     & -     & -     & -     & -     & 0.2400 & -     & WG11-full \\
    HR9B  & EPINARBO & CU1   & FEH   & 0.1247046739 & 0.2199904770 & 1.6158947945 & 5.83671379 & -0.34 & 0.32  & 0.1590 & 8.3430 & WG11-full \\
    HR9B  & EPINARBO & DY2   & OFFSET & 1.5720 & -     & -     & -     & -     & -     & 1.5720 & -     & Scaled Solar \\
    HR9B  & EPINARBO & FE1   & FEH\_G & 0.0232848290 & -0.0875269150 & -     & -     & -0.11 & 0.35  & 0.0110 & 0.1808 & WG11xmat \\
    HR9B  & EPINARBO & FE1   & FEH\_D & -0.0204007620 & -0.2323460920 & -0.1628028780 & -     & -1.30 & 0.10  & -0.0080 & 0.1993 & WG11xmat \\
    HR9B  & EPINARBO & FE2   & FEH   & 0.0273802910 & -0.3660831260 & -0.2672870260 & -     & -1.30 & 0.38  & 0.0340 & 0.2372 & WG11xmat \\
    HR9B  & EPINARBO & LA2   & FEH   & 0.0754013062 & -0.7898668051 & 0.5343078971 & 2.14124775 & -0.45 & 0.39  & 0.0910 & 2.4200 & WG11-full \\
    HR9B  & EPINARBO & ND2   & LOGG\_D & -0.3105725040 & 0.1304330020 & -     & -     & 3.52  & 4.41  & 0.2590 & 0.3008 & WG11xmat \\
    HR9B  & EPINARBO & ND2   & LOGG\_G & 1.5173952460 & -0.4037327140 & -     & -     & 2.24  & 3.28  & 0.4210 & 0.2801 & WG11xmat \\
    HR9B  & EPINARBO & NI1   & TEFF  & -0.9498372340 & 0.0000613150 & -     & -     & 4701  & 6117  & -0.6290 & 0.2130 & WG11xmat \\
    HR9B  & EPINARBO & PR2   & OFFSET & 0.4590 & -     & -     & -     & -     & 0     & 0.4590 & -     & WG11-full \\
    HR9B  & EPINARBO & TI1   & OFFSET\_D & 0.0980 & -     & -     & -     & 3.46  & 4.50  & 0.0980 & 0.2925 & WG11xmat \\
    HR9B  & EPINARBO & TI1   & FEH\_G & 0.0361210337 & -0.2220528902 & -     & -     & -1.50 & 0.28  & -0.0550 & 0.2197 & WG11xmat \\
    HR9B  & EPINARBO & TI2   & TEFF\_D & 1.8107969524 & -0.0003057979 & -     & -     & 5700  & 6364  & 0.0020 & 0.2637 & WG11xmat \\
    HR9B  & EPINARBO & TI2   & TEFF\_G & 0.1834844010 & -0.0000514757 & -     & -     & 4521  & 5000  & -0.0820 & 0.1278 & WG11xmat \\
    HR9B  & EPINARBO & Y2    & LOGG  & -1.7901607240 & 0.8767585000 & -0.1081206540 & -     & 2.24  & 4.57  & -0.1990 & 0.3211 & WG11xmat \\
\hline\hline

\end{longtable}
%\noalign{%
\parbox{\textwidth}{}%}
\tablefoot{For each element per node per SETUP,  the independent parameter (IP), polynomial fit coefficients, and parameter limits, of the fit of the difference between the WG10 node values and the reference values. The median and $\sigma$ of the difference is also given.  For the IP, \_G refers to giant sample, \_D refers to dwarf sample.}

}
\end{spacing}
%\twocolumn

\end{appendix}
\end{document}